\documentclass[review]{elsarticle}

\pdfoutput=1
\usepackage[hyphens]{url}
\usepackage{lineno, hyperref}
\modulolinenumbers[5]

\usepackage{caption}
\usepackage{subcaption}
\usepackage{graphicx}
\usepackage{amsmath}
\usepackage{algorithm}
\usepackage{algorithmic}
\usepackage{rotating}
\usepackage{adjustbox}
\usepackage{arydshln}
\usepackage{multirow}

\makeatletter
\newcommand*{\rom}[1]{\expandafter\@slowromancap\romannumeral #1@}
\makeatother

\journal{arXiv}

\begin{document}

\begin{frontmatter}

\title{Constructing Trajectory and Predicting Estimated Time of Arrival for Long Distance Travelling Vessels: A Probability Density-based Scanning Approach}

\author[a,b]{Deqing Zhai\corref{mycorrespondingauthor}}
\ead{dzhai001@e.ntu.edu.sg}

\author[a]{Xiuju Fu\corref{mycorrespondingauthor}}
\ead{fuxj@ihpc.a-star.edu.sg}

\author[a]{Xiao Feng Yin}
\author[a]{Haiyan Xu}
\author[a]{Wanbing Zhang}
\author[a]{Ning Li}

\cortext[mycorrespondingauthor]{Corresponding author}
\address[a]{1 Fusionopolis Way, Institute of High Performance Computing,\\ Agency for Science, Technology and Research, Singapore 138632}
\address[b]{50 Nanyang Avenue, School of Electrical and Electronic Engineering, \\ Nanyang Technological University, Singapore 639798}

\begin{abstract}
In this study, a probability density-based approach for constructing trajectories is proposed and validated through an typical use-case application: Estimated Time of Arrival (ETA) prediction given origin-destination pairs. The ETA prediction is based on physics and mathematical laws given by the extracted information of probability density-based trajectories constructed. The overall ETA prediction errors are about 0.106 days (i.e. 2.544 hours) on average with 0.549 days (i.e. 13.176 hours) standard deviation, and the proposed approach has an accuracy of 92.08\% with 0.959 R-Squared value for overall trajectories between Singapore and Australia ports selected.
\end{abstract}


\end{frontmatter}


\section{Introduction}
Overseas and international shipping highly depends on maritime logistics networks globally. According to Review of Maritime Transport 2021, over 80\% of the volume of international trade in goods is carried by sea, and the percentage is even higher for most developing countries \cite{UNCTAD2021}. Therefore, it is important to predict estimated time of arrival (ETA) of vessels accurately, so that the whole logistics networks run efficiently and smoothly with advanced scheduling and better operation arrangement. In this study, a probability density-based approach is proposed for constructing trajectories from historical ones, and physics-mathematical laws are used for calculating ETA given current dynamic and static status of target vessels.

In this study, the work is organized as follows: Section 2 presents literature reviews on this topic. Problem statement follows in Section 3. Methodology is presented in Section 4 for both trajectory construction and ETA prediction. Section 5 shows experimental results and discussion, and lastly conclusion is drawn in Section 6.

\section{Literature Review}
After literature reviews, studies of trajectories and ETA predictions basically can be classified into two categories. One is machine learning based (e.g. \cite{Bodunov2018, Myung2011, VanLint2005, Wu2003}), and another is physics and mathematical based (e.g. \cite{Fei2011, Tay2012}) for simplicity. In the study of Bodunov $et. al.$\cite{Bodunov2018}, feed-forward neural networks (FFNN) were proposed to predict ETA of vessels, and ensemble machine learning technique was used for destination prediction of vessels. The ensemble machine learning technique was based on Random Forest, Gradient Boosting Decision Trees, XGBoost Trees and Extremely Randomized Trees. The machine learning based approach achieved 90\% and 97\% in accuracy for ETA prediction (in minutes) and destination prediction, respectively. The approach was to train the FFNN models using a target variable time in minutes between the cut-off time point of training snapshot and the timestamp associated to the last point of the trajectory with normalized features between 0 and 1. However, the problem of this approach was not robust when vessels diverge from historical geo-datasets, and the prediction of n-timestamp ahead was not covered due to lack of trajectory behaviour. Additionally, the computational efforts of training these FFNN models were expensive and time consuming. On the other side, according to the study of Fei $et. al.$\cite{Fei2011}, a short-term travelling time prediction approach was proposed by using Bayesian inference-based dynamic linear model. By introducing Bayesian approach, the approach was more robust than sole machine learning models without sacrificing accuracy. The proposed approach achieved as low as 0.26 minutes in MAE. However, this work's point was to find travelling time, not exactly ETA due to unconditional and conditional time coefficients.

\section{Problem Statement}
Given the background knowledge as mentioned above, the problem here is ``How to predict ETA accurately and effectively, given object vessels' profiles?''. The vessel's profile consists of dynamic status and static information. The dynamic status refers to instantaneous speed, heading, course, position, etc. While static information is about origin, destination, vessel type, cargo type, etc.

\section{Methodology}
In this section, the methodology of solving the problem stated is presented. To solve the problem stated, there are two steps. The first step is to construct possible trajectories and relevant information that object vessels could encounter according to static and dynamic information. The second step is to predict ETA of object vessels incorporated with the extracted trajectories and relevant information. 

\subsection{Trajectory Construction}

Based on the historical journeys of vessels, their generic trajectories can be extracted and constructed with density scanning approaches, such as Lat-scanning, Lon-scanning and LatLon-scanning. Assume the historical trajectories of a particular origin-destination pair have Latitude-Longitude coordinates ($\mathbf{K}$) as follows:
	\begin{align}
		\mathbf{K} = \begin{bmatrix}
						\phi_1 & \lambda_1 \\
						\phi_2 & \lambda_2 \\
						\vdots & \vdots    \\
						\phi_m & \lambda_m
					\end{bmatrix}
	\end{align}
, where $\phi_i$ and $\lambda_i$ are the $i^{th}$ latitude and longitude of the historical trajectories in radian.

Taking Lat-scanning approach as a typical example, the idea is to scan latitudes across the historical trajectories ($\mathbf{K}$) from $\phi_{min}$ to $\phi_{max}$ with a scanning interval ($\eta$), where $\phi_{min} = \min(\{\phi_i, i = 1, 2, 3, ..., m\})$ and $\phi_{max} = \max(\{\phi_i, i = 1, 2, 3, ..., m\})$. At $k^{th}$ scanning latitude $\phi_k$, the longitude values within corresponding scanning interval are captured and the set of captured longitudes is denoted as $\mathbf{\Lambda_k}$, then the probability density of the set of captured longitudes is calculated. Thus, the longitude value with densest probability (i.e. $\lambda_k \in \mathbf{\Lambda_k}$ ) is returned as the corresponding longitude at the scanning latitude $\phi_k$. After iterations of scanning from $\phi_{min}$ to $\phi_{max}$, the constructed trajectories ($\mathbf{L}$)  will be further sorted and Kalman filtered according to its trajectory smoothness from origin to destination. Similarly, this whole process can be applied for Lon-scanning and LatLon-scanning. Thus, the constructed trajectory ($\mathbf{L}$) given a particular origin-destination pair is obtained and denoted as follows: 

	\begin{align}
		\mathbf{L} = \begin{bmatrix}
			\phi_1 & \lambda_1 \\
			\phi_2 & \lambda_2 \\
			\vdots & \vdots    \\
			\phi_n & \lambda_n
		\end{bmatrix}
		\label{Eq: L}
	\end{align}

\subsection{ETA Prediction}

Here are the mathematical steps for calculating ETA of an particular vessel, given that current position, current speed and origin-destination. As we have defined the trajectory constructed ($\mathbf{L}$) given by the origin-destination as before in Eq. \ref{Eq: L}. Thus, current position ($\mathbf{X_c}$) is defined as follows:
	\begin{align}
		\mathbf{X_c} = \begin{bmatrix}
			\phi_c & \lambda_c
		\end{bmatrix}
	\end{align}
, where $\phi_c$ and $\lambda_c$ are the current latitude and longitude of the vessel in radian. 

Given $\mathbf{X_c}$ and $\mathbf{L}$, the closest next position ($\mathbf{X_k}$) in the constructed trajectory can be easily located. Therefore, the total distance to the destination ($D$) can be formulated as follows:
	\begin{align}
		D({\mathbf{X_c}, \mathbf{L})} = \psi(\mathbf{X_c}, \mathbf{X_k}) + \sum_{i=k \in \mathbf{L}}^{n-1} \psi(\mathbf{X_i}, \mathbf{X_{i+1}}) 
	\end{align}
, where function $\psi(\cdot)$ is the function to calculate distance under great-circle of sphere given by two coordinates on the sphere with Haversine formula \cite{GreatCircleDist}. 

Any two coordinates on the Earth are denoted as $\mathbf{X_i}, \mathbf{X_j} \equiv (\phi_i, \lambda_i), (\phi_j, \lambda_j)$. According to the Haversine formula, we have equations:
	\begin{align}
		H(\theta) & = \sin^{2}(\frac{\theta}{2})  
		\label{Eq. Haversine 1} 
		\\
		H(\theta) & = H(\phi_j-\phi_i) + H(\lambda_j-\lambda_i) \cdot \cos(\phi_i) \cdot \cos(\phi_j) 
		\label{Eq. Haversine 2}
	\end{align}
, where $\theta$ is the great-circle central angle between two coordinates on Earth in radian. Thus, we derive the following equations from Eq. \ref{Eq. Haversine 1} and \ref{Eq. Haversine 2}:
	\begin{align}
		\sin^{2}(\frac{\theta}{2}) & = \sin^{2}(\frac{\phi_j-\phi_i}{2}) + \sin^{2}(\frac{\lambda_j-\lambda_i}{2}) \cdot \cos(\phi_i) \cdot \cos(\phi_j) \\
		\Rightarrow \theta & = 2 \sin^{-1} \left[  \sqrt{\sin^{2}(\frac{\phi_j-\phi_i}{2}) + \sin^{2}(\frac{\lambda_j-\lambda_i}{2}) \cdot \cos(\phi_i) \cdot \cos(\phi_j)}  \right]
	\end{align}

Hence, the function $\psi(\cdot)$ is derived as follows:
	\begin{align}
		& \psi(\mathbf{X_i}, \mathbf{X_j}) \equiv \psi((\phi_i, \lambda_i), (\phi_j, \lambda_j)) = R \cdot \theta  \\
		& = 2R \cdot \sin^{-1} \left[  \sqrt{\sin^{2}(\frac{\phi_j-\phi_i}{2}) + \sin^{2}(\frac{\lambda_j-\lambda_i}{2}) \cdot \cos(\phi_i) \cdot \cos(\phi_j)}  \right]
	\end{align}
, where $R$ is the radius of the Earth in km.

Given the total distance to the destination ($D$) and current speed ($V$) of vessel (assuming at this constant average speed), the ETA ($T$) can be easily calculated as follows:
	\begin{align}
		T = \frac{D}{V}
	\end{align}

\section{Result and Discussion}

Based on the methodologies above, numerous rounds of experiments on ETA prediction were conducted. The experiments of ETA prediction are based on the routes between three ports in Australia and one port in Singapore. The three selected ports in Australia are Adelaide, Brisbane and Perth. The reason of the three ports selected is that the trajectories of these ports to Singapore (vice versa) are typical routes covering western, southern and eastern parts of Australia in general. For each particular trajectory, there are 10 timestamps randomly selected and validated across different vessels and journeys to demonstrate the proposed approach has homogeneous and generic capability of ETA predicting without biased conditions of object vessels.

\begin{figure}[htbp]
		\centering
		\includegraphics[width=\linewidth]{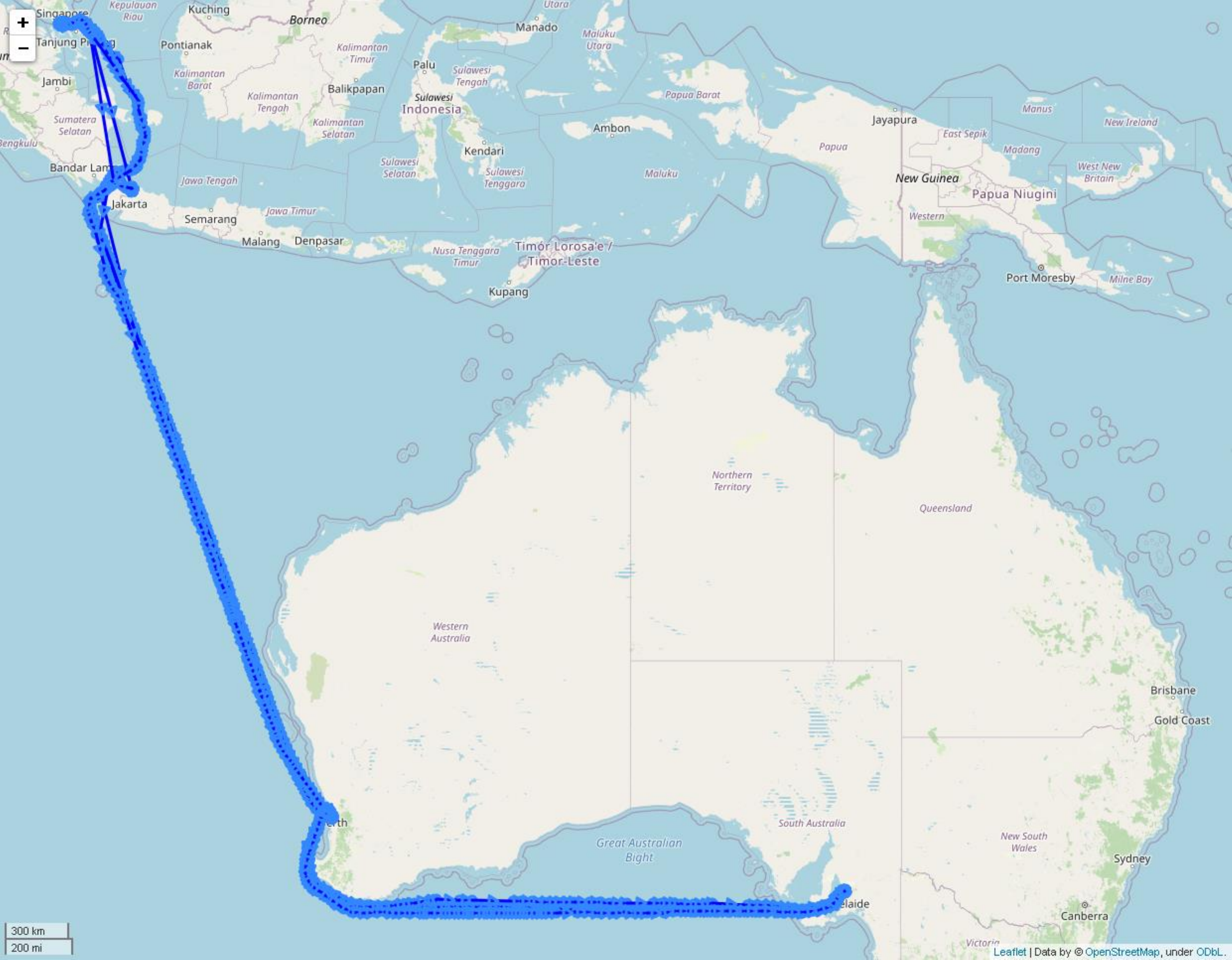}
		\caption{Actual Trajectories (Singapore $\rightarrow$ Adelaide)}
		\label{Fig: Actual Trajectories of Singapore to Adelaide}
	\end{figure}

\begin{figure}[htbp]
		\begin{subfigure}{.3\textwidth}
			\centering
			\includegraphics[width=0.98\linewidth]{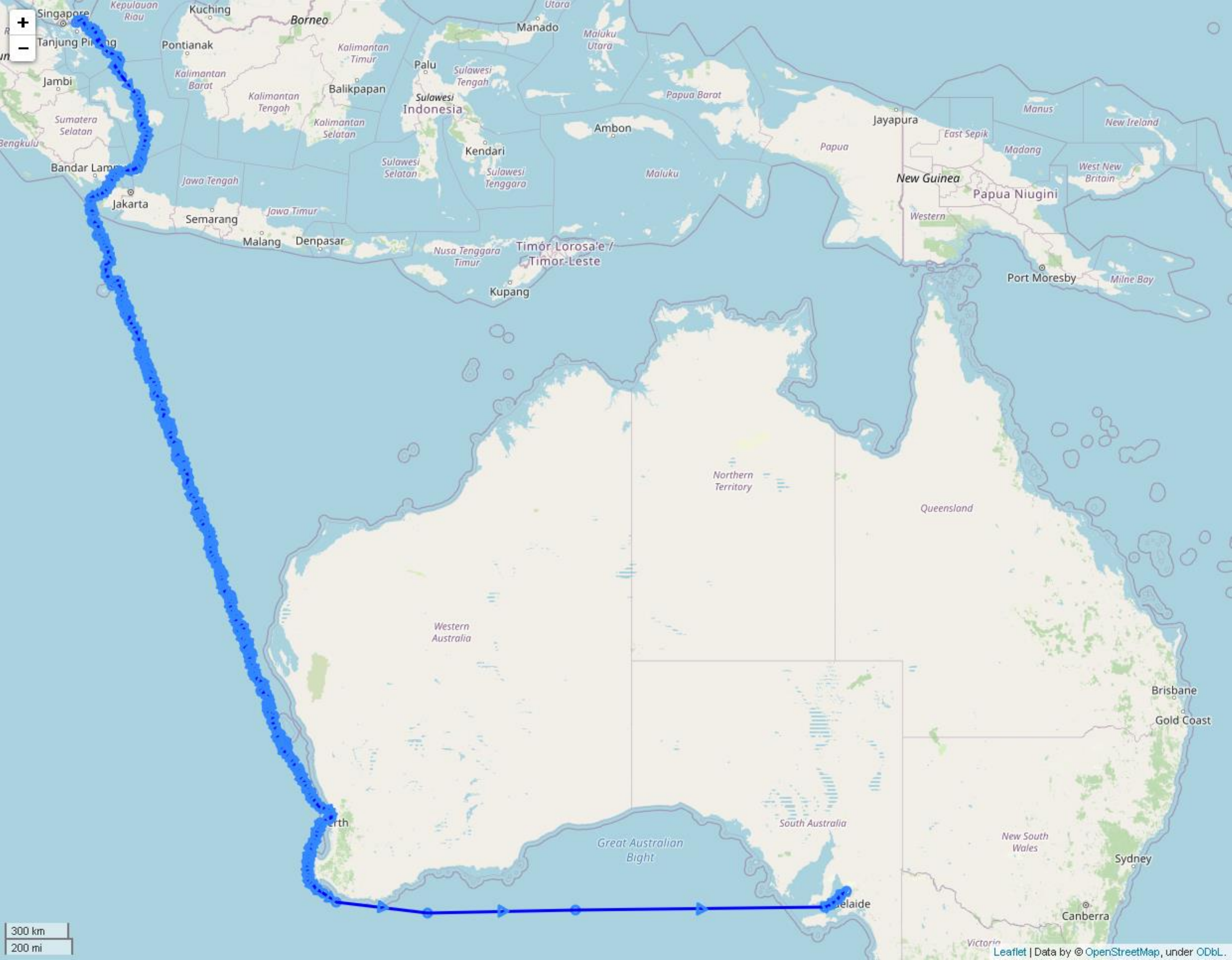}
			\caption{$\eta$=0.1, Lat-Scan}
		\end{subfigure}%
		\begin{subfigure}{.3\textwidth}
			\centering
			\includegraphics[width=0.98\linewidth]{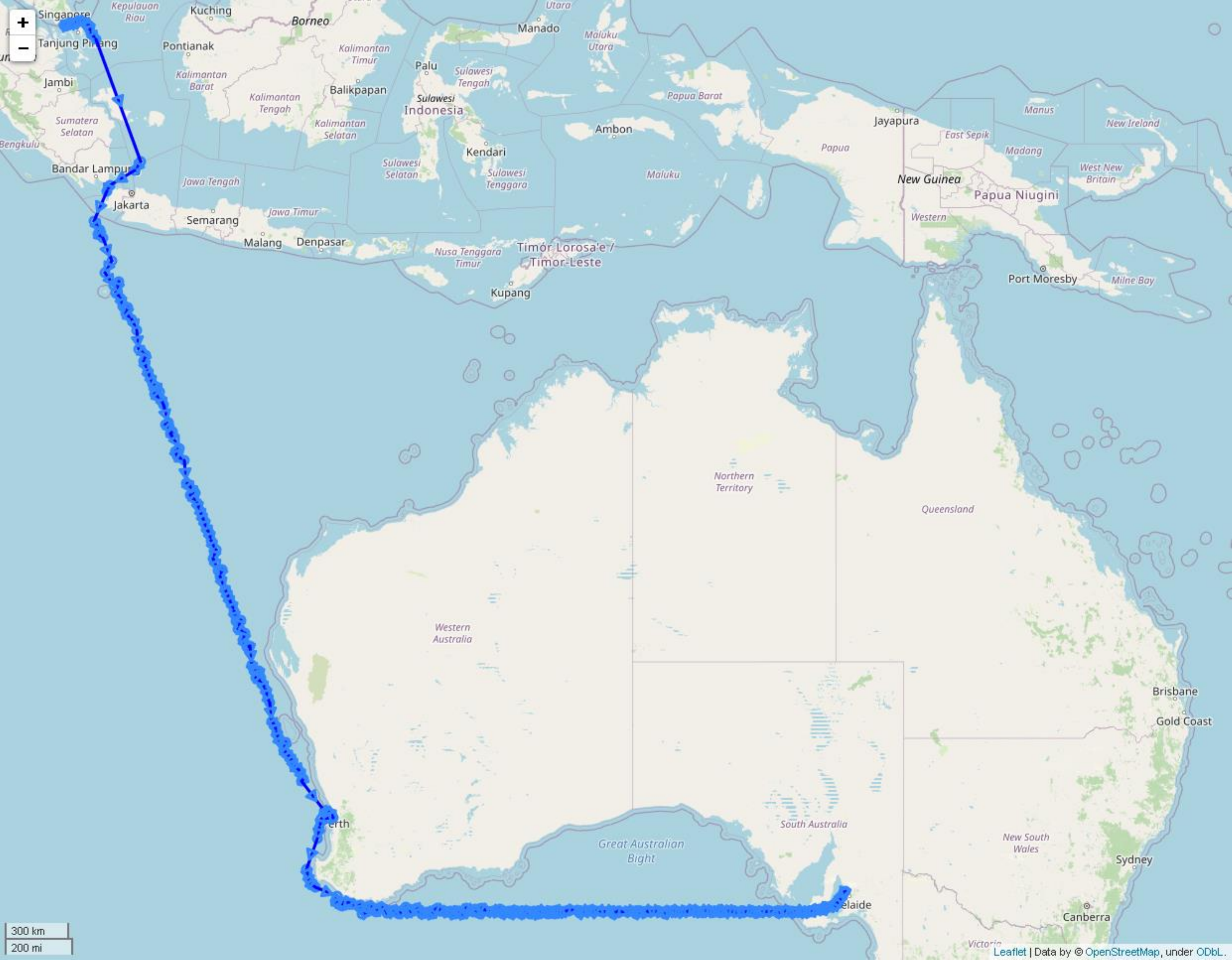}
			\caption{$\eta$=0.1, Lon-Scan}
		\end{subfigure}%
		\begin{subfigure}{.3\textwidth}
			\centering
			\includegraphics[width=0.98\linewidth]{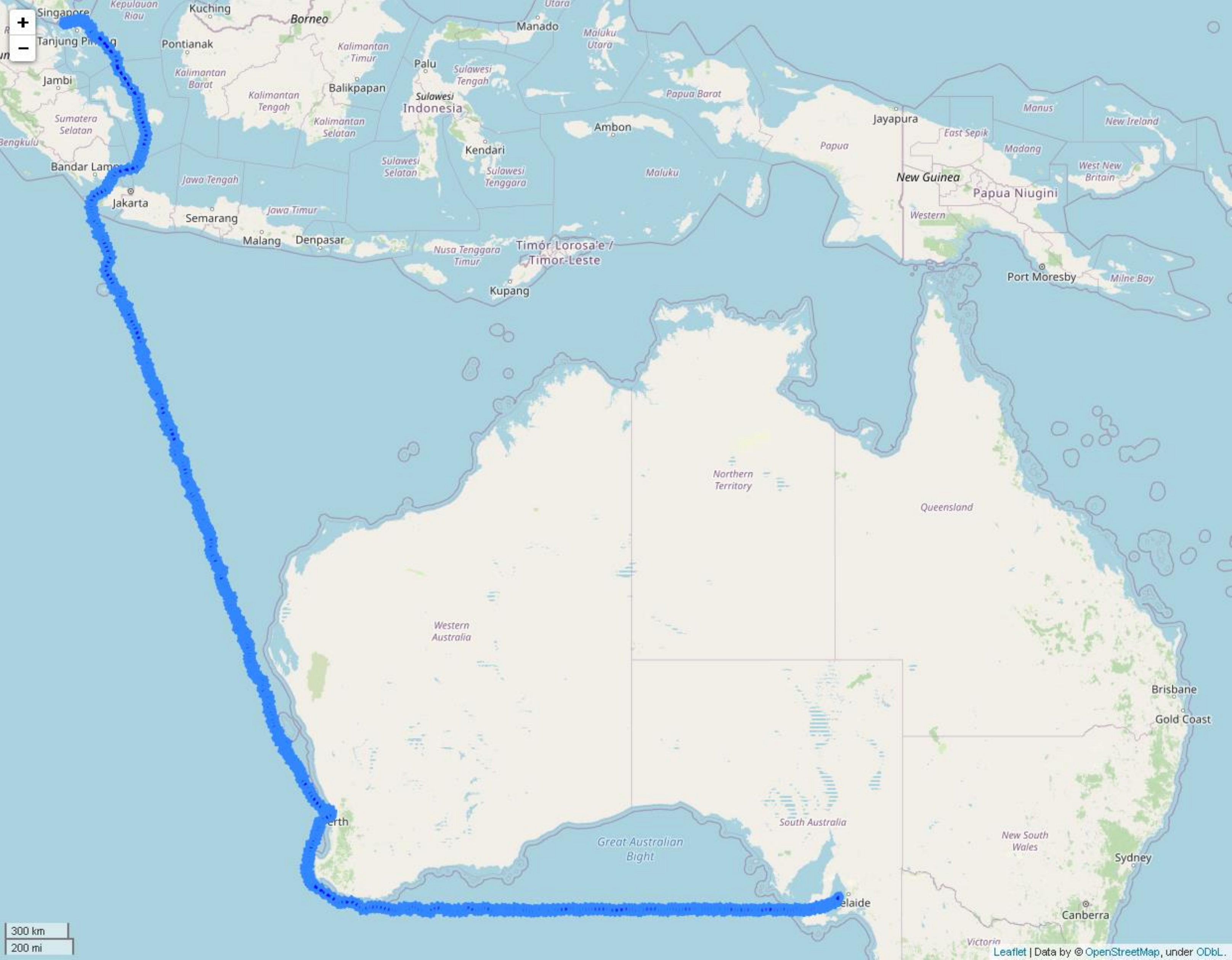}
			\caption{$\eta$=0.1, LatLon-Scan}
		\end{subfigure}

		\begin{subfigure}{.3\textwidth}
			\centering
			\includegraphics[width=0.98\linewidth]{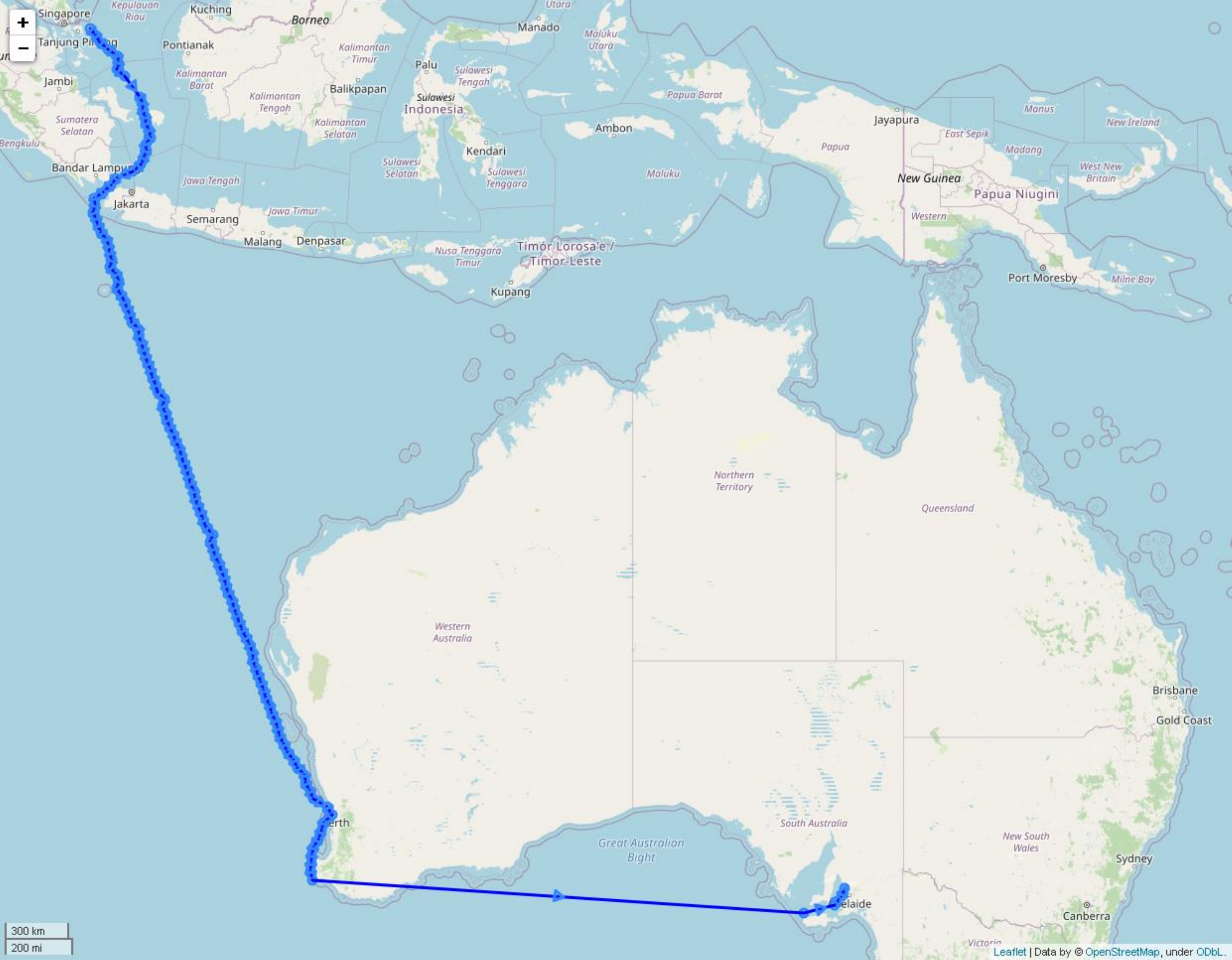}
			\caption{$\eta$=0.3, Lat-Scan}
		\end{subfigure}%
		\begin{subfigure}{.3\textwidth}
			\centering
			\includegraphics[width=0.98\linewidth]{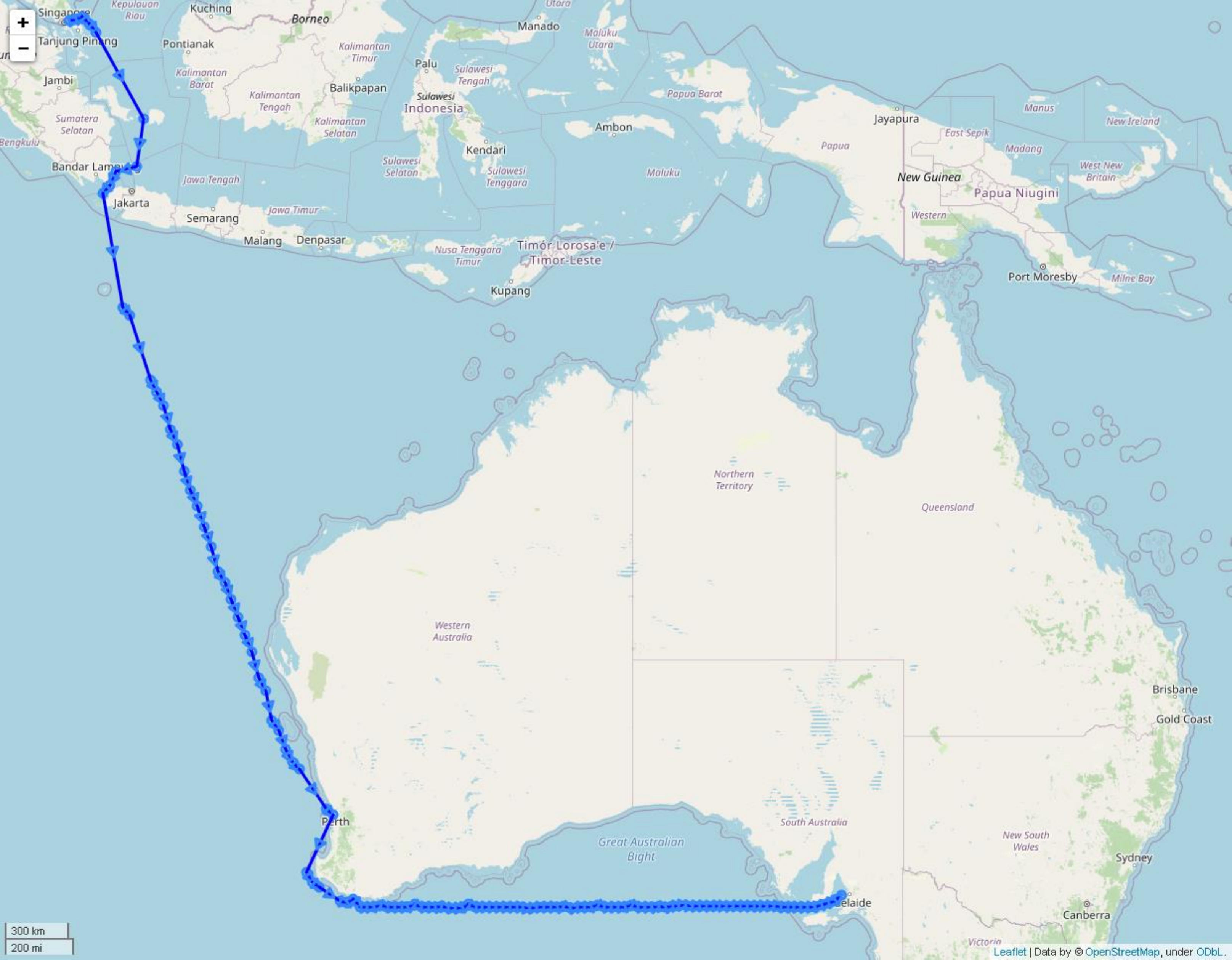}
			\caption{$\eta$=0.3, Lon-Scan}
		\end{subfigure}%
		\begin{subfigure}{.3\textwidth}
			\centering
			\includegraphics[width=0.98\linewidth]{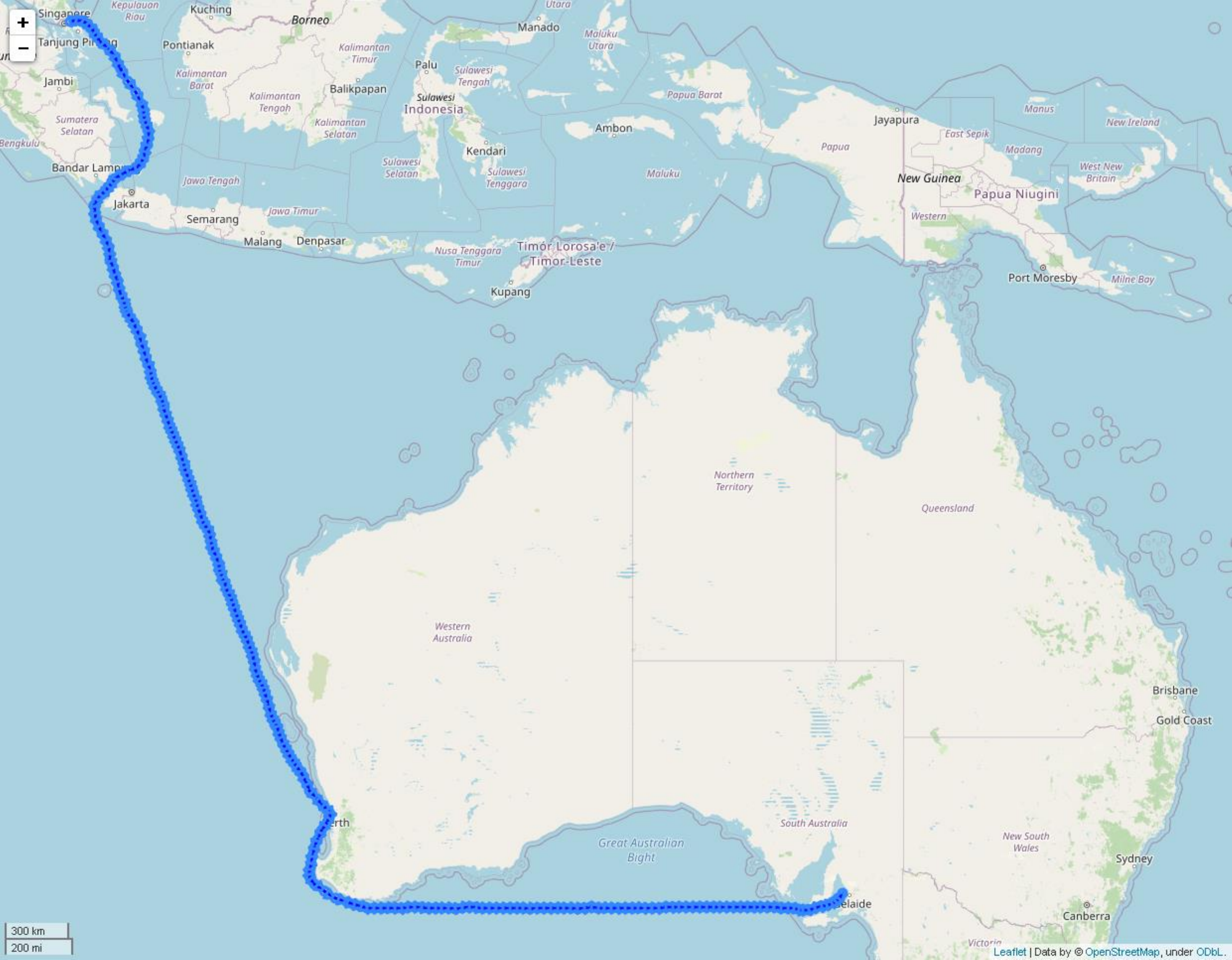}
			\caption{$\eta$=0.3, LatLon-Scan}
		\end{subfigure}

		\begin{subfigure}{.3\textwidth}
			\centering
			\includegraphics[width=0.98\linewidth]{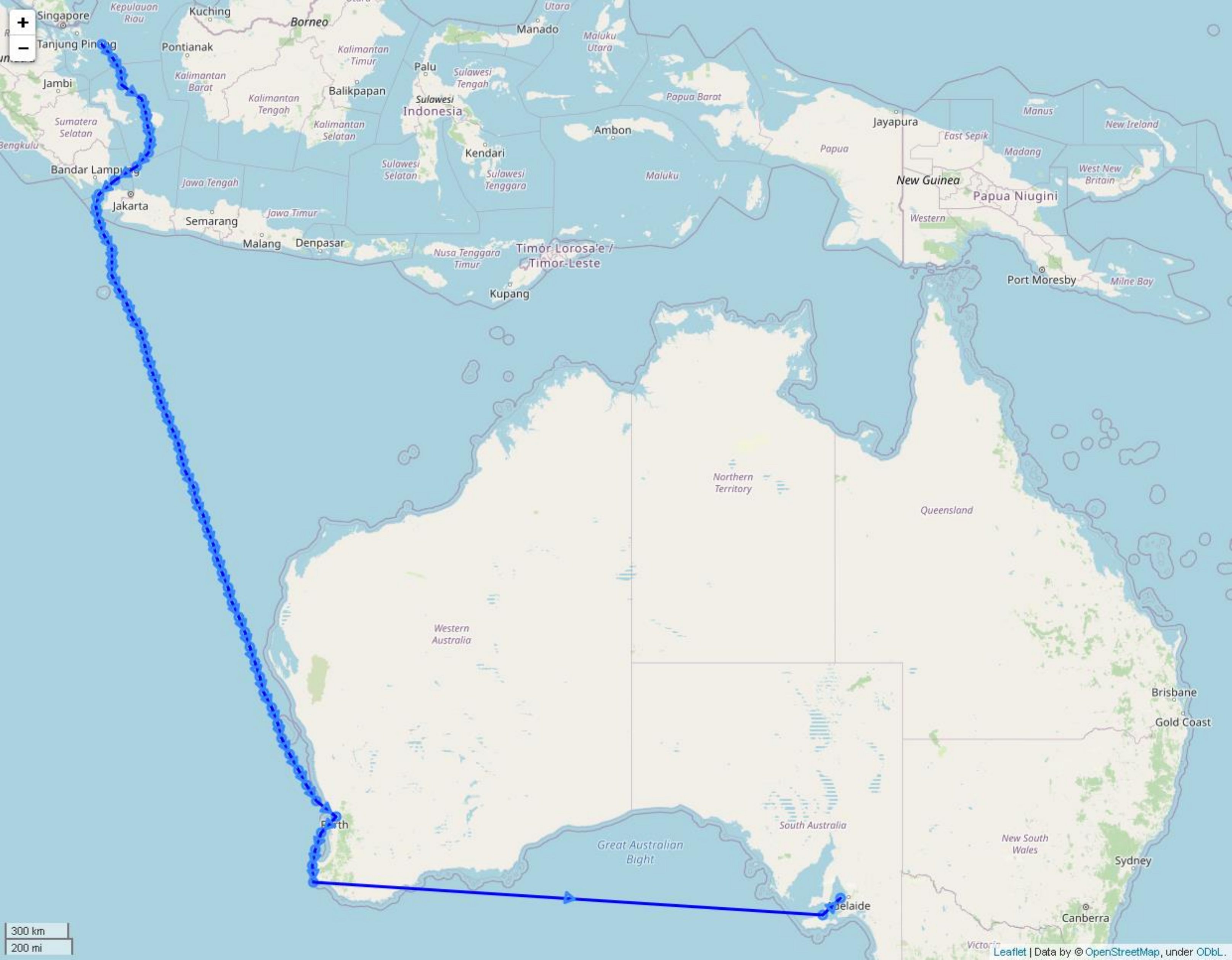}
			\caption{$\eta$=0.6, Lat-Scan}
		\end{subfigure}%
		\begin{subfigure}{.3\textwidth}
			\centering
			\includegraphics[width=0.98\linewidth]{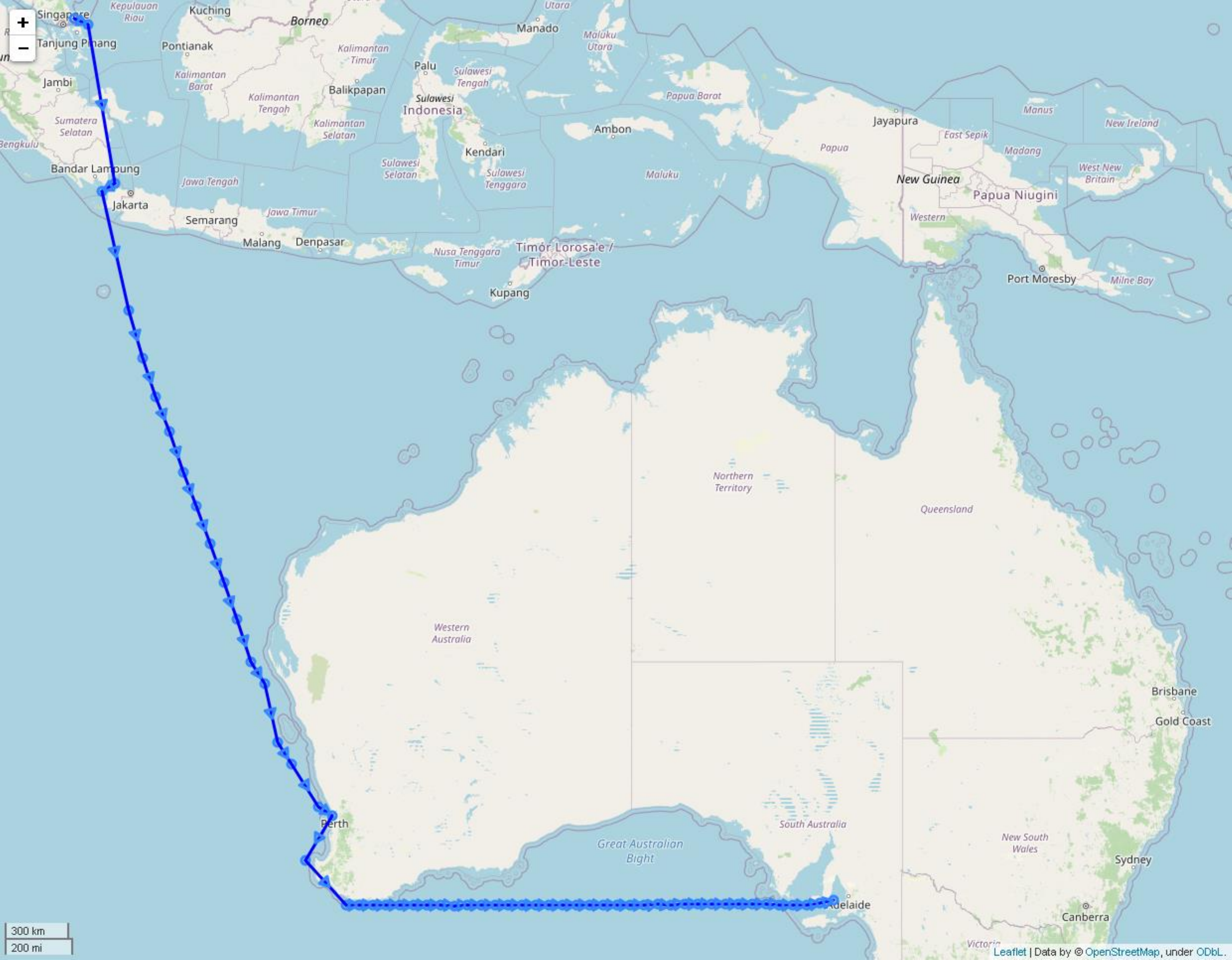}
			\caption{$\eta$=0.6, Lon-Scan}
		\end{subfigure}%
		\begin{subfigure}{.3\textwidth}
			\centering
			\includegraphics[width=0.98\linewidth]{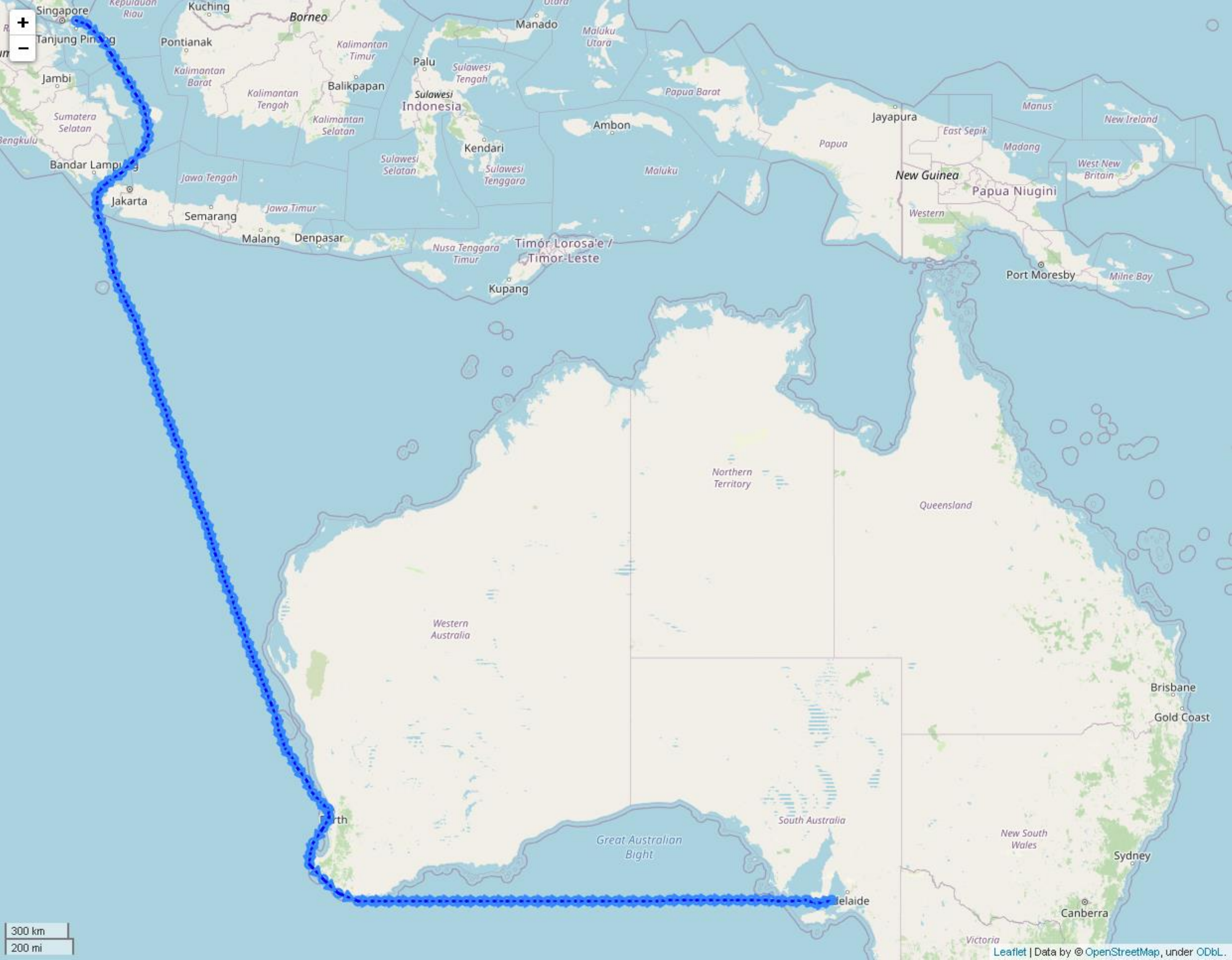}
			\caption{$\eta$=0.6, LatLon-Scan}
		\end{subfigure}

		\begin{subfigure}{.3\textwidth}
			\centering
			\includegraphics[width=0.98\linewidth]{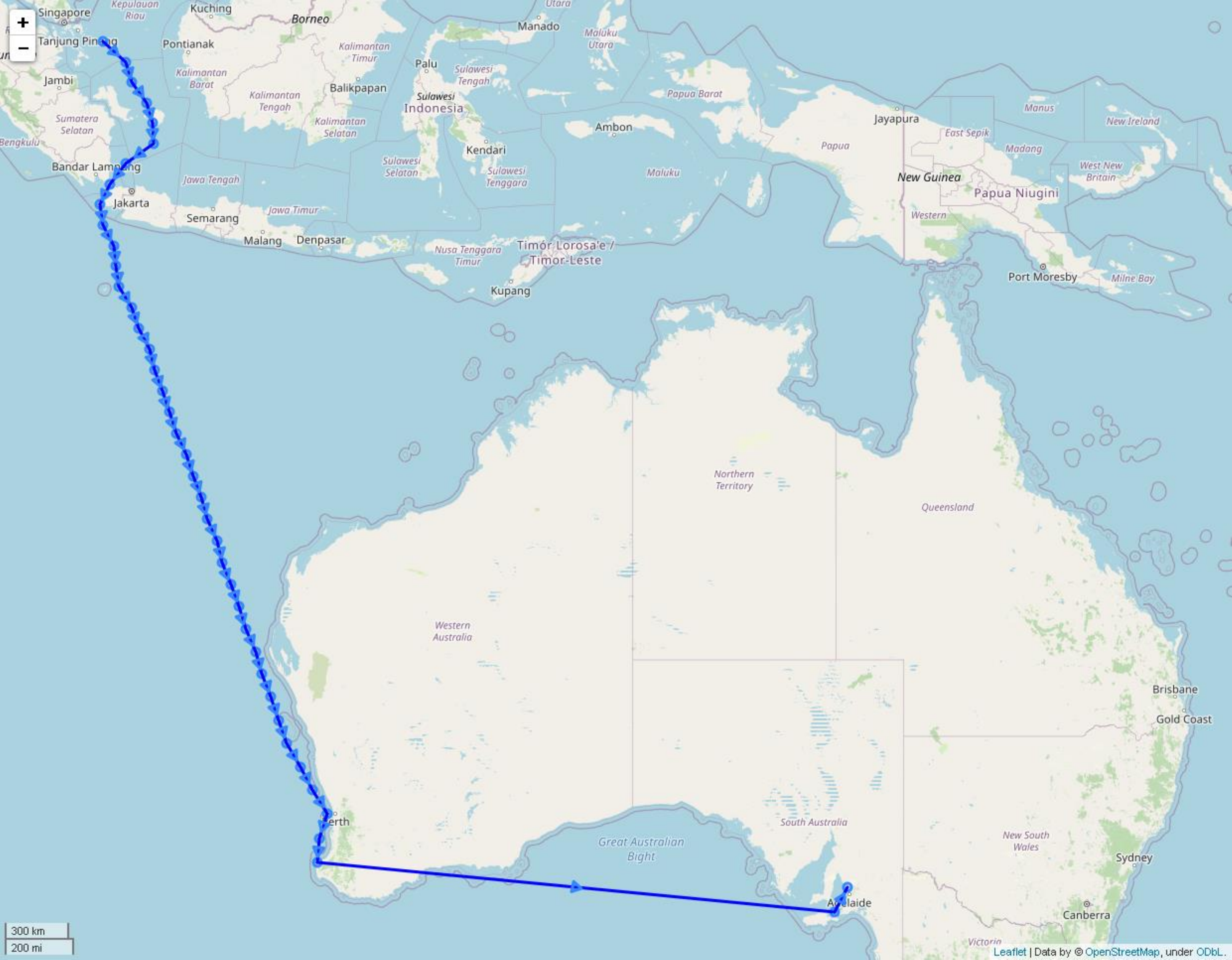}
			\caption{$\eta$=0.9, Lat-Scan}
		\end{subfigure}%
		\begin{subfigure}{.3\textwidth}
			\centering
			\includegraphics[width=0.98\linewidth]{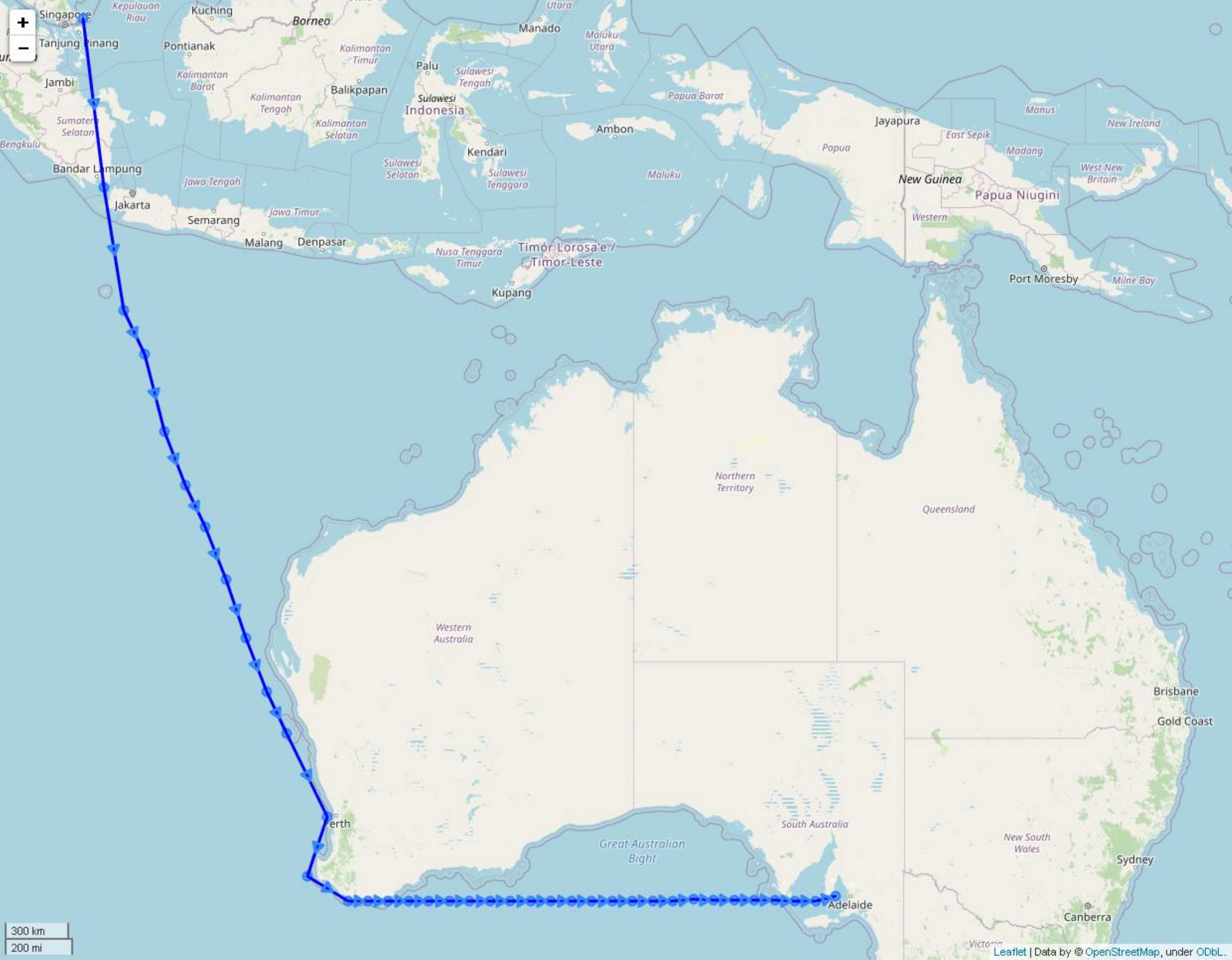}
			\caption{$\eta$=0.9, Lon-Scan}
		\end{subfigure}%
		\begin{subfigure}{.3\textwidth}
			\centering
			\includegraphics[width=0.98\linewidth]{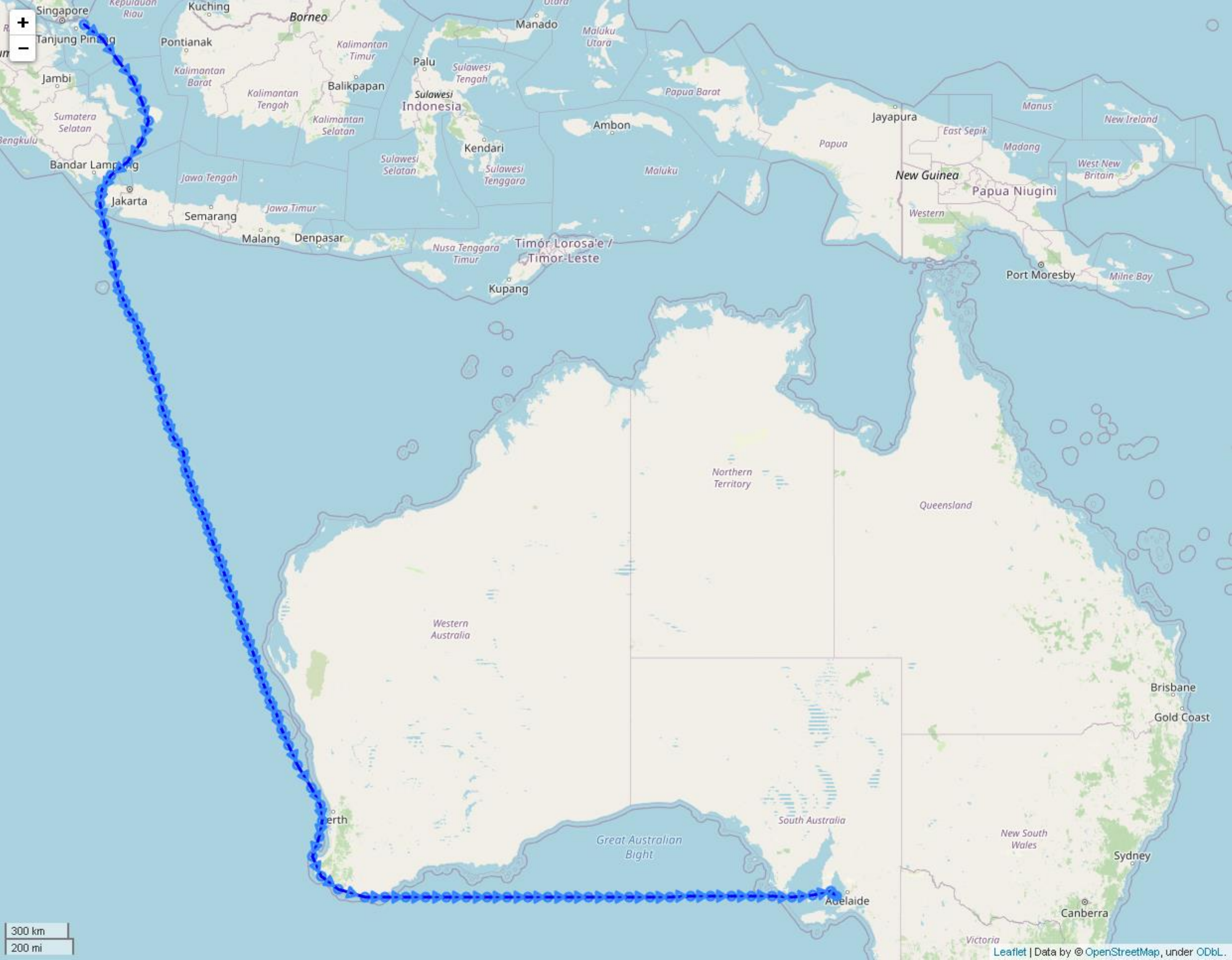}
			\caption{$\eta$=0.9, LatLon-Scan}
		\end{subfigure}

		\begin{subfigure}{.3\textwidth}
			\centering
			\includegraphics[width=0.98\linewidth]{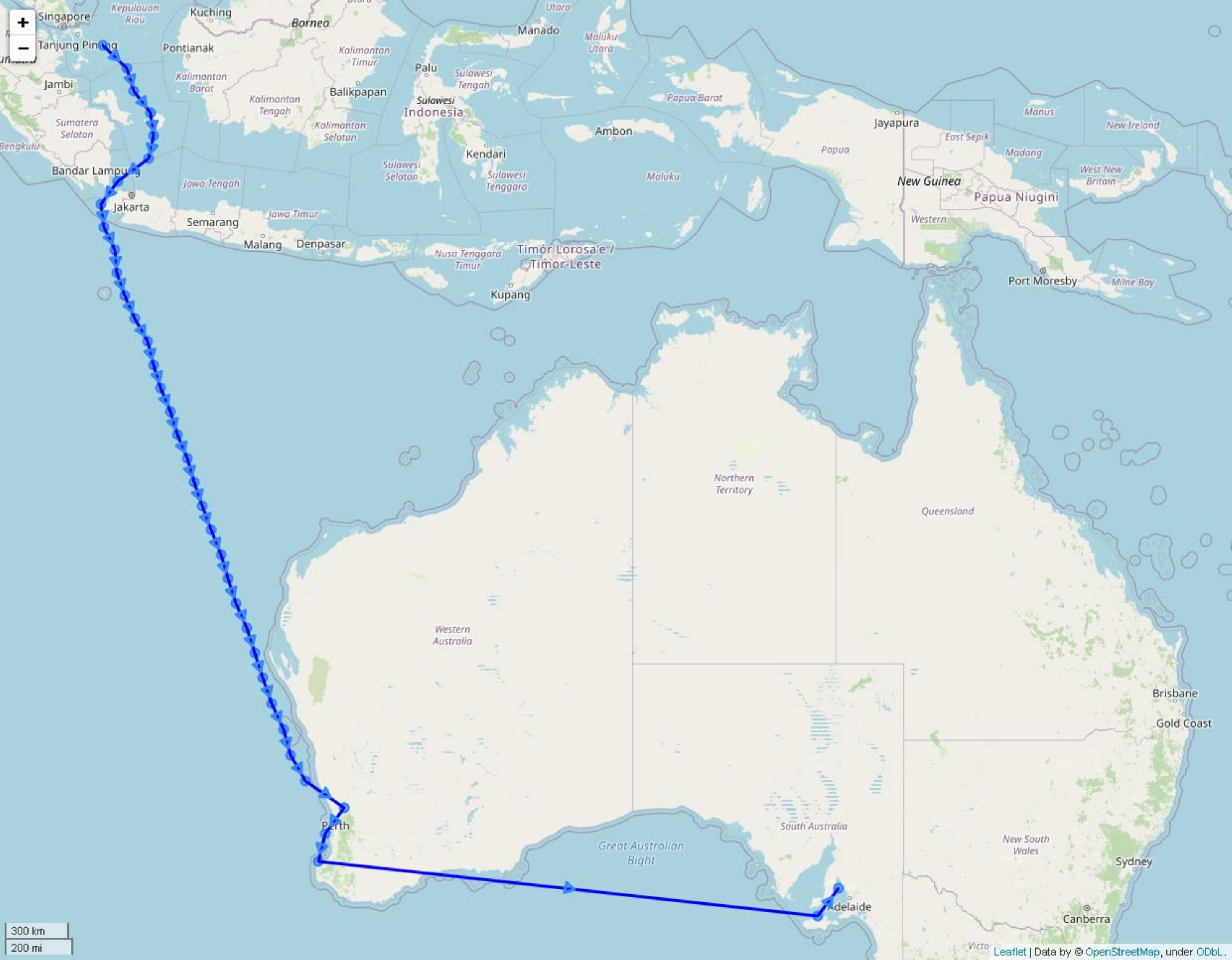}
			\caption{$\eta$=1.0, Lat-Scan}
		\end{subfigure}%
		\begin{subfigure}{.3\textwidth}
			\centering
			\includegraphics[width=0.98\linewidth]{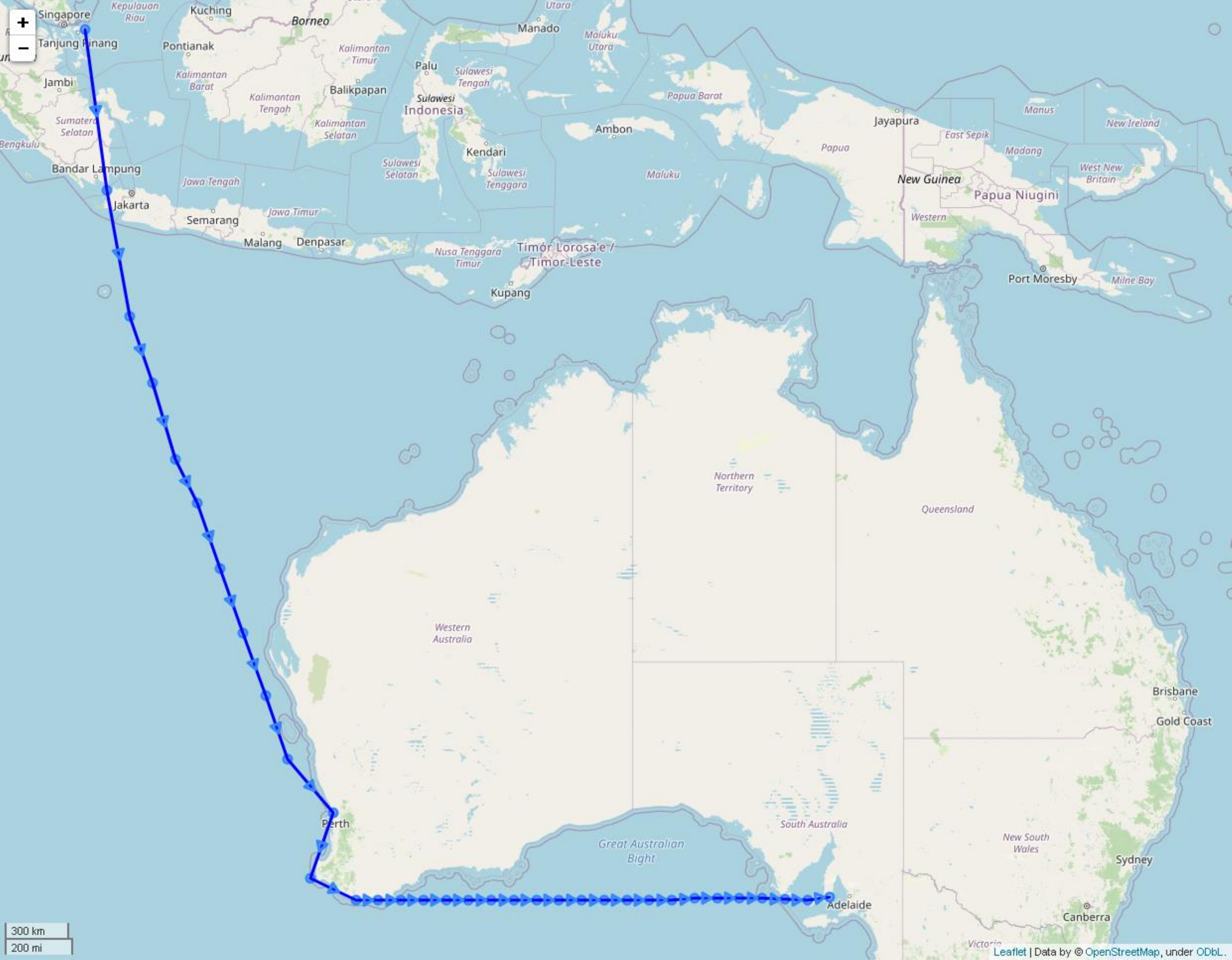}
			\caption{$\eta$=1.0, Lon-Scan}
		\end{subfigure}%
		\begin{subfigure}{.3\textwidth}
			\centering
			\includegraphics[width=0.98\linewidth]{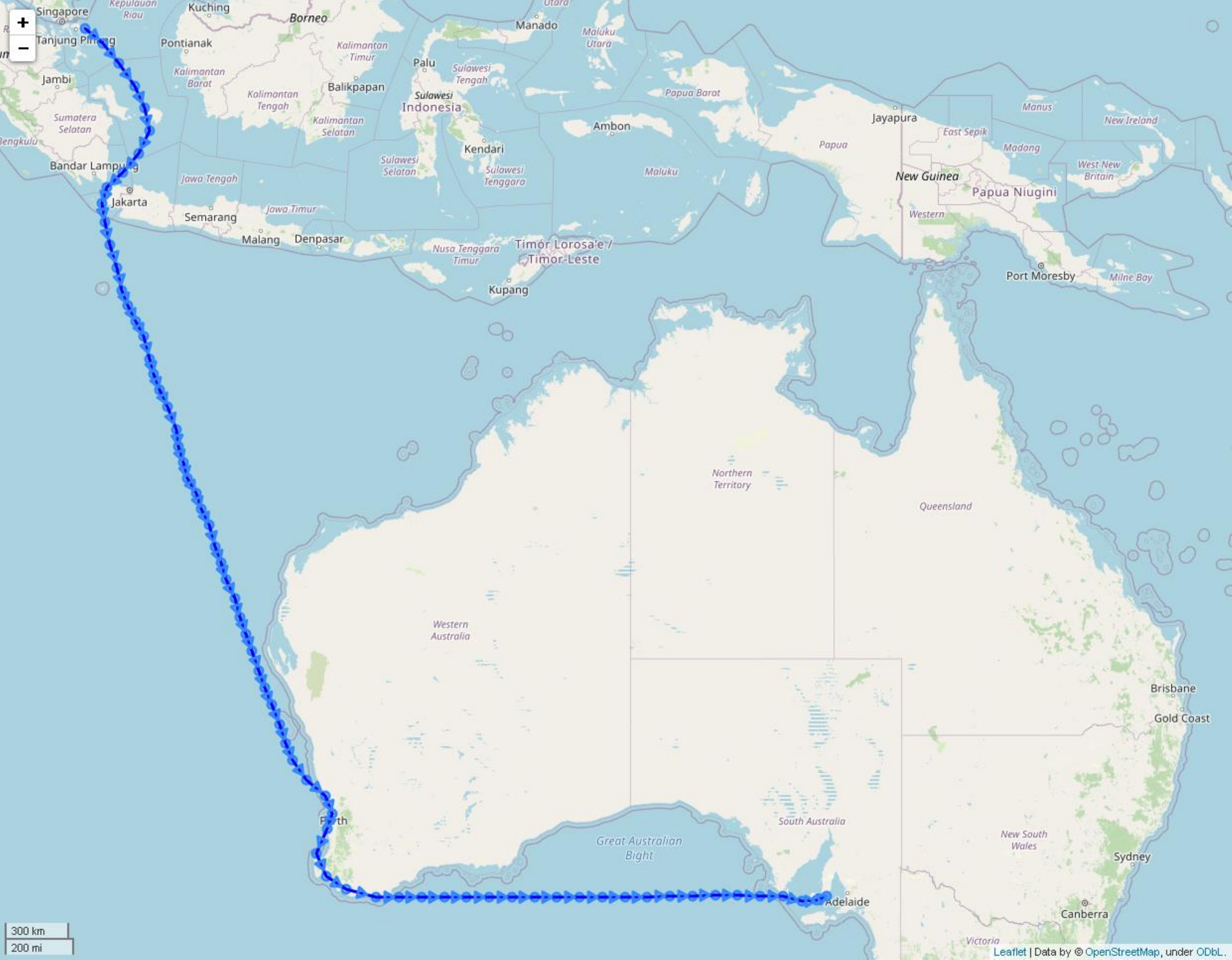}
			\caption{$\eta$=1.0, LatLon-Scan}
		\end{subfigure}

		\caption{Constructed Trajectories (Singapore $\rightarrow$ Adelaide)}
		\label{Fig: Constructed Trajectories of Singapore to Adelaide}
	\end{figure}

\newpage
Here are the historical (Figure \ref{Fig: Actual Trajectories of Singapore to Adelaide}) and constructed (Figure \ref{Fig: Constructed Trajectories of Singapore to Adelaide}) trajectories by Lat-scanning, Lon-scanning and LatLon-scanning for Singapore-Adelaide journey with different scanning internals ($\eta$) in degree.

According to the historical trajectories from Singapore to Adelaide in Figure \ref{Fig: Actual Trajectories of Singapore to Adelaide}, it is clear to demonstrate that there is no direct route. Generally, vessels will have a stop at Perth in between. The stop may introduce uncertainties in ETA prediction, which may make the prediction less accurate in general. The uncertainties may be due to loading or unloading cargoes, supplies or other services, which duration may be flexible and unpredictable. This stop delay would propagate to impact on ETA prediction toward a corresponding final destination.

According to the constructed trajectories from Singapore to Adelaide in Figure \ref{Fig: Constructed Trajectories of Singapore to Adelaide} across different scanning internals ($\eta$) and scanning methods, it is clear to note that LatLon-scanning can perform better in latitude and longitude directions, compared to merely Lat-scanning or Lon-scanning. For Lat-scanning, it is accurately extracting movements along latitude direction, but there are distortions and missing points along longitude direction. Conversely, Lon-scanning provides detailed movements along longitude direction, but likely fails along latitude direction. Both Lat-scanning and Lon-scanning are losing fidelity with larger scanning intervals ($\eta$), while the issue can be resolved by applying LatLon-scanning.

\begin{figure}[htbp]
		\centering
		\includegraphics[width=\linewidth]{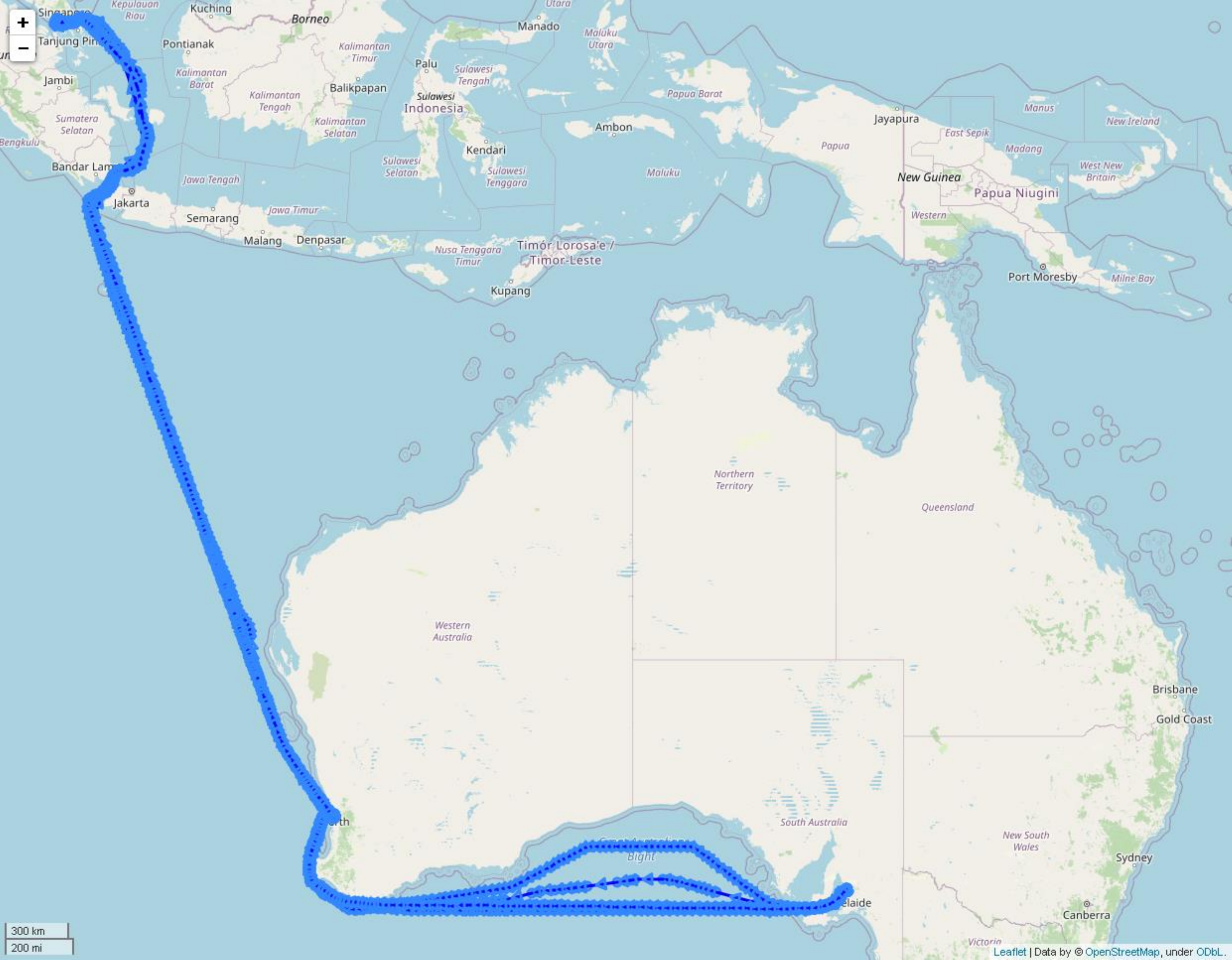}
		\caption{Actual Trajectories (Adelaide $\rightarrow$ Singapore)}
		\label{Fig: Actual Trajectories of Adelaide to Singapore}
	\end{figure}

\begin{figure}[htbp]
		\begin{subfigure}{.3\textwidth}
			\centering
			\includegraphics[width=0.98\linewidth]{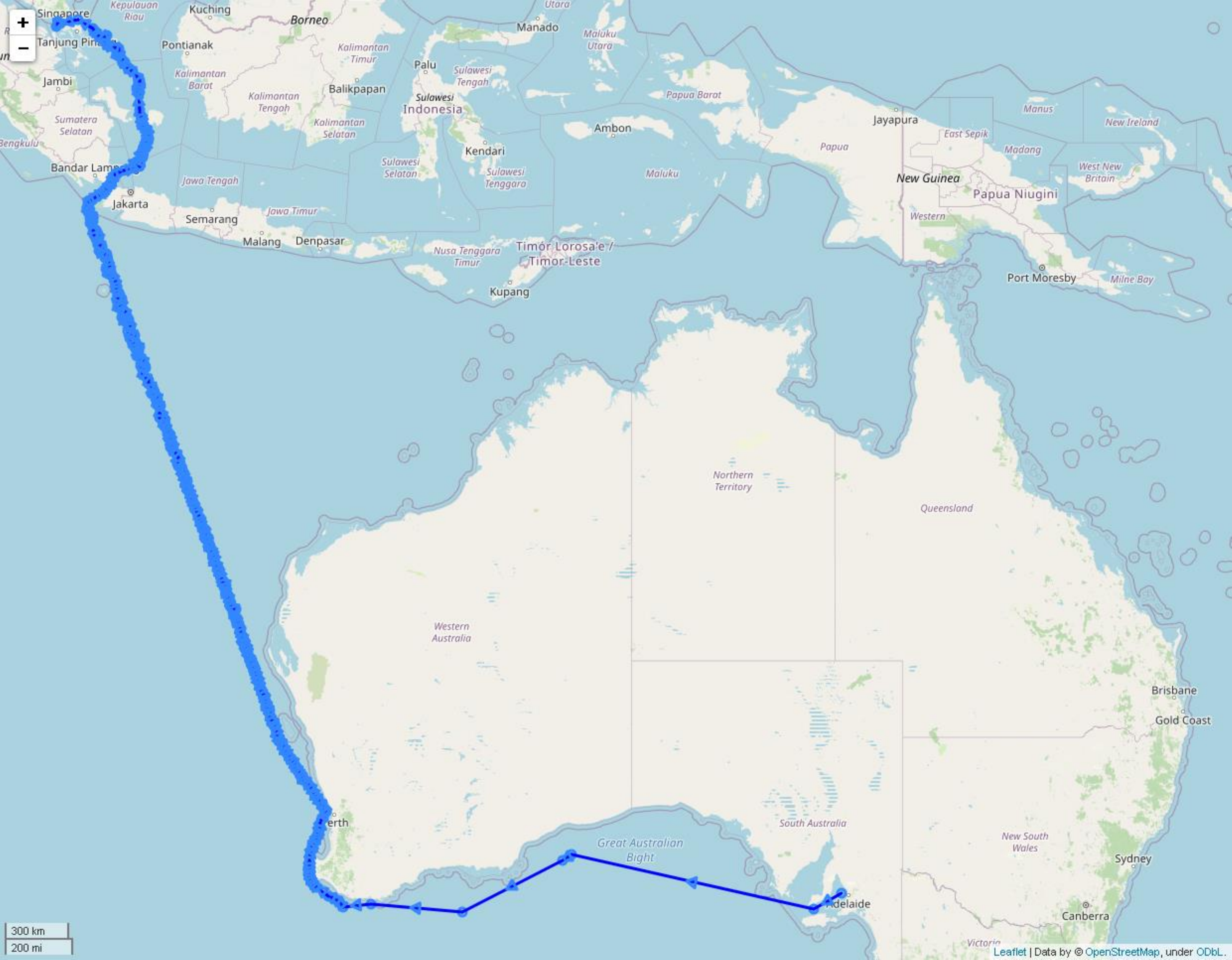}
			\caption{$\eta$=0.1, Lat-Scan}
		\end{subfigure}%
		\begin{subfigure}{.3\textwidth}
			\centering
			\includegraphics[width=0.98\linewidth]{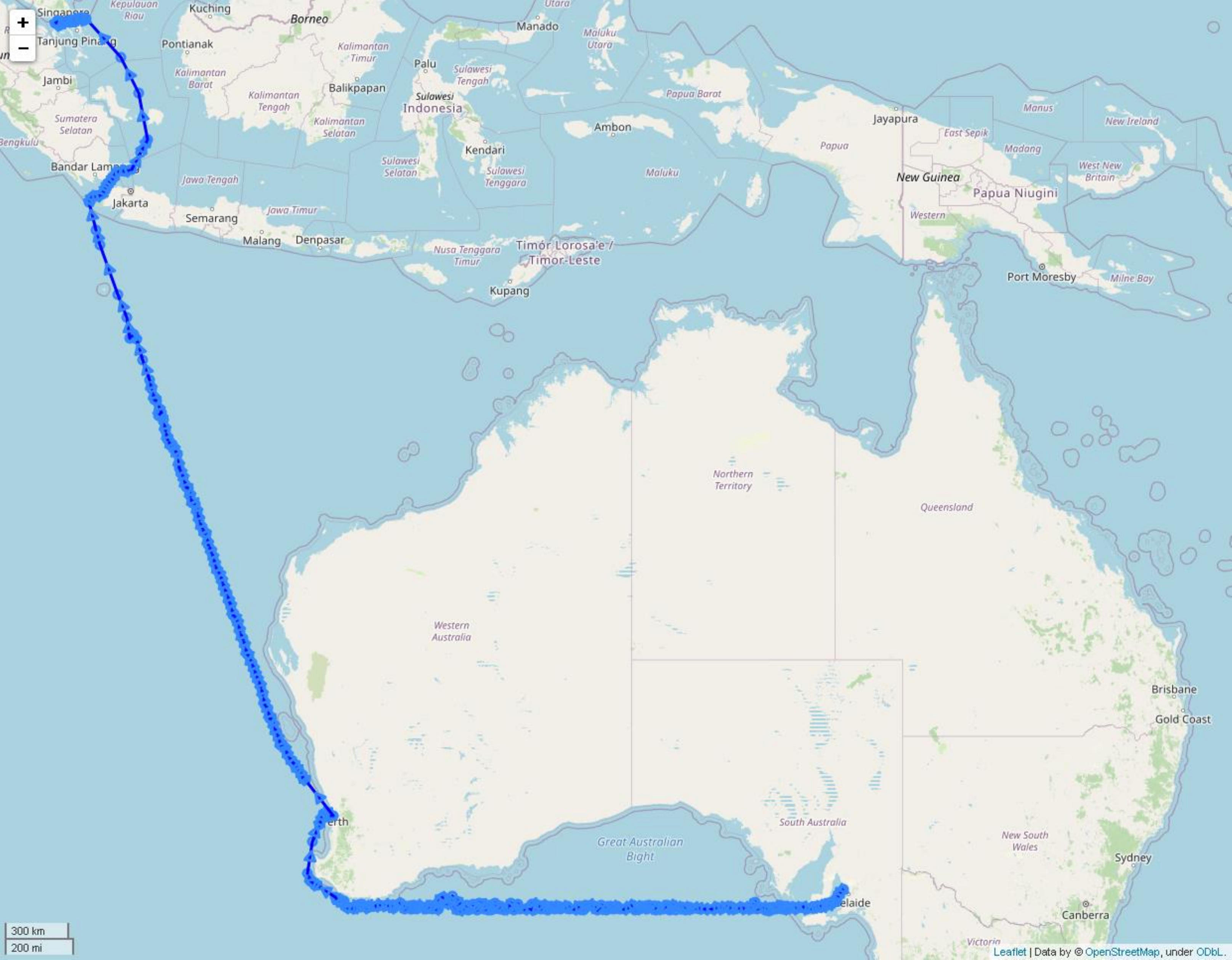}
			\caption{$\eta$=0.1, Lon-Scan}
		\end{subfigure}%
		\begin{subfigure}{.3\textwidth}
			\centering
			\includegraphics[width=0.98\linewidth]{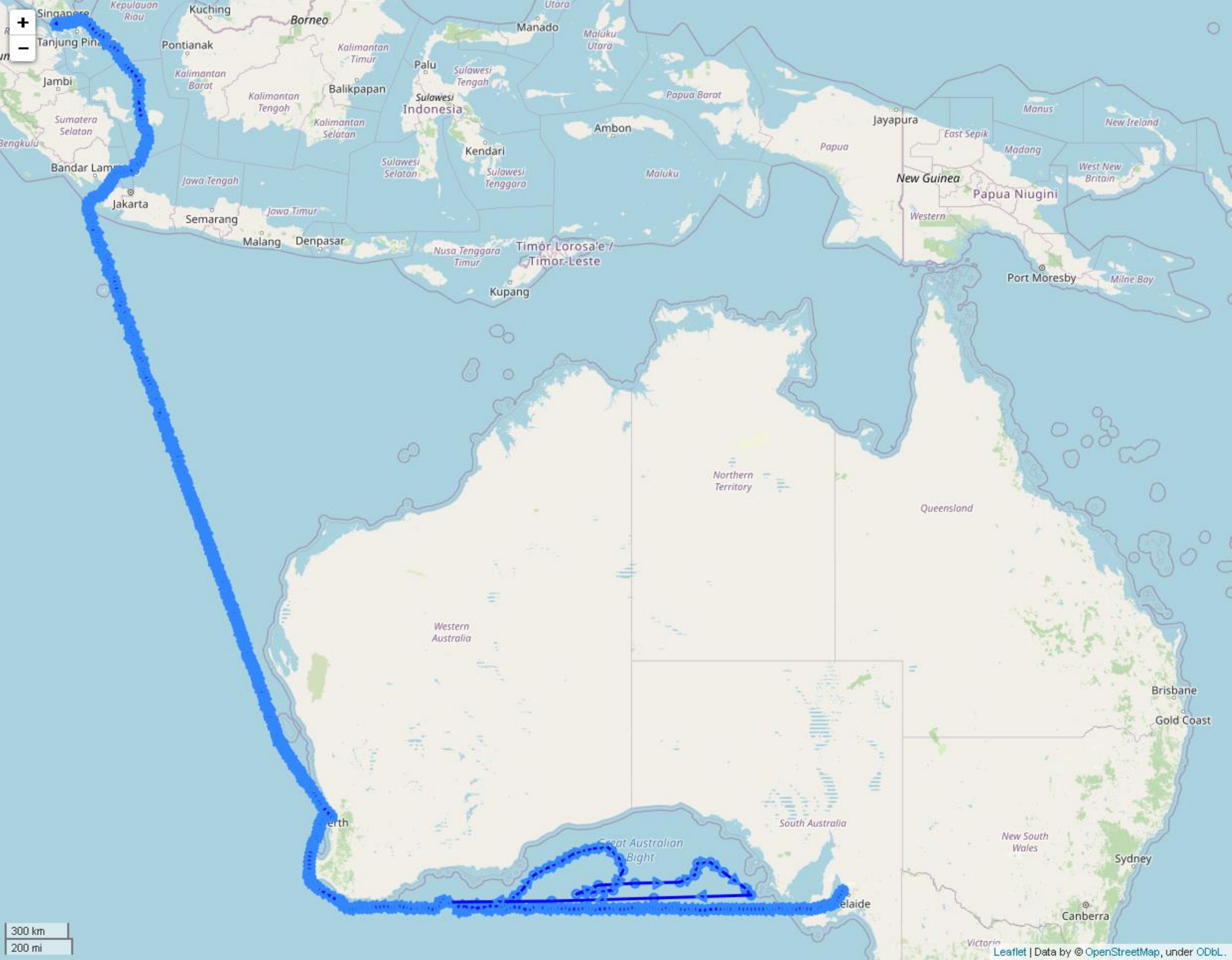}
			\caption{$\eta$=0.1, LatLon-Scan}
		\end{subfigure}

		\begin{subfigure}{.3\textwidth}
			\centering
			\includegraphics[width=0.98\linewidth]{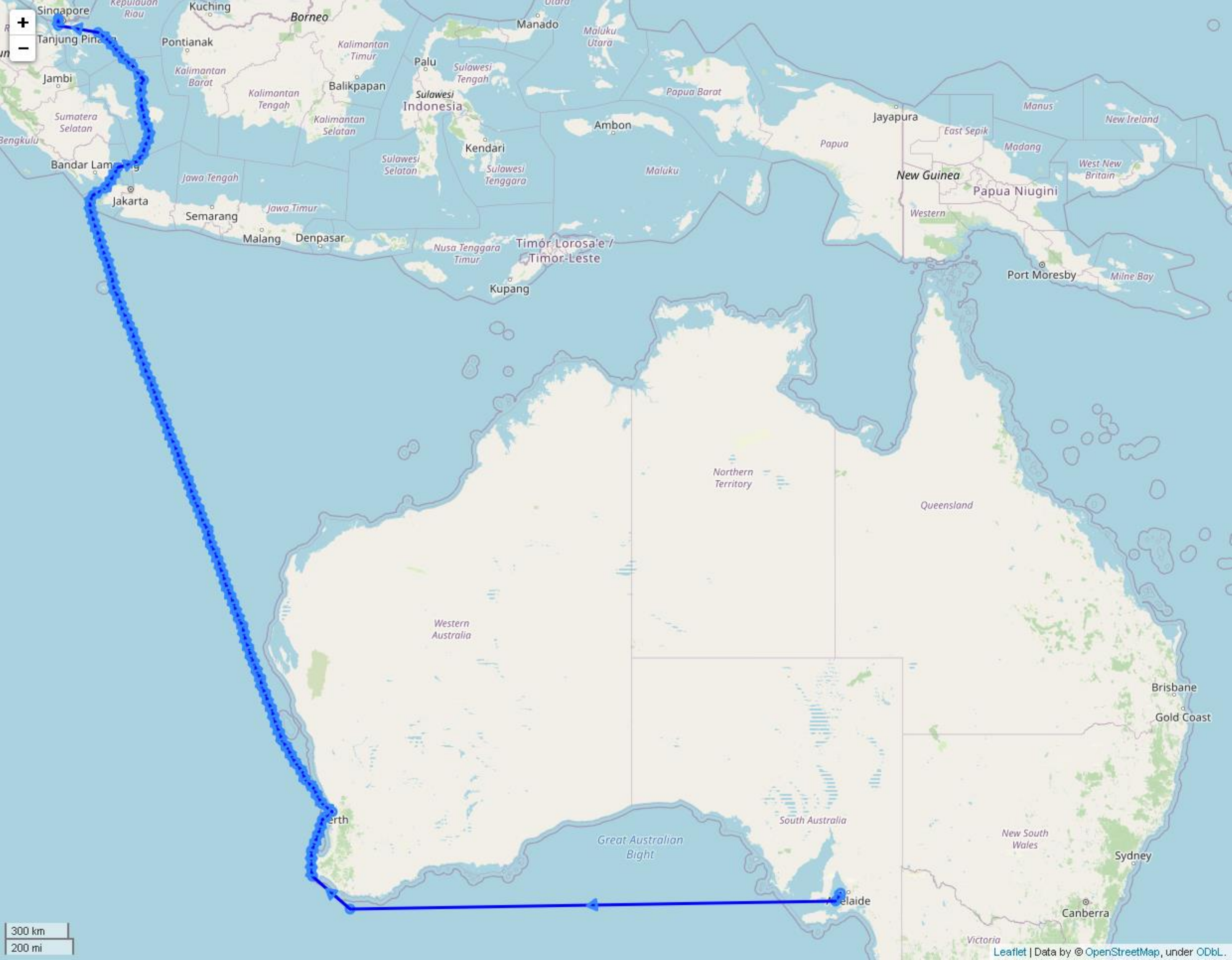}
			\caption{$\eta$=0.3, Lat-Scan}
		\end{subfigure}%
		\begin{subfigure}{.3\textwidth}
			\centering
			\includegraphics[width=0.98\linewidth]{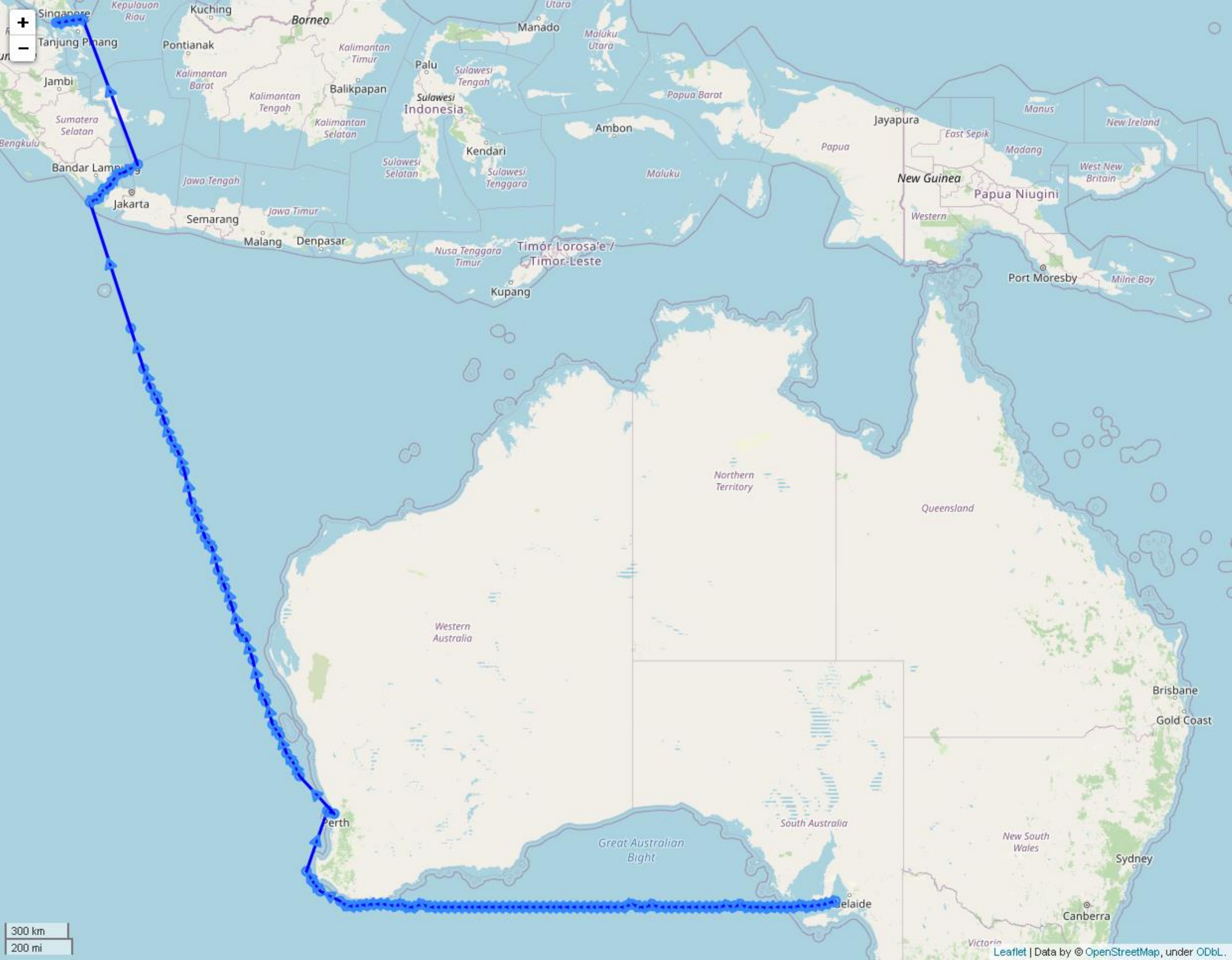}
			\caption{$\eta$=0.3, Lon-Scan}
		\end{subfigure}%
		\begin{subfigure}{.3\textwidth}
			\centering
			\includegraphics[width=0.98\linewidth]{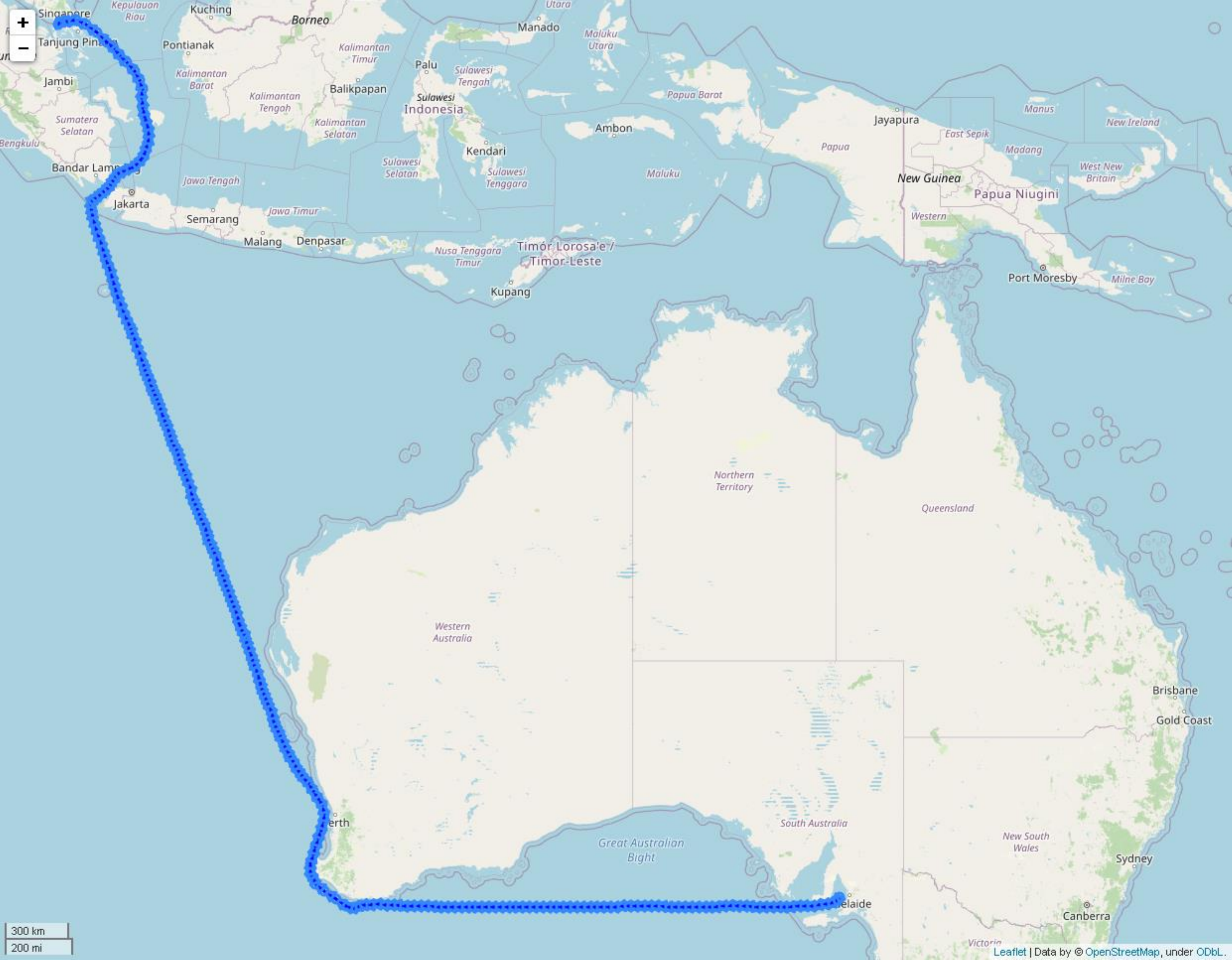}
			\caption{$\eta$=0.3, LatLon-Scan}
		\end{subfigure}

		\begin{subfigure}{.3\textwidth}
			\centering
			\includegraphics[width=0.98\linewidth]{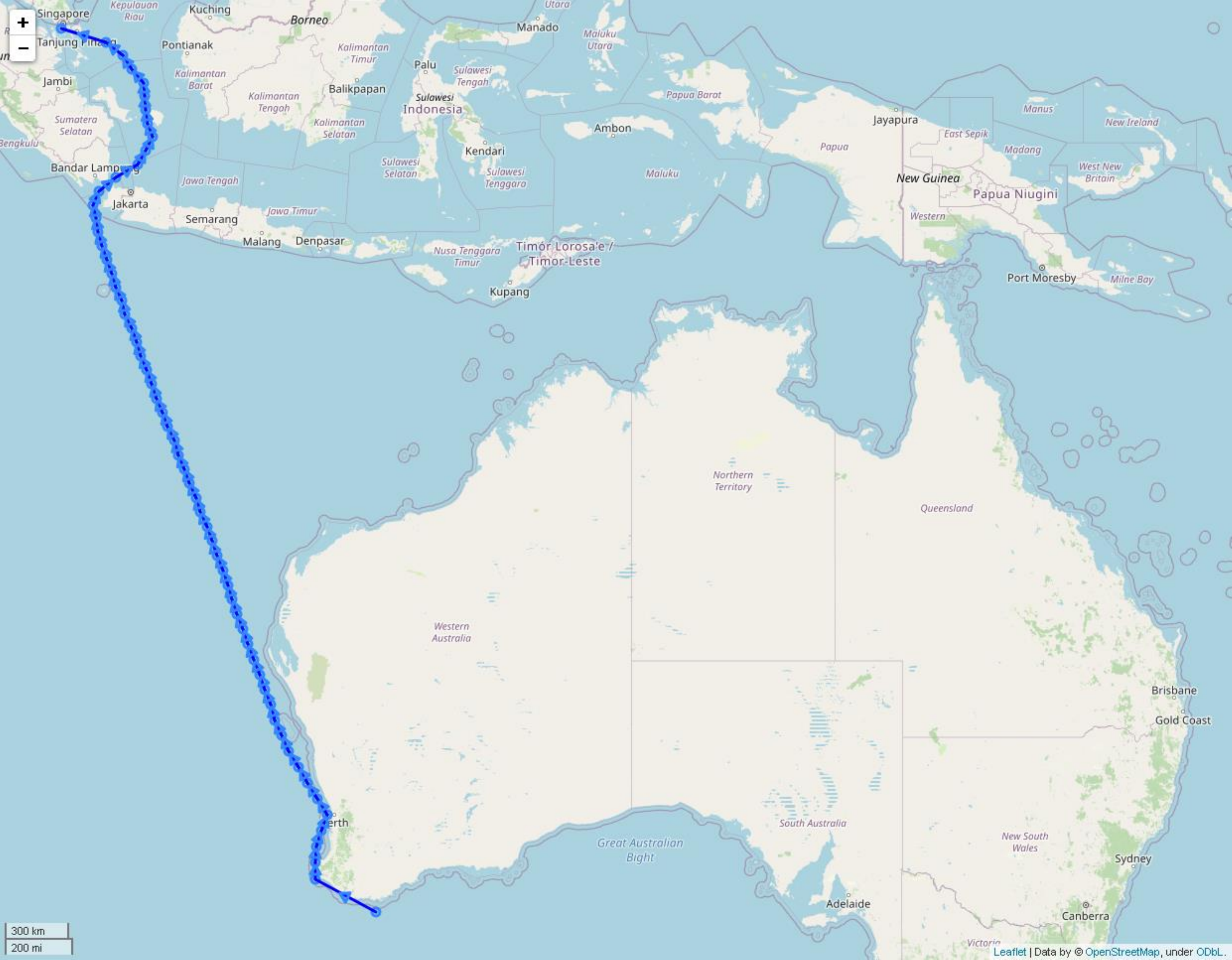}
			\caption{$\eta$=0.6, Lat-Scan}
		\end{subfigure}%
		\begin{subfigure}{.3\textwidth}
			\centering
			\includegraphics[width=0.98\linewidth]{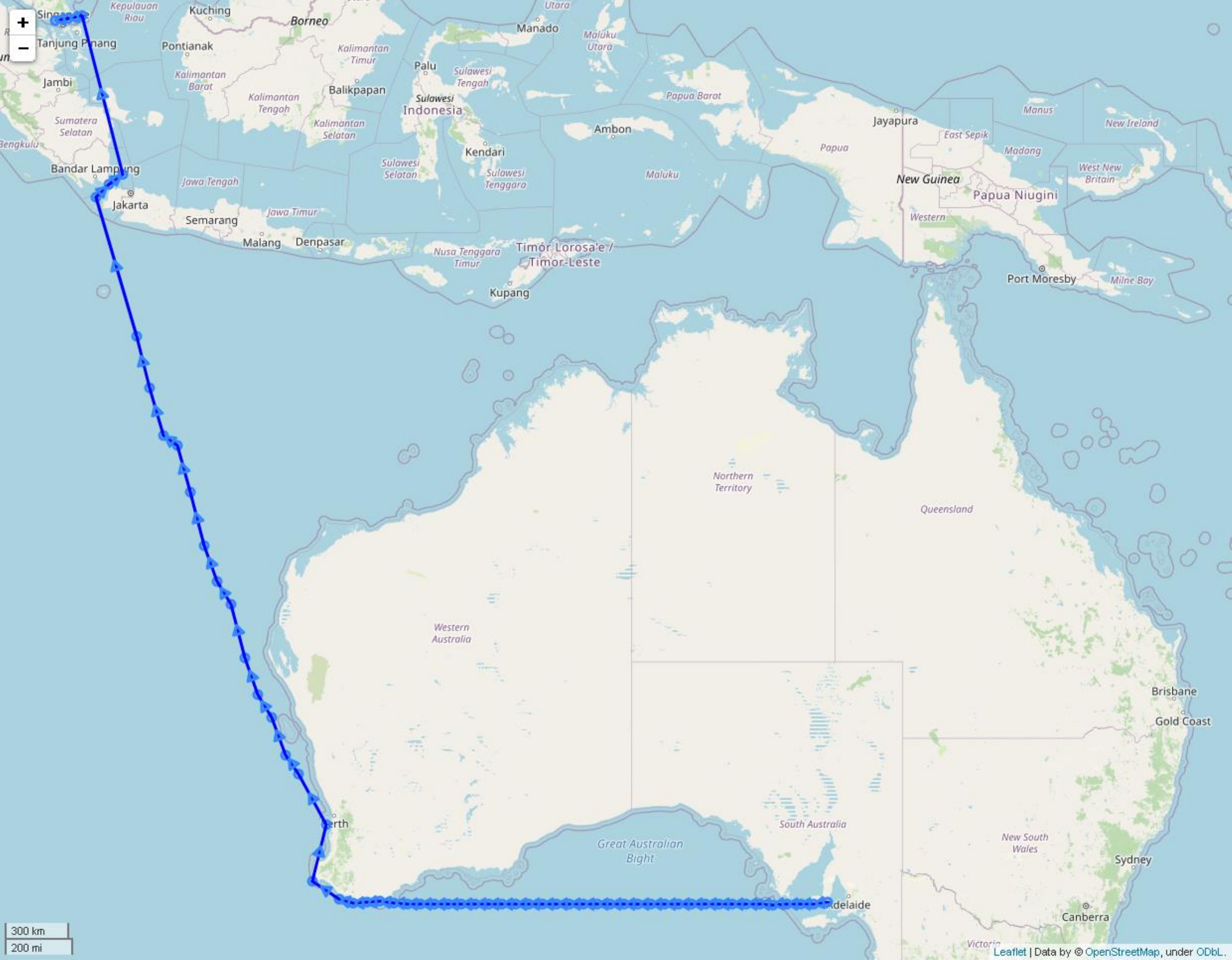}
			\caption{$\eta$=0.6, Lon-Scan}
		\end{subfigure}%
		\begin{subfigure}{.3\textwidth}
			\centering
			\includegraphics[width=0.98\linewidth]{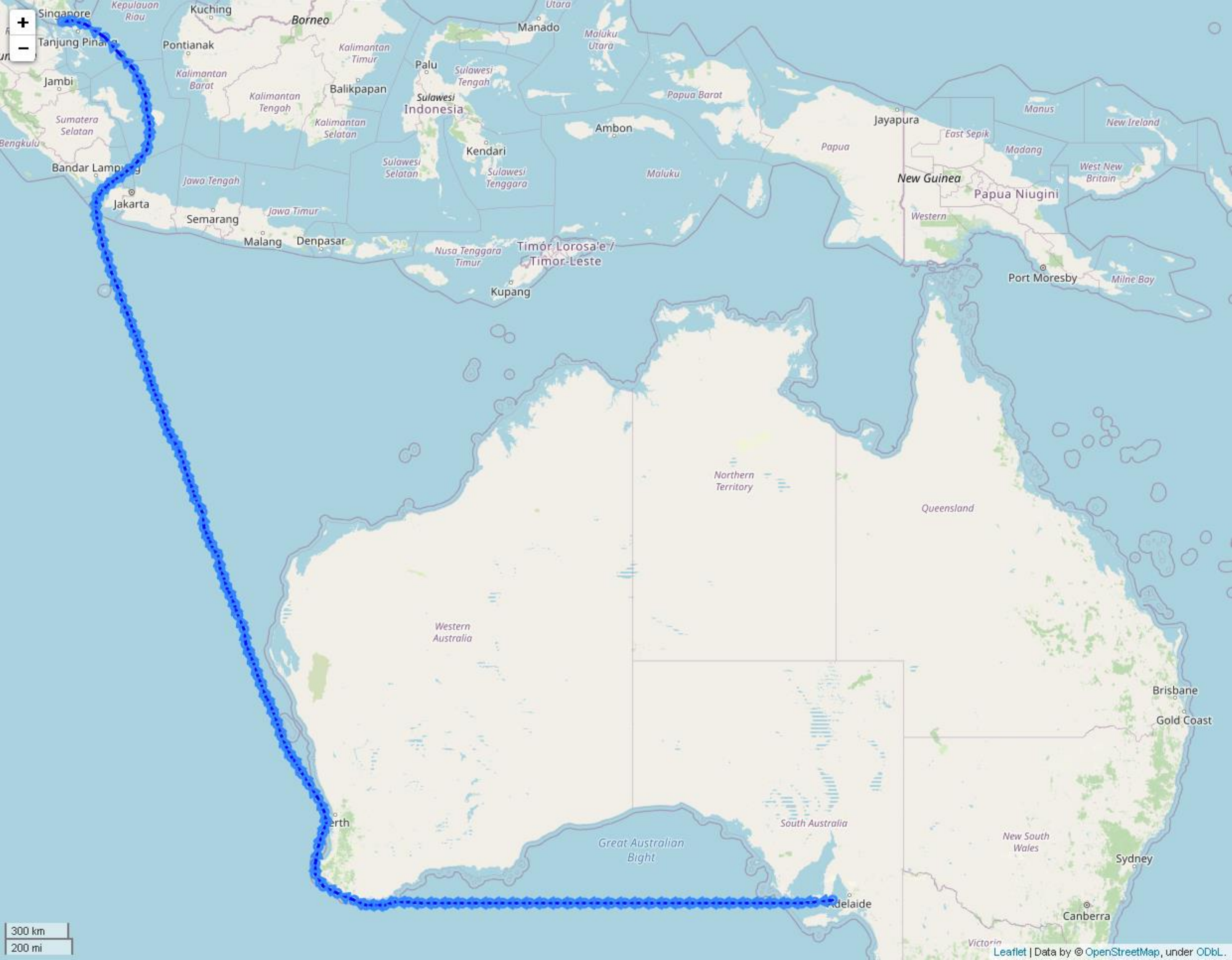}
			\caption{$\eta$=0.6, LatLon-Scan}
		\end{subfigure}

		\begin{subfigure}{.3\textwidth}
			\centering
			\includegraphics[width=0.98\linewidth]{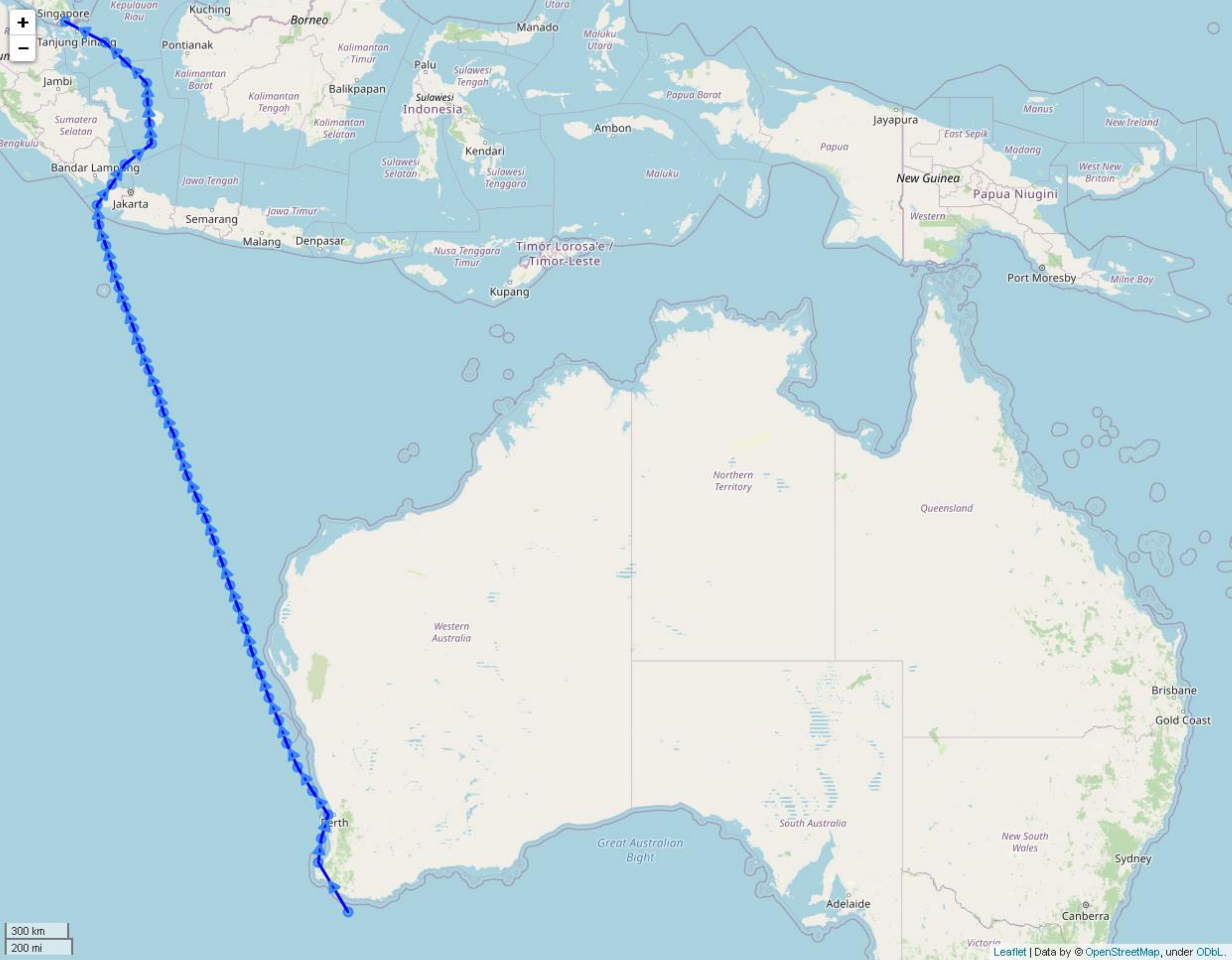}
			\caption{$\eta$=0.9, Lat-Scan}
		\end{subfigure}%
		\begin{subfigure}{.3\textwidth}
			\centering
			\includegraphics[width=0.98\linewidth]{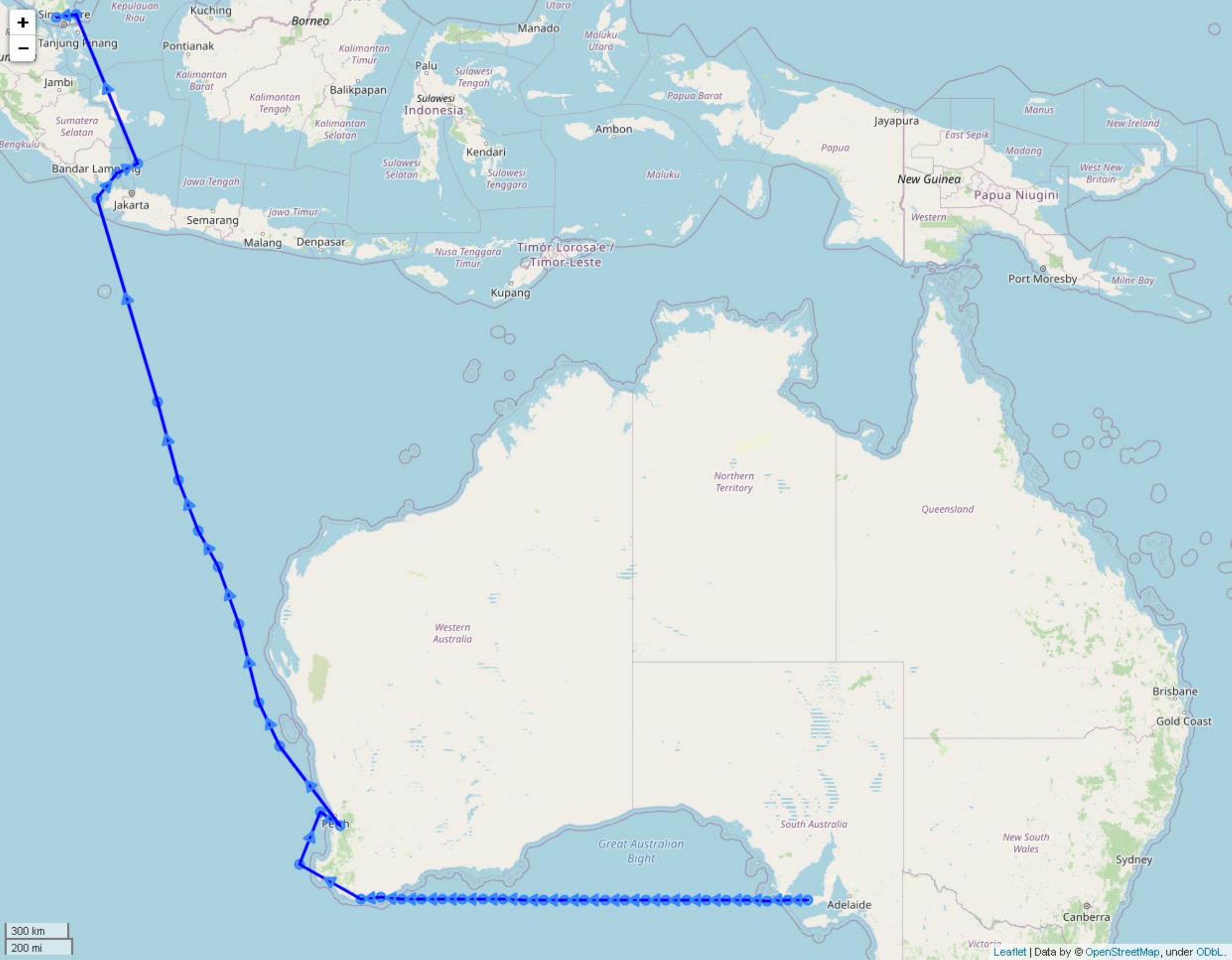}
			\caption{$\eta$=0.9, Lon-Scan}
		\end{subfigure}%
		\begin{subfigure}{.3\textwidth}
			\centering
			\includegraphics[width=0.98\linewidth]{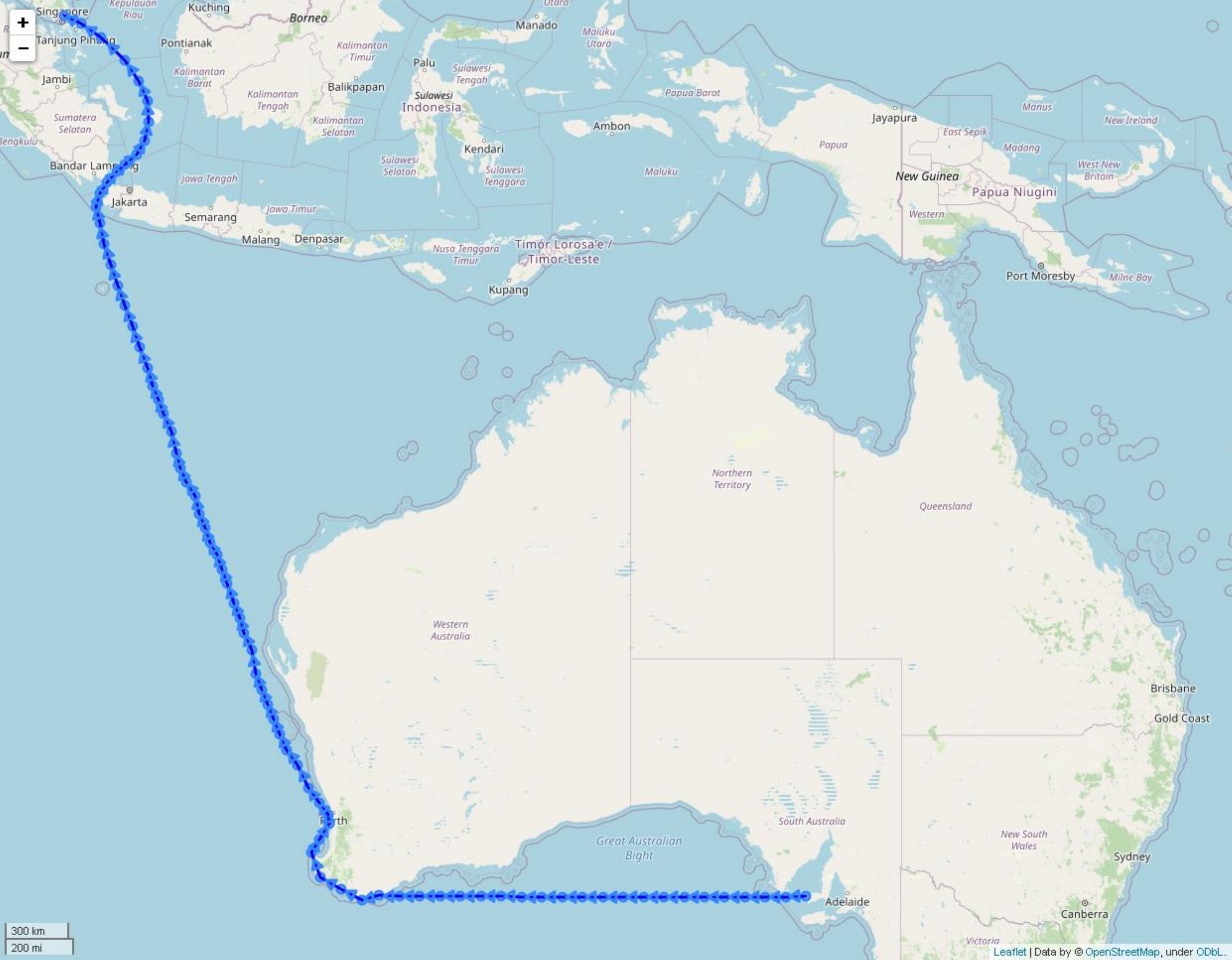}
			\caption{$\eta$=0.9, LatLon-Scan}
		\end{subfigure}

		\begin{subfigure}{.3\textwidth}
			\centering
			\includegraphics[width=0.98\linewidth]{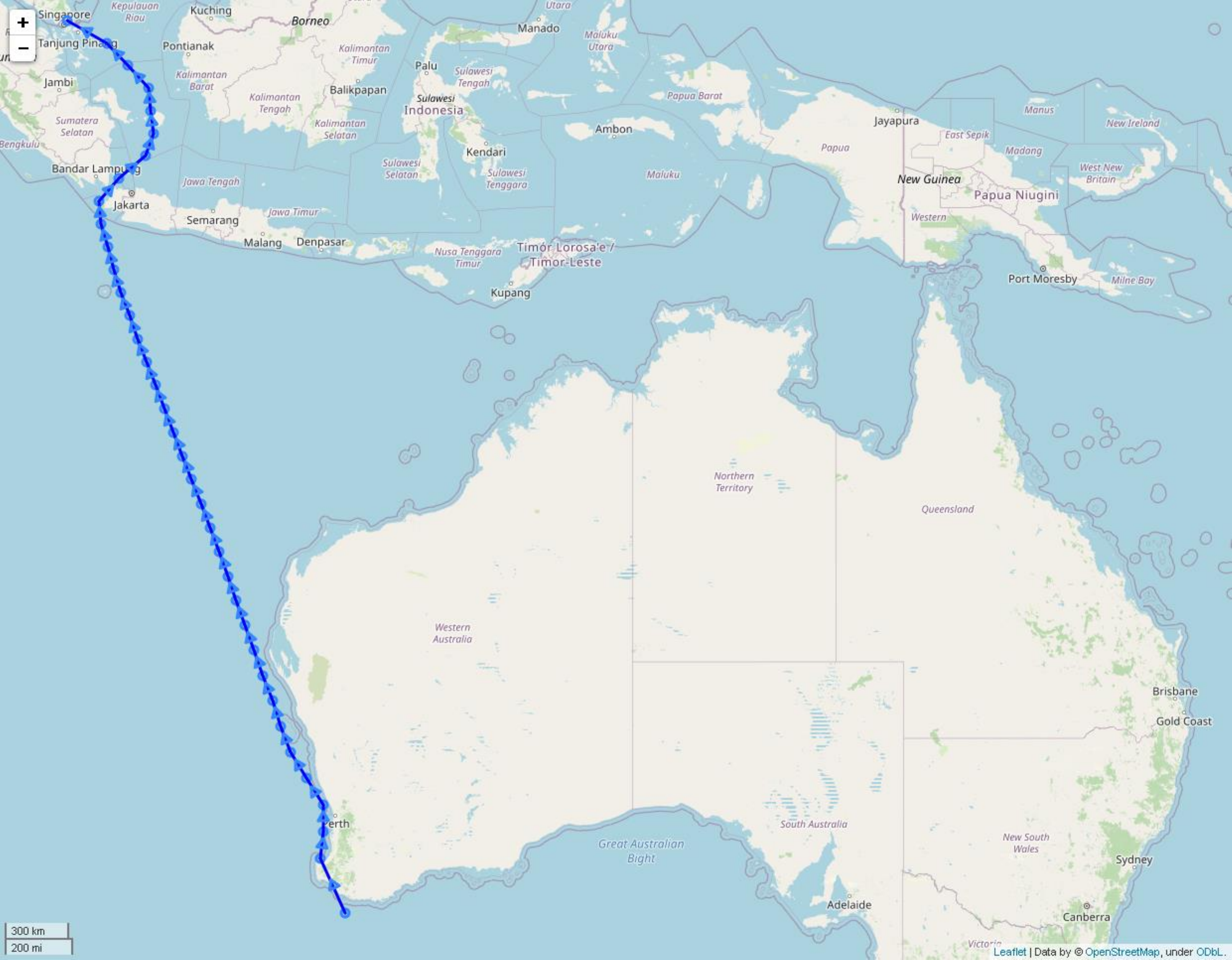}
			\caption{$\eta$=1.0, Lat-Scan}
		\end{subfigure}%
		\begin{subfigure}{.3\textwidth}
			\centering
			\includegraphics[width=0.98\linewidth]{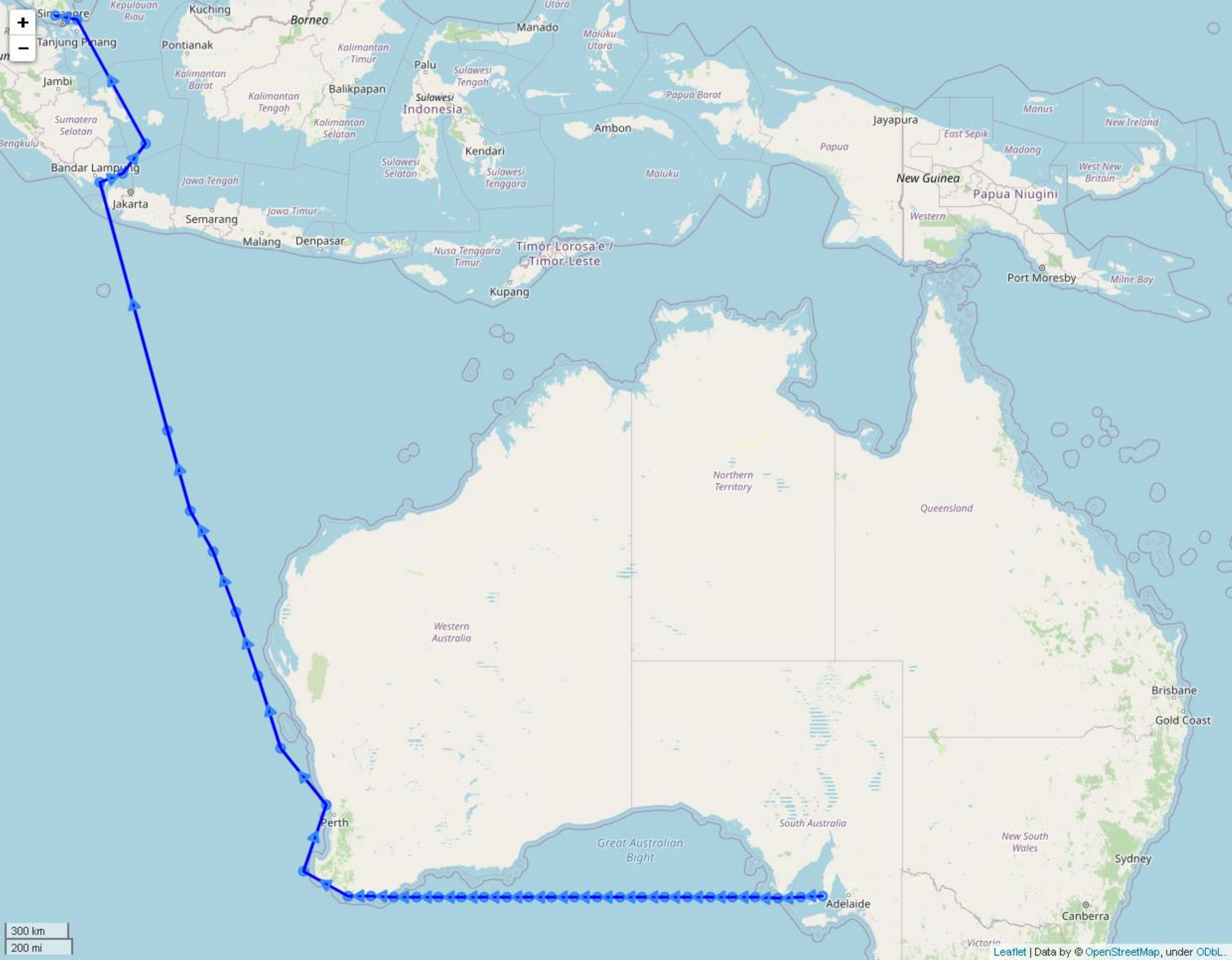}
			\caption{$\eta$=1.0, Lon-Scan}
		\end{subfigure}%
		\begin{subfigure}{.3\textwidth}
			\centering
			\includegraphics[width=0.98\linewidth]{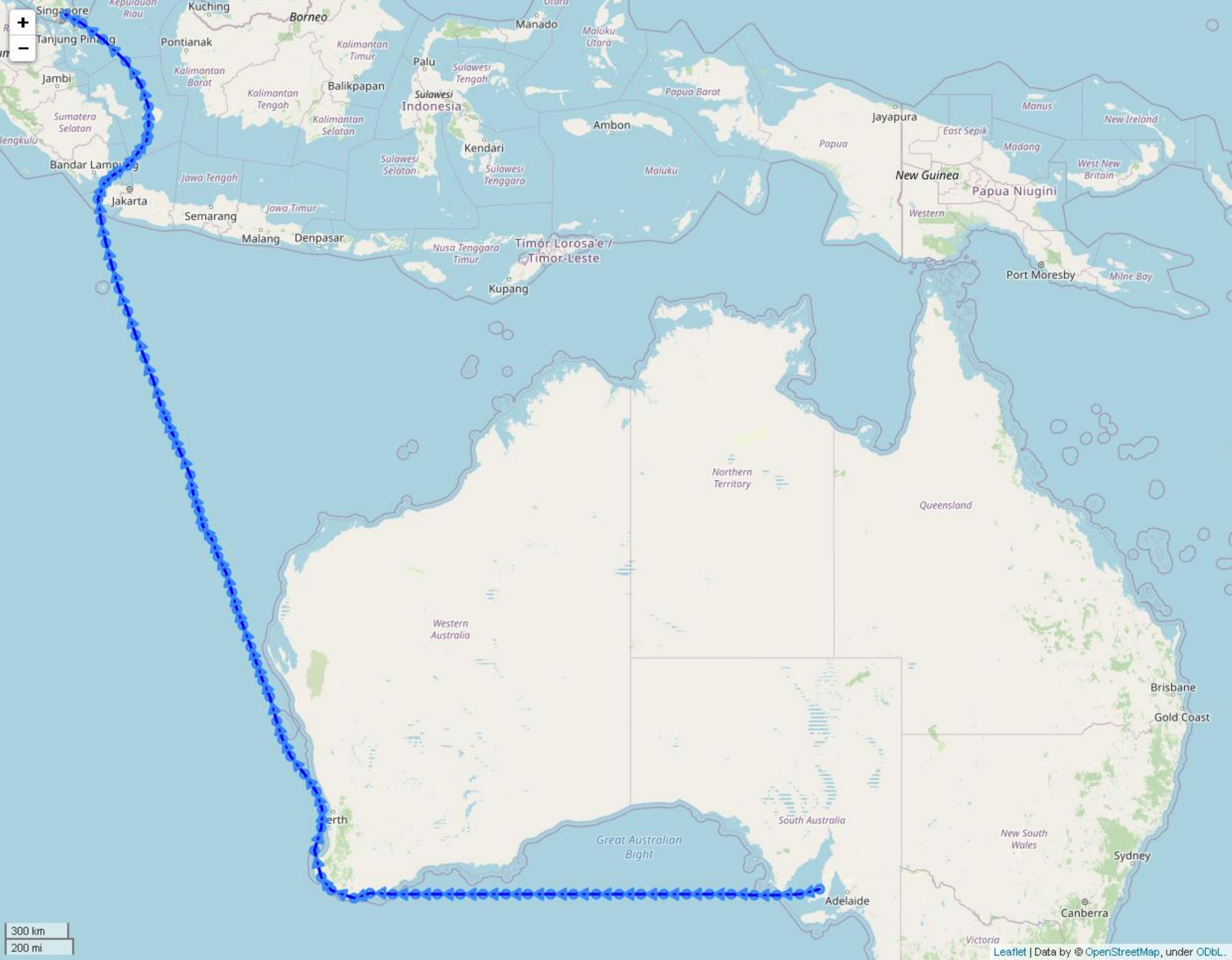}
			\caption{$\eta$=1.0, LatLon-Scan}
		\end{subfigure}

		\caption{Constructed Trajectories (Adelaide $\rightarrow$ Singapore)}
		\label{Fig: Constructed Trajectories of Adelaide to Singapore}
	\end{figure}

\newpage
Here are the historical (Figure \ref{Fig: Actual Trajectories of Adelaide to Singapore}) and constructed (Figure \ref{Fig: Constructed Trajectories of Adelaide to Singapore}) trajectories by Lat-scan, Lon-Scan and LatLon-Scan for Adelaide-Singapore journey with different scanning internals ($\eta$) in degree.

According to the historical trajectories from Adelaide to Singapore in Figure \ref{Fig: Actual Trajectories of Adelaide to Singapore}, it is clear to illustrate that there is no direct route too. Generally, vessels will have a stop at Perth in between. In this case, it is similar to the trajectories above from Singapore to Adelaide, and there are also uncertainties introduced by this stop to influence on ETA prediction toward a corresponding final destination. 

According to the constructed trajectories from Adelaide to Singapore in Figure \ref{Fig: Constructed Trajectories of Adelaide to Singapore} across different scanning internals ($\eta$) and scanning methods, it is clear to note that LatLon-scanning can perform better in latitude and longitude directions, compared to merely Lat-scanning or Lon-scanning. For Lat-scanning, the movements are missing significantly along longitude direction as the scanning interval ($\eta$) increases. While, there are missing points along latitude direction for Lon-scanning. 
	
\begin{figure}[htbp]
		\centering
		\includegraphics[width=\linewidth]{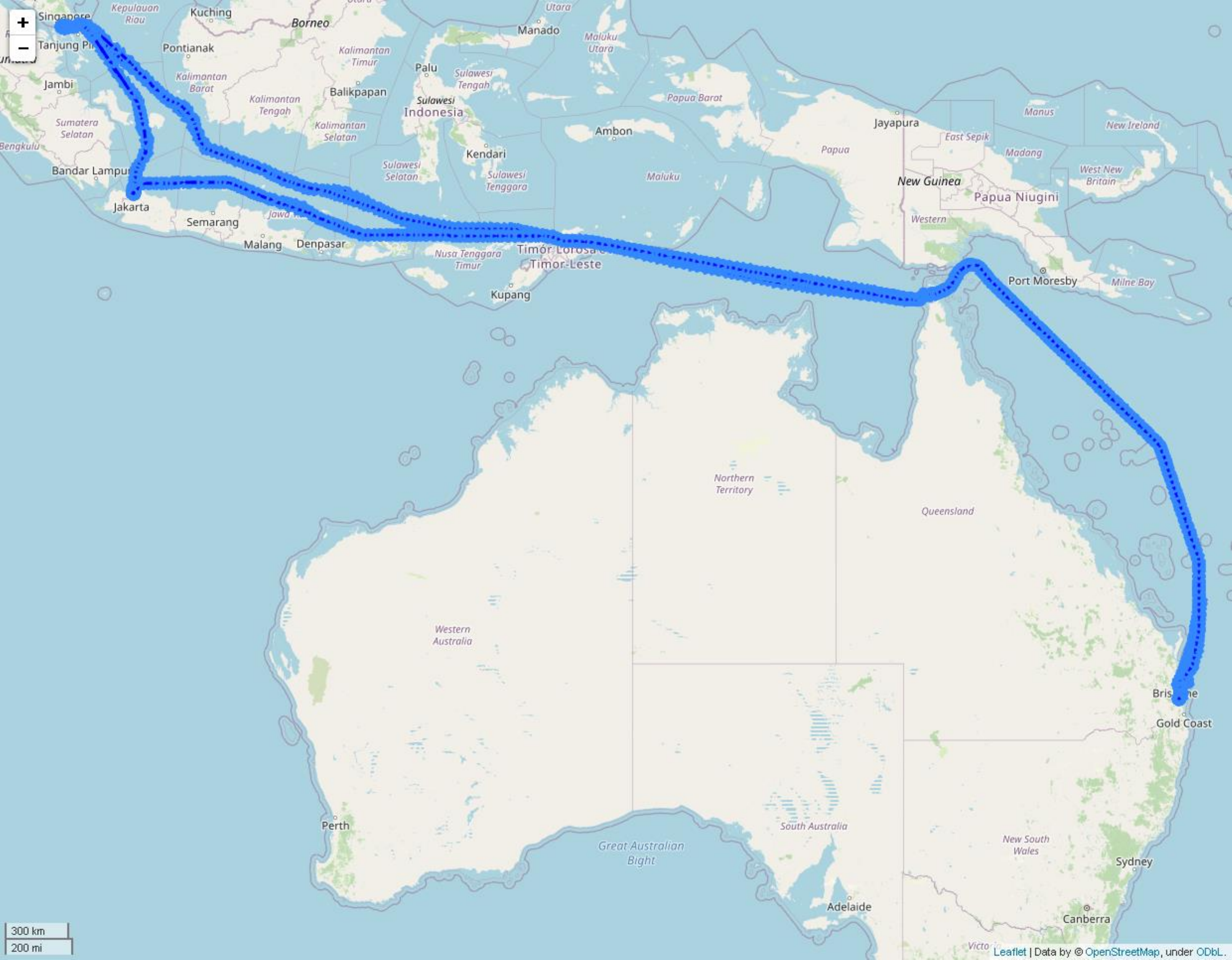}
		\caption{Actual Trajectories (Singapore $\rightarrow$ Brisbane)}
		\label{Fig: Actual Trajectories of Singapore to Brisbane}
	\end{figure}

\begin{figure}[htbp]
		\begin{subfigure}{.3\textwidth}
			\centering
			\includegraphics[width=0.98\linewidth]{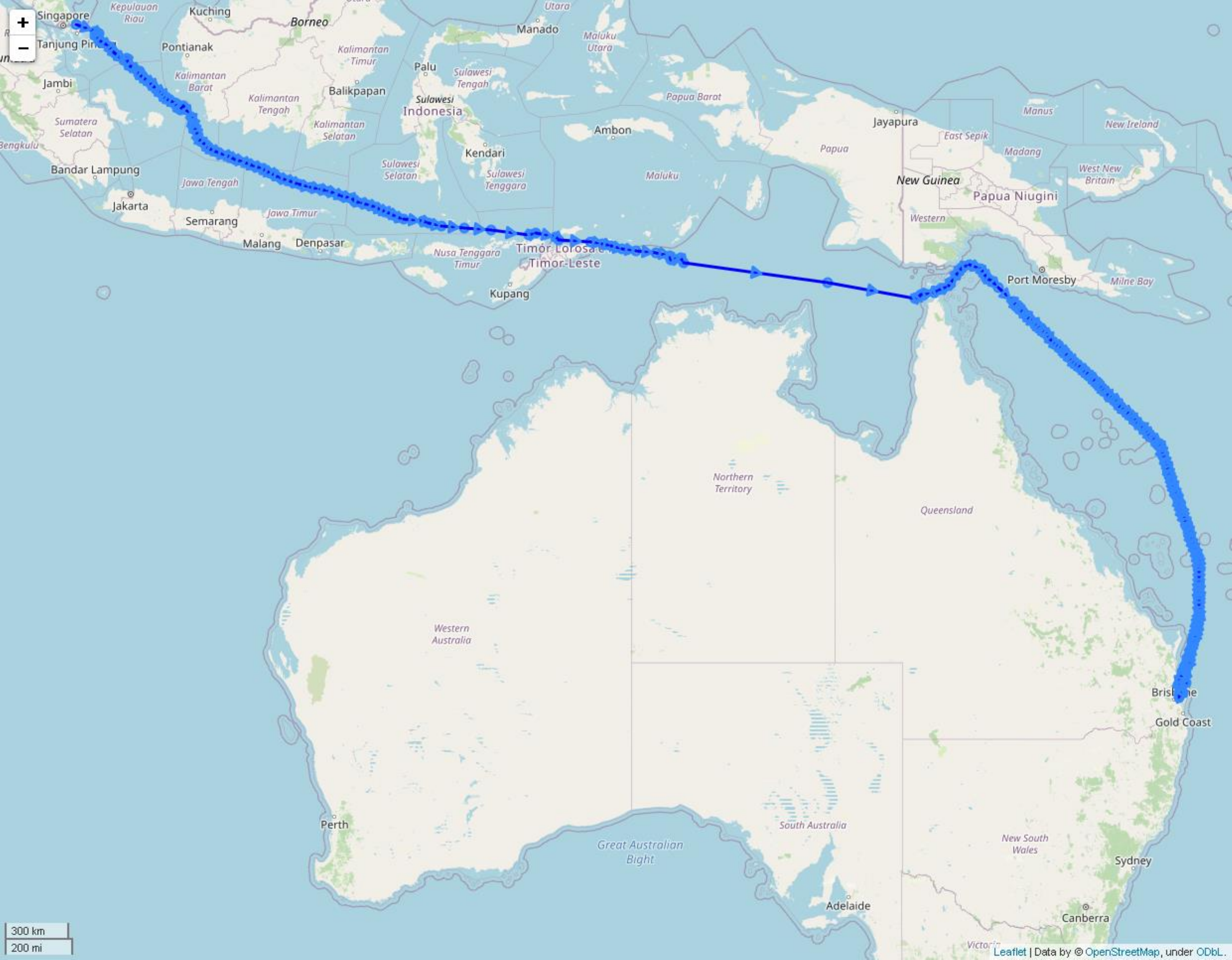}
			\caption{$\eta$=0.1, Lat-Scan}
		\end{subfigure}%
		\begin{subfigure}{.3\textwidth}
			\centering
			\includegraphics[width=0.98\linewidth]{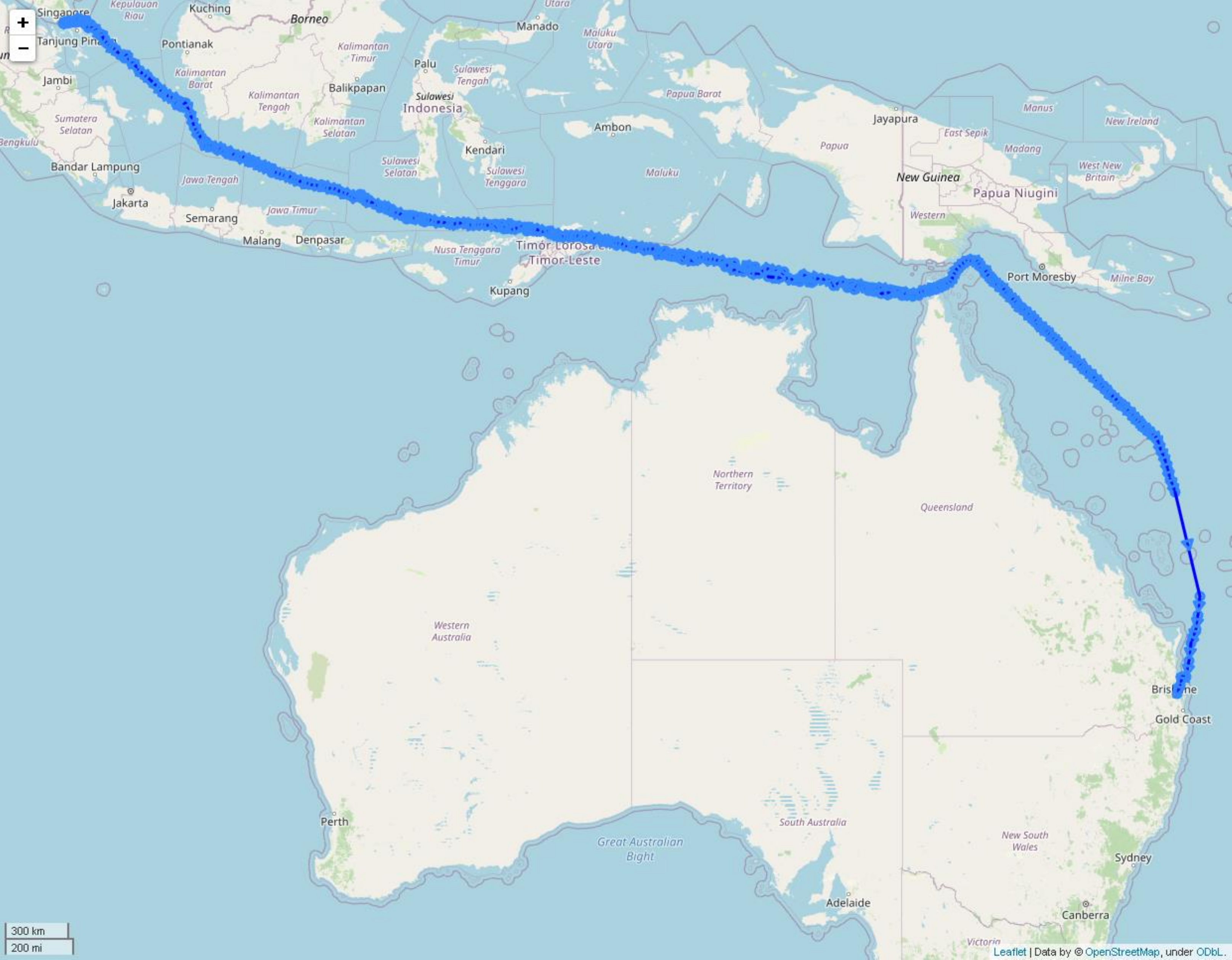}
			\caption{$\eta$=0.1, Lon-Scan}
		\end{subfigure}%
		\begin{subfigure}{.3\textwidth}
			\centering
			\includegraphics[width=0.98\linewidth]{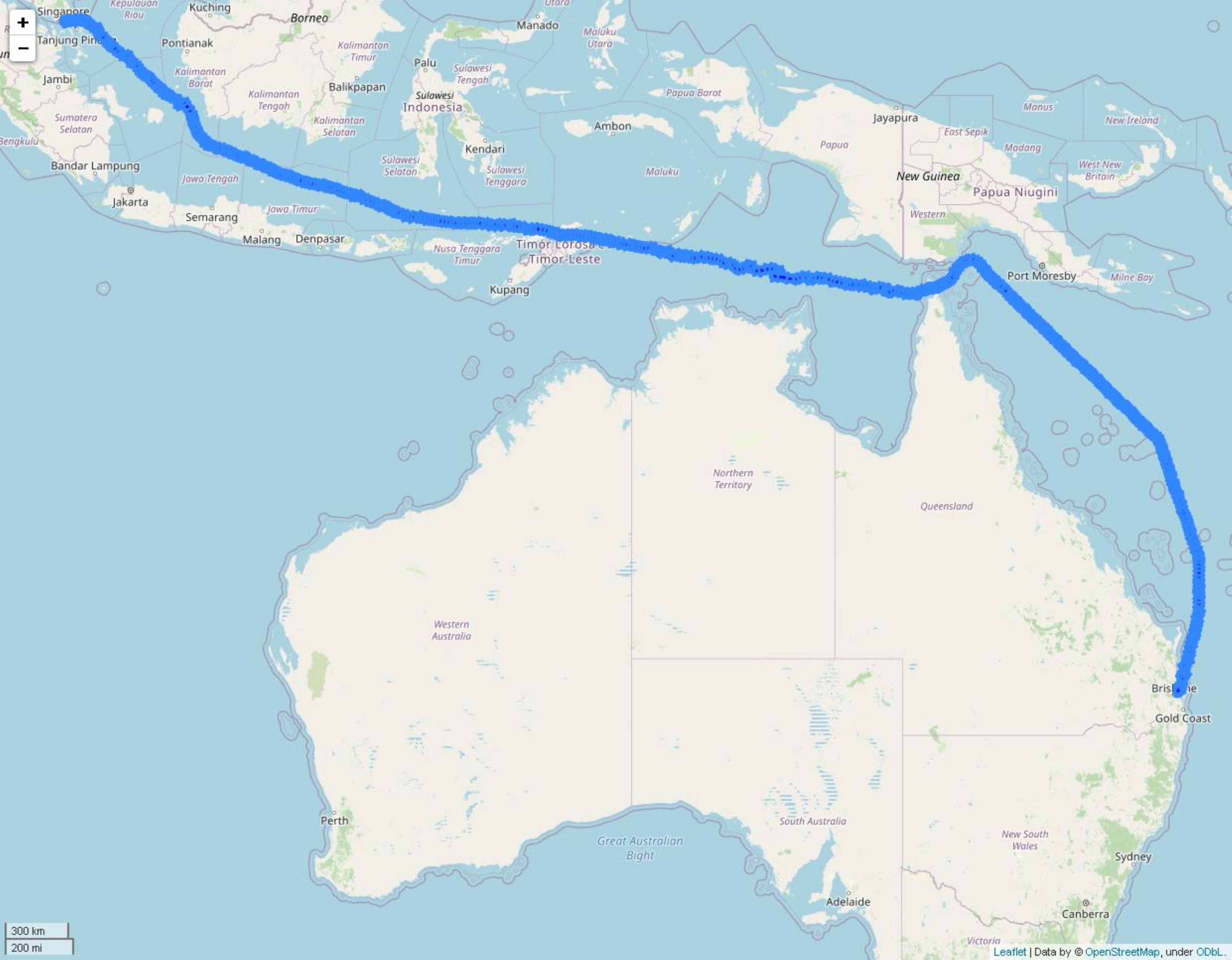}
			\caption{$\eta$=0.1, LatLon-Scan}
		\end{subfigure}

		\begin{subfigure}{.3\textwidth}
			\centering
			\includegraphics[width=0.98\linewidth]{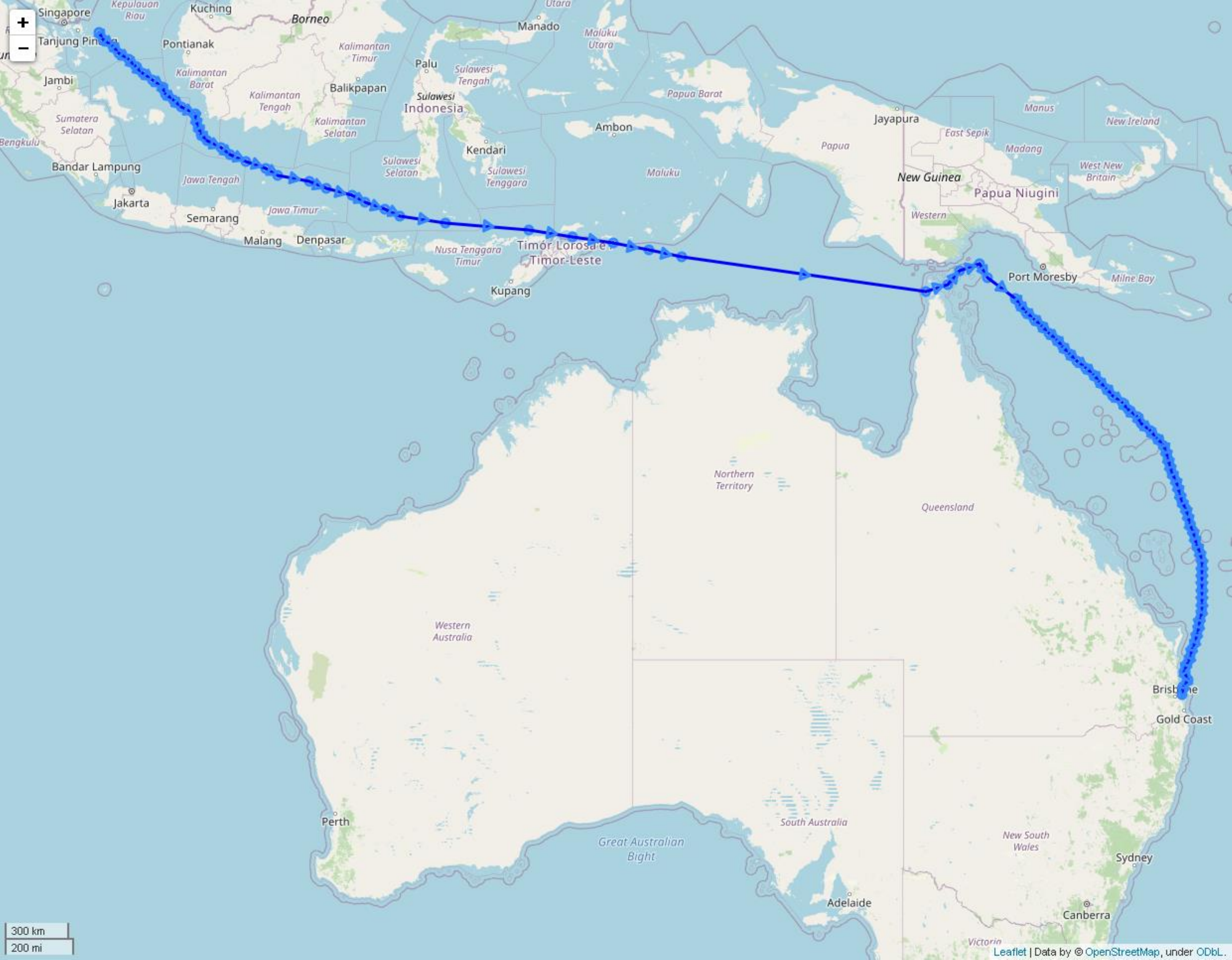}
			\caption{$\eta$=0.3, Lat-Scan}
		\end{subfigure}%
		\begin{subfigure}{.3\textwidth}
			\centering
			\includegraphics[width=0.98\linewidth]{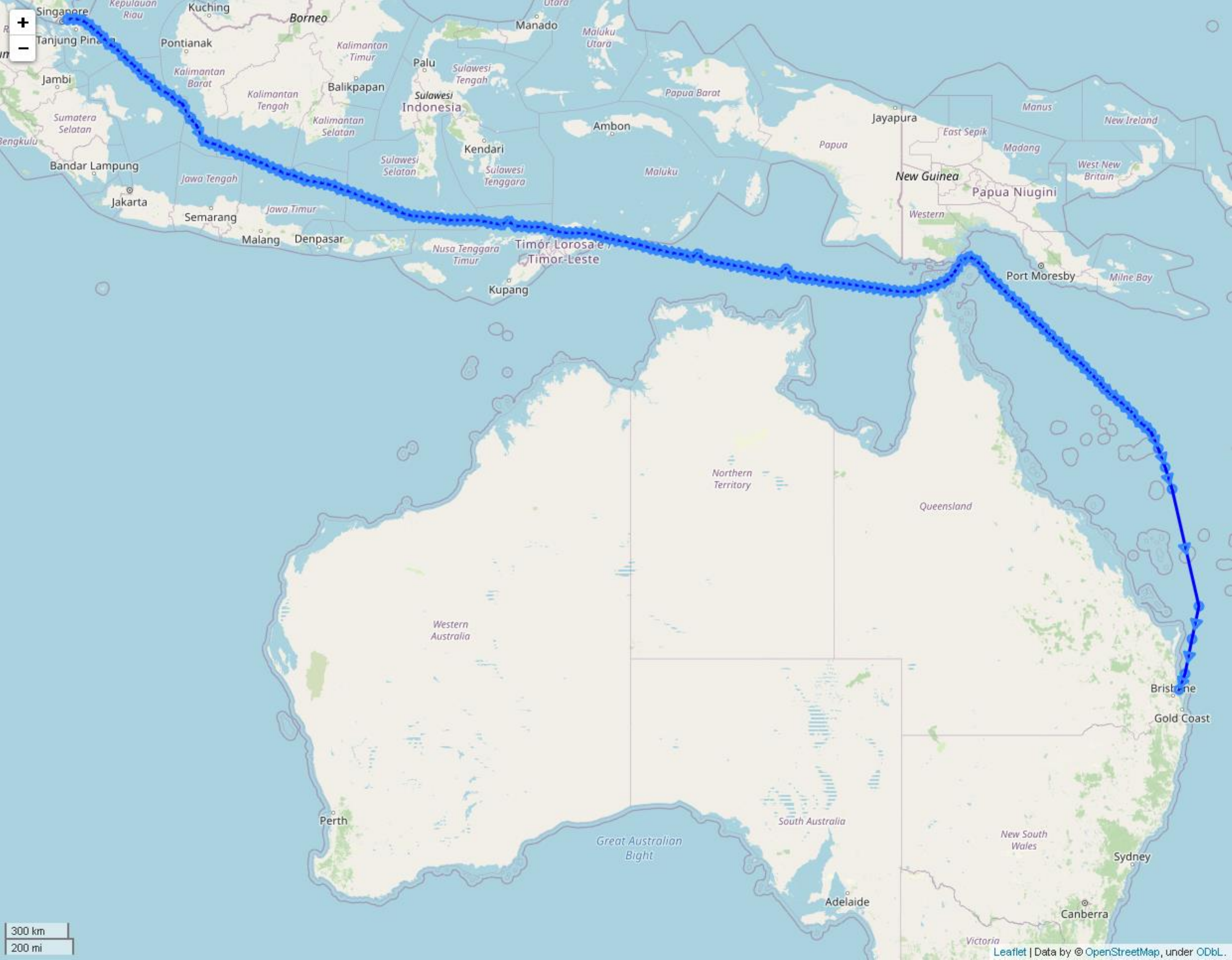}
			\caption{$\eta$=0.3, Lon-Scan}
		\end{subfigure}%
		\begin{subfigure}{.3\textwidth}
			\centering
			\includegraphics[width=0.98\linewidth]{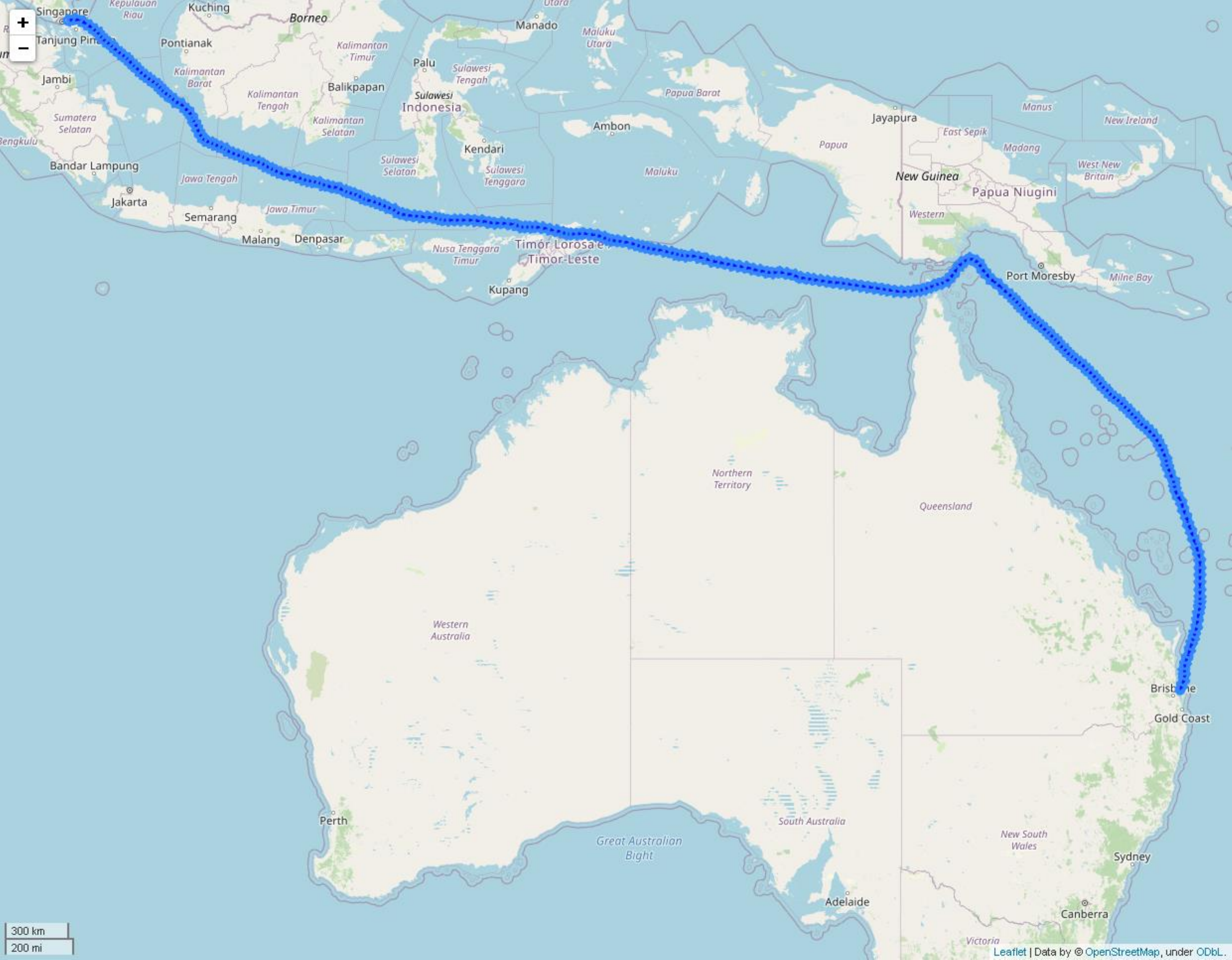}
			\caption{$\eta$=0.3, LatLon-Scan}
		\end{subfigure}

		\begin{subfigure}{.3\textwidth}
			\centering
			\includegraphics[width=0.98\linewidth]{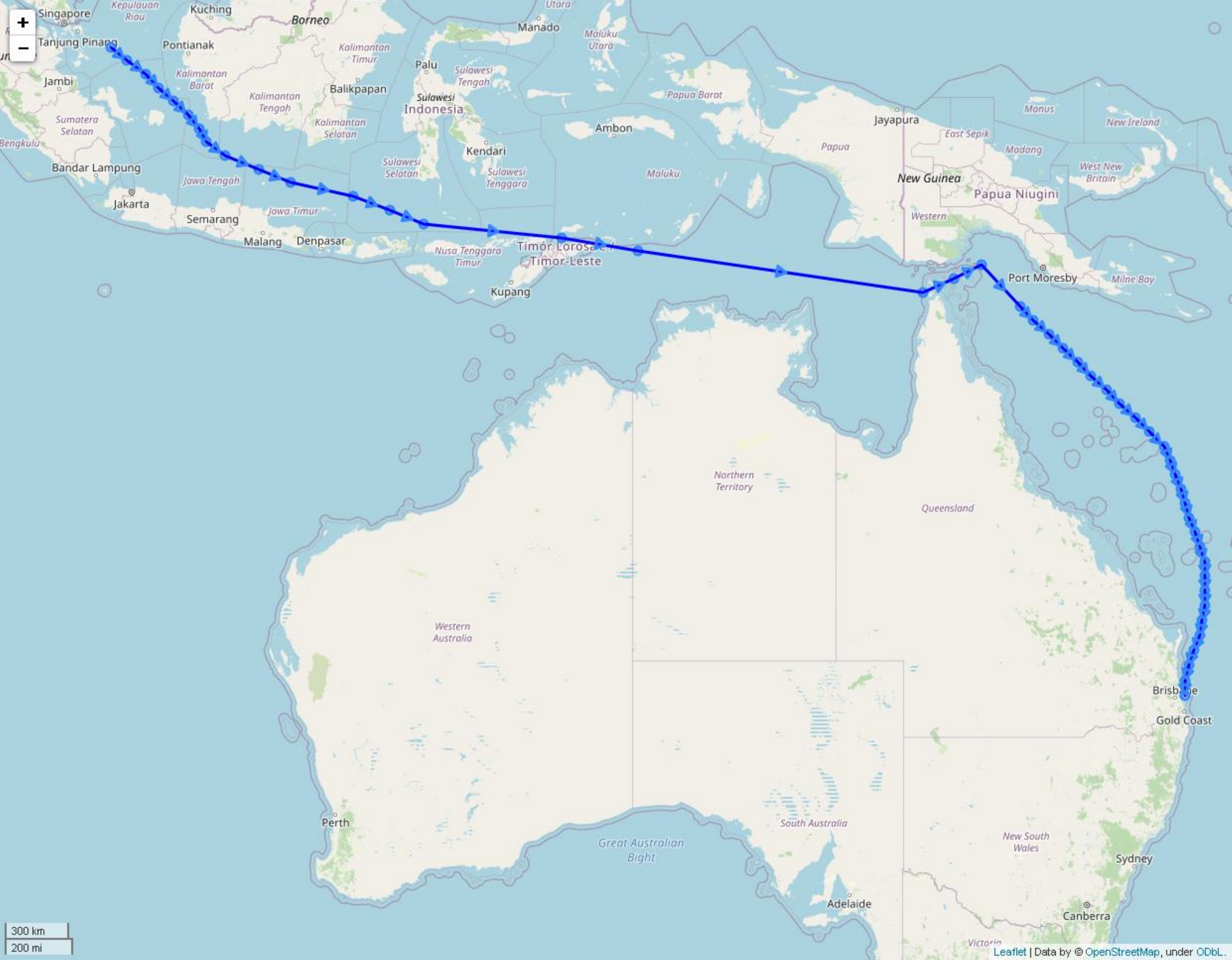}
			\caption{$\eta$=0.6, Lat-Scan}
		\end{subfigure}%
		\begin{subfigure}{.3\textwidth}
			\centering
			\includegraphics[width=0.98\linewidth]{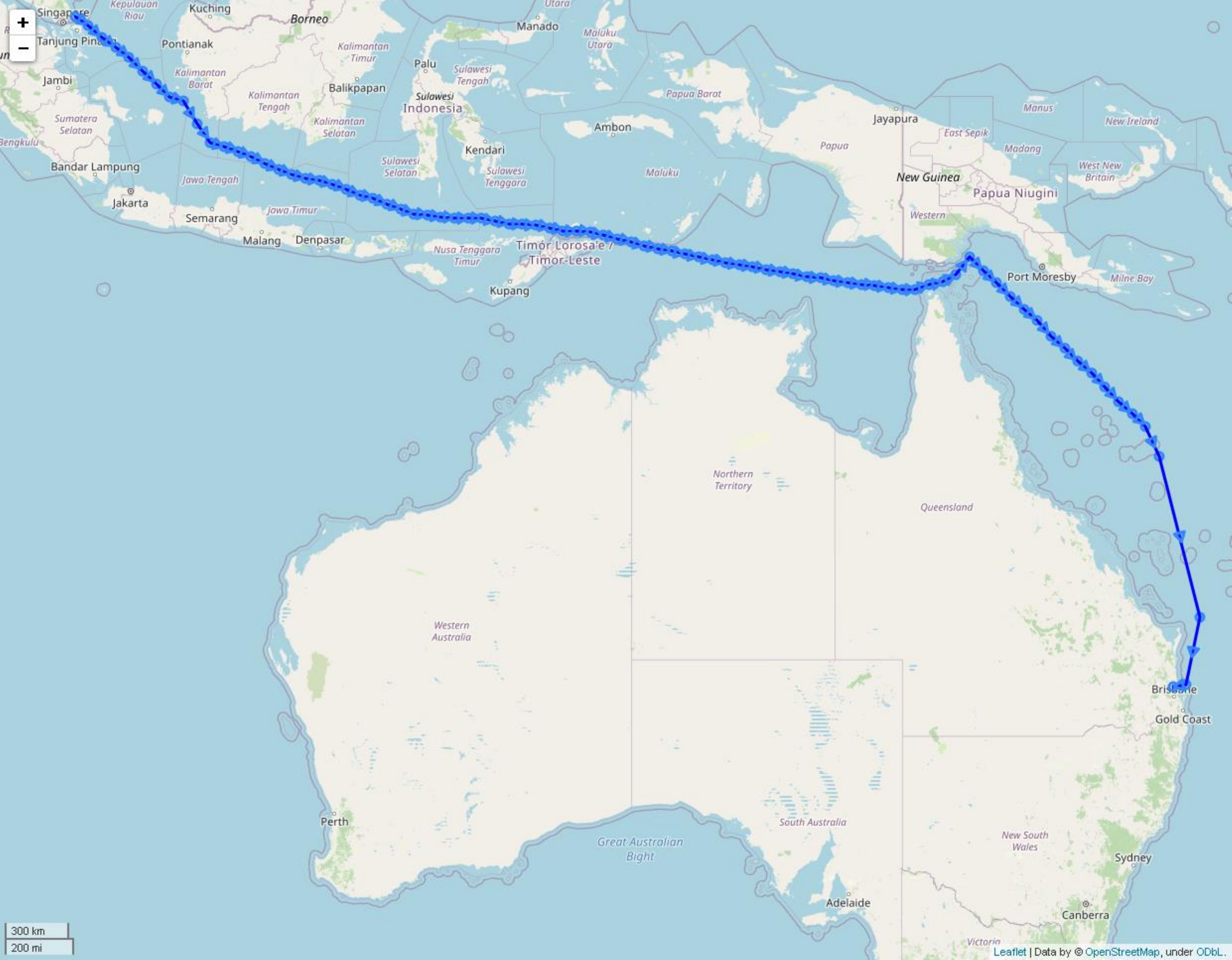}
			\caption{$\eta$=0.6, Lon-Scan}
		\end{subfigure}%
		\begin{subfigure}{.3\textwidth}
			\centering
			\includegraphics[width=0.98\linewidth]{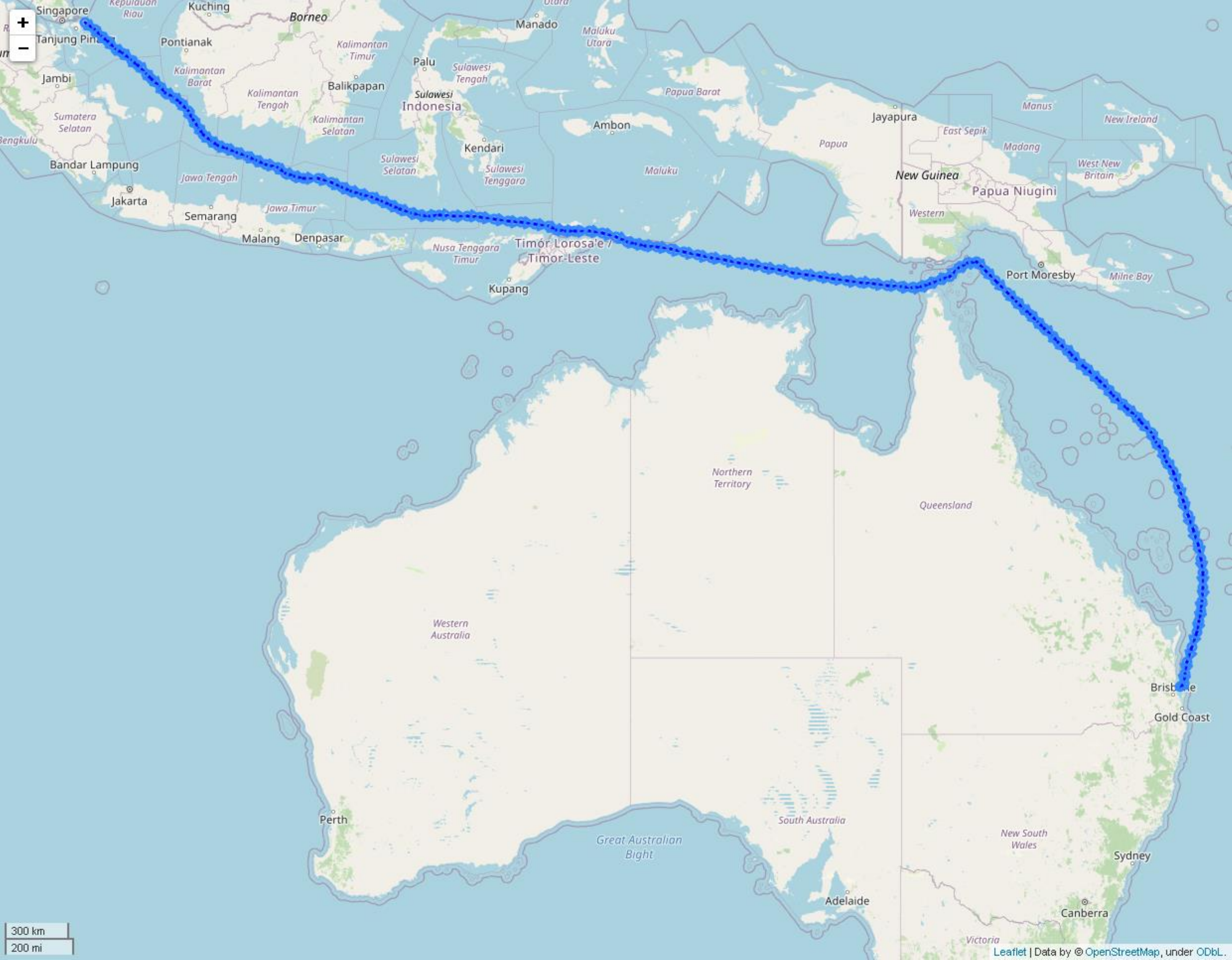}
			\caption{$\eta$=0.6, LatLon-Scan}
		\end{subfigure}

		\begin{subfigure}{.3\textwidth}
			\centering
			\includegraphics[width=0.98\linewidth]{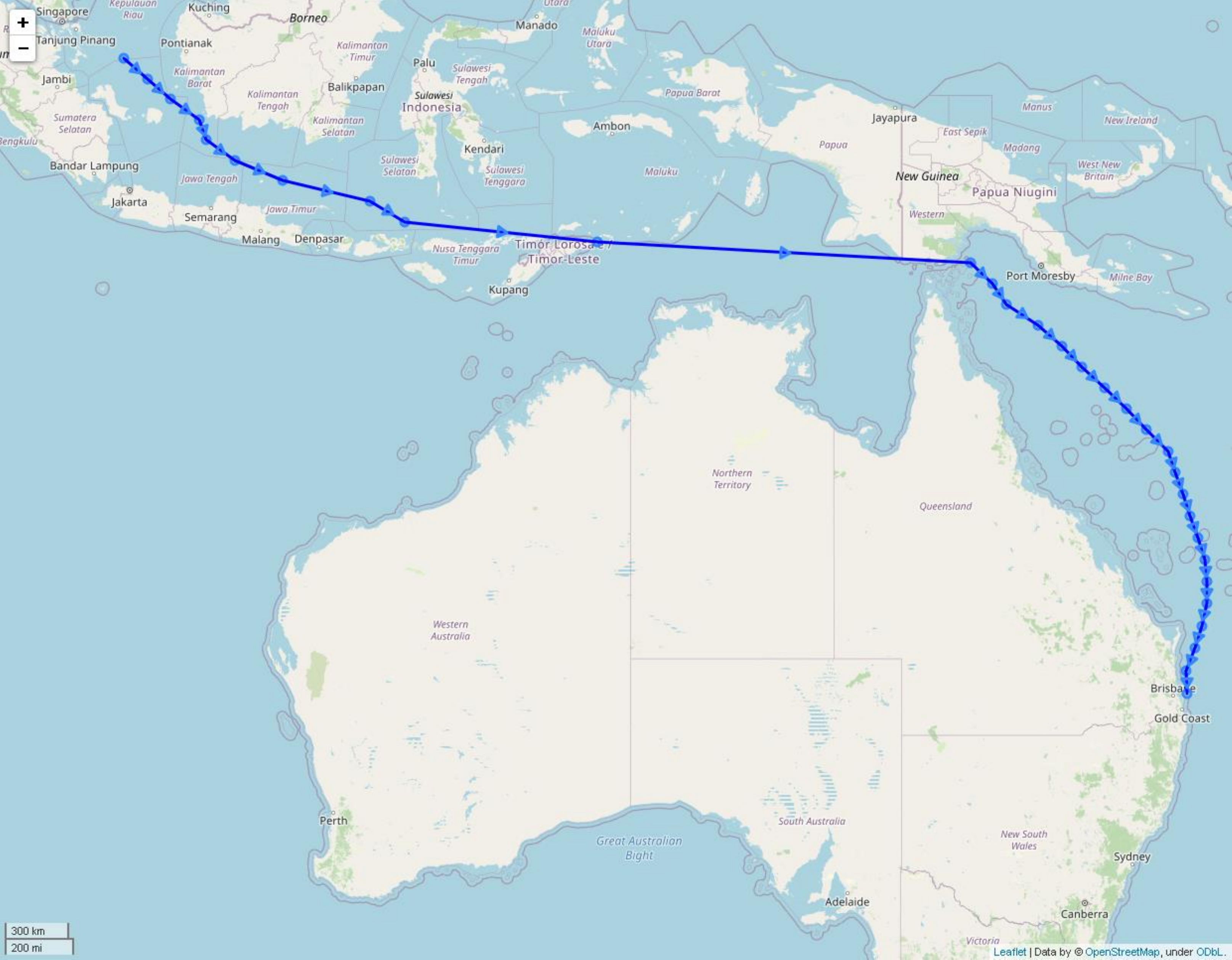}
			\caption{$\eta$=0.9, Lat-Scan}
		\end{subfigure}%
		\begin{subfigure}{.3\textwidth}
			\centering
			\includegraphics[width=0.98\linewidth]{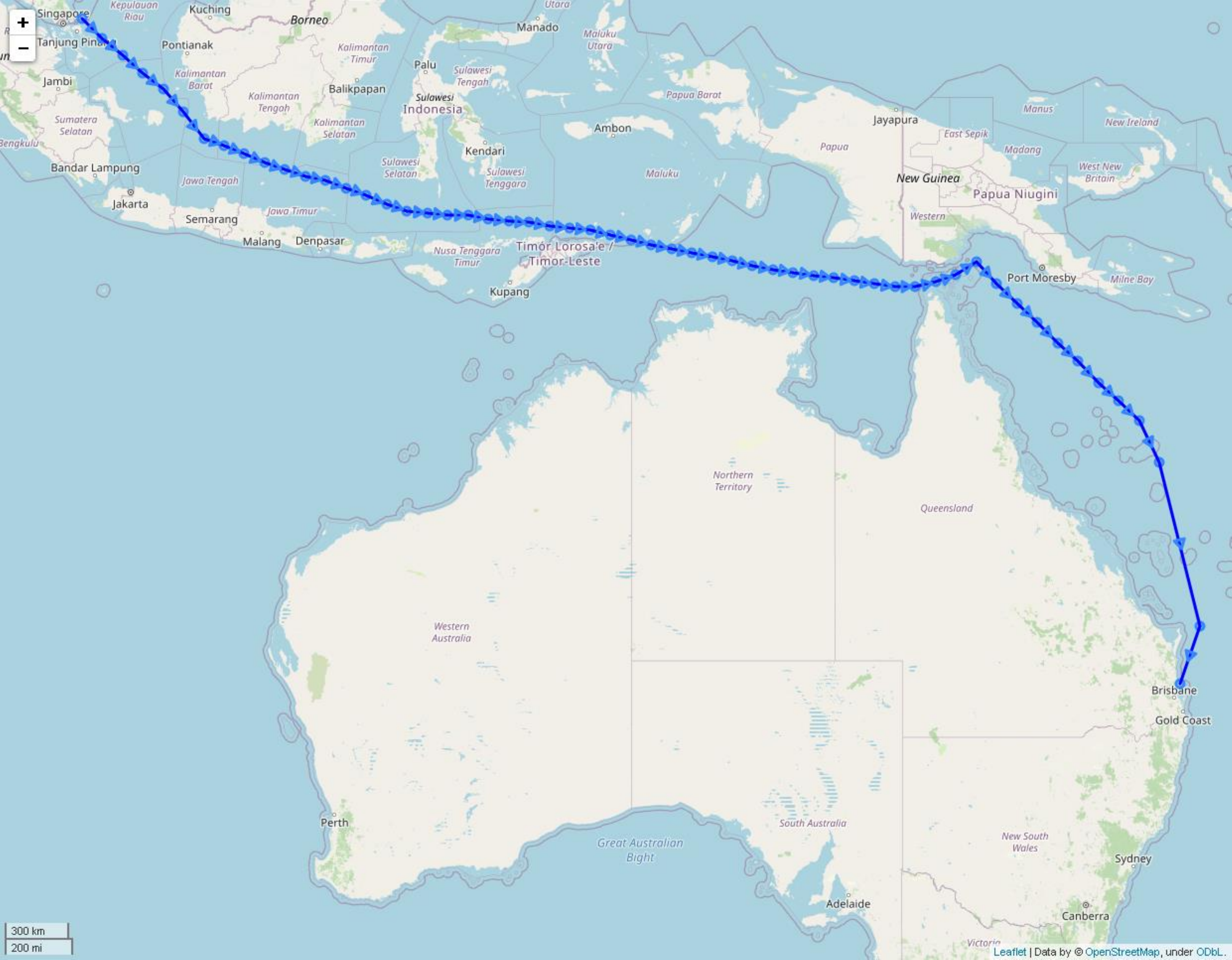}
			\caption{$\eta$=0.9, Lon-Scan}
		\end{subfigure}%
		\begin{subfigure}{.3\textwidth}
			\centering
			\includegraphics[width=0.98\linewidth]{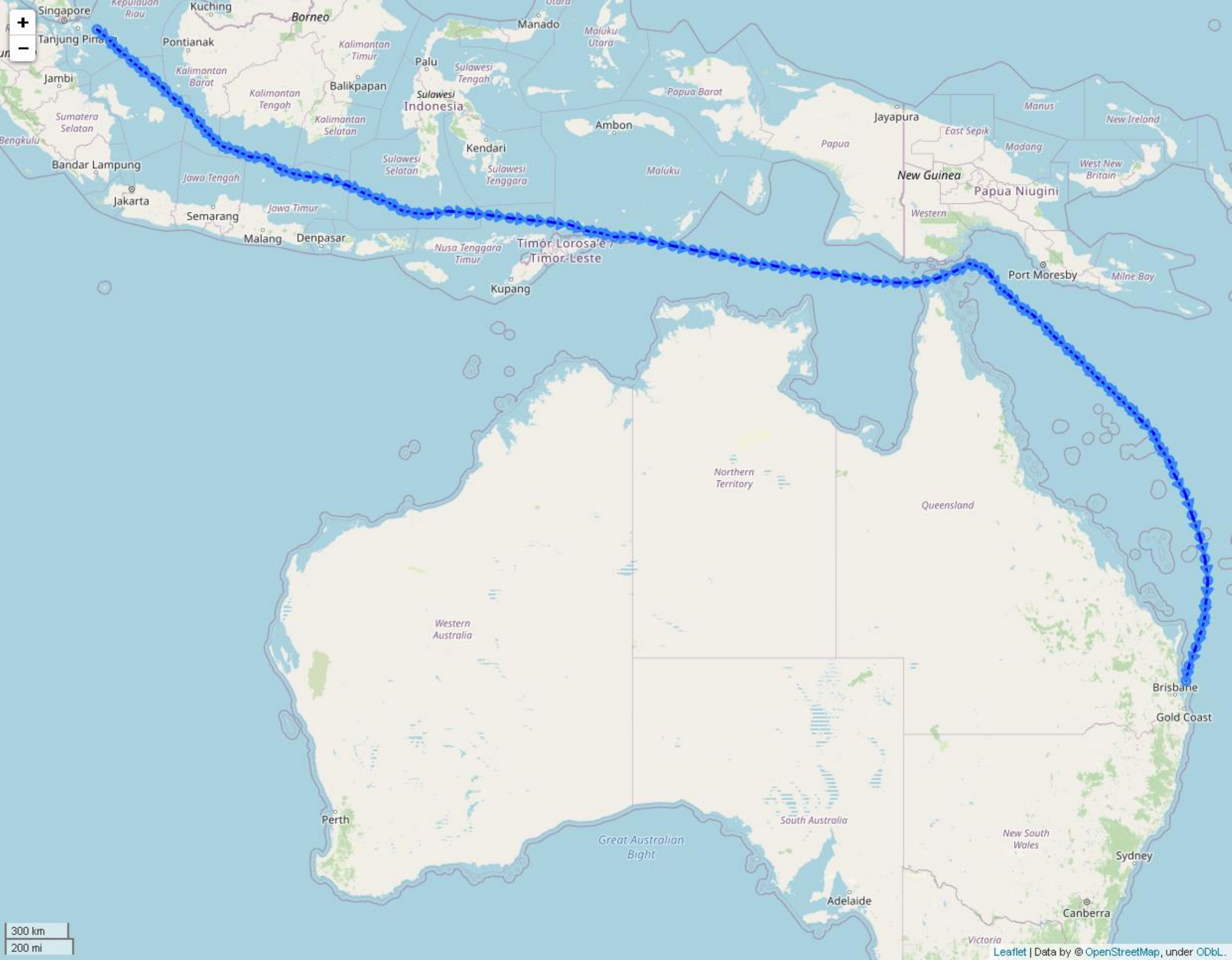}
			\caption{$\eta$=0.9, LatLon-Scan}
		\end{subfigure}

		\begin{subfigure}{.3\textwidth}
			\centering
			\includegraphics[width=0.98\linewidth]{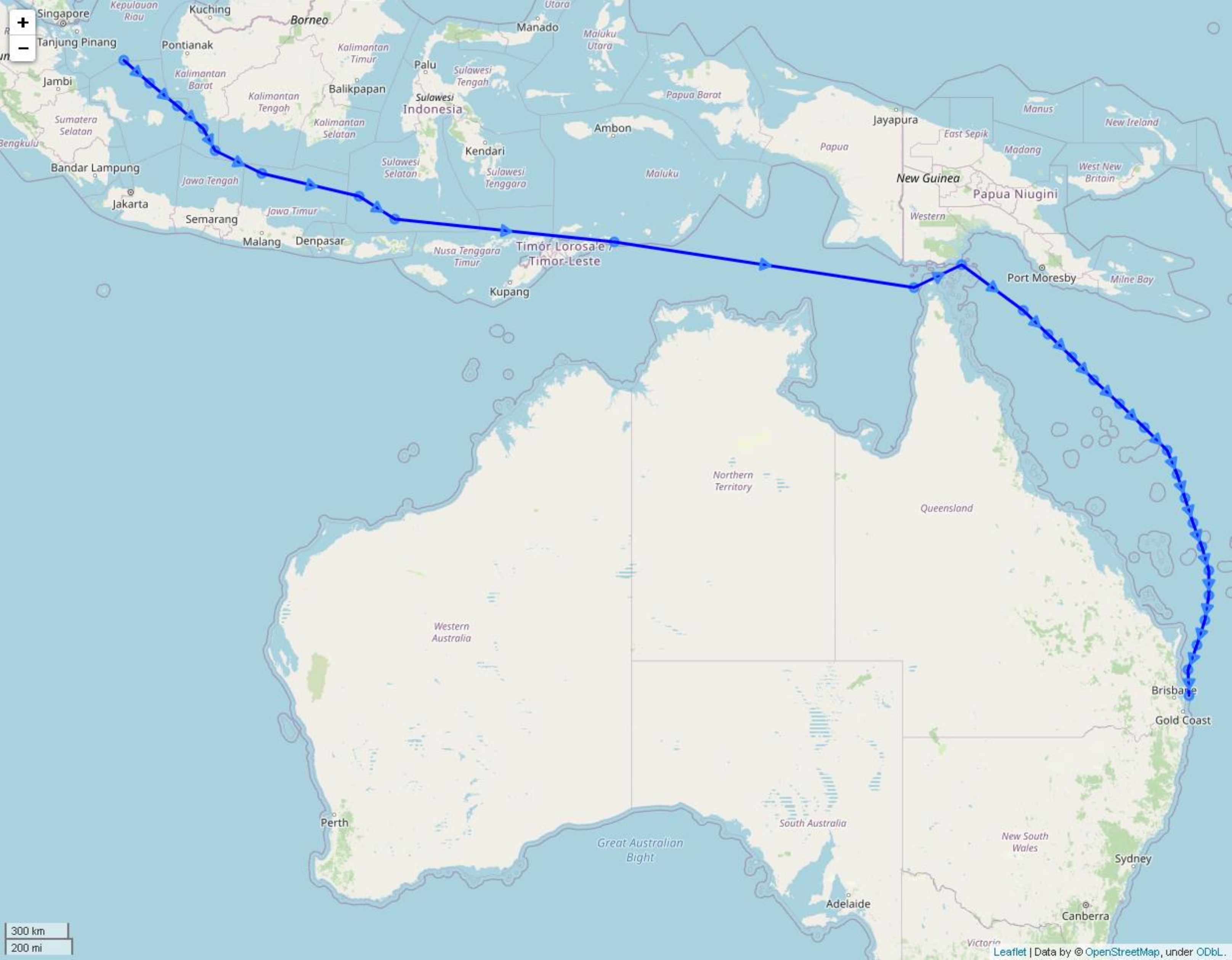}
			\caption{$\eta$=1.0, Lat-Scan}
		\end{subfigure}%
		\begin{subfigure}{.3\textwidth}
			\centering
			\includegraphics[width=0.98\linewidth]{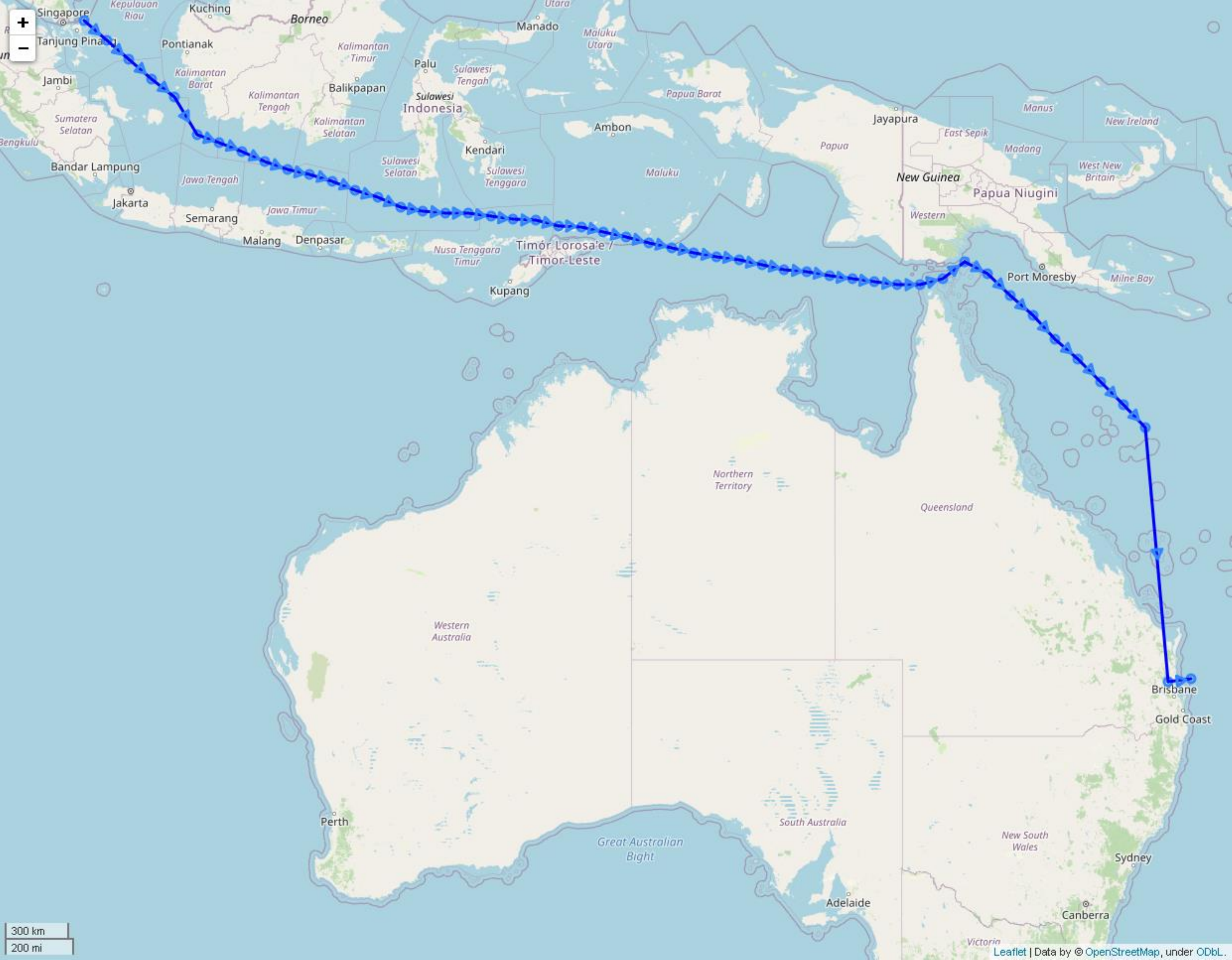}
			\caption{$\eta$=1.0, Lon-Scan}
		\end{subfigure}%
		\begin{subfigure}{.3\textwidth}
			\centering
			\includegraphics[width=0.98\linewidth]{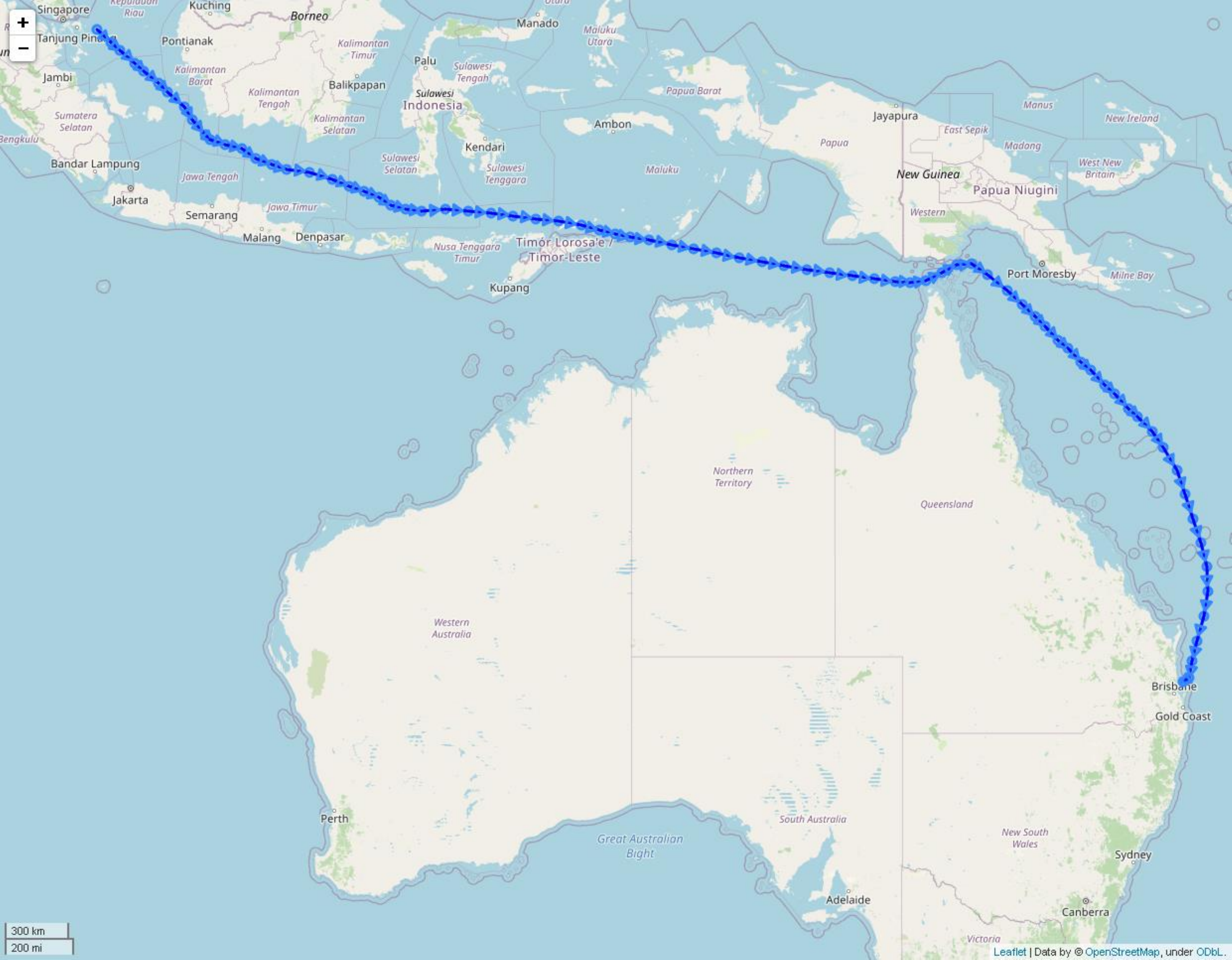}
			\caption{$\eta$=1.0, LatLon-Scan}
		\end{subfigure}

		\caption{Constructed Trajectories (Singapore $\rightarrow$ Brisbane)}
		\label{Fig: Constructed Trajectories of Singapore to Brisbane}
	\end{figure}

\newpage
Here are the historical (Figure \ref{Fig: Actual Trajectories of Singapore to Brisbane}) and constructed (Figure \ref{Fig: Constructed Trajectories of Singapore to Brisbane}) trajectories by Lat-scan, Lon-Scan and LatLon-Scan for Singapore-Brisbane journey with different scanning internals ($\eta$) in degree.

According to the historical trajectories from Singapore to Brisbane in Figure \ref{Fig: Actual Trajectories of Singapore to Brisbane}, it is clear to illustrate that there is a direct route. In this case, it could be more straight forward and accurate to predict ETA of vessels.

According to the constructed trajectories from Singapore to Brisbane in Figure \ref{Fig: Constructed Trajectories of Singapore to Brisbane} across different scanning internals ($\eta$) and scanning methods, it is clear to note that LatLon-scanning can perform better in latitude and longitude directions, compared to merely Lat-scanning or Lon-scanning. For Lat-scanning, the movements are missing significantly near Singapore along longitude direction as the scanning interval ($\eta$) increases. While Lon-scanning has similar missing issue where it approaches Brisbane along latitude direction.
	
\begin{figure}[htbp]
		\centering
		\includegraphics[width=\linewidth]{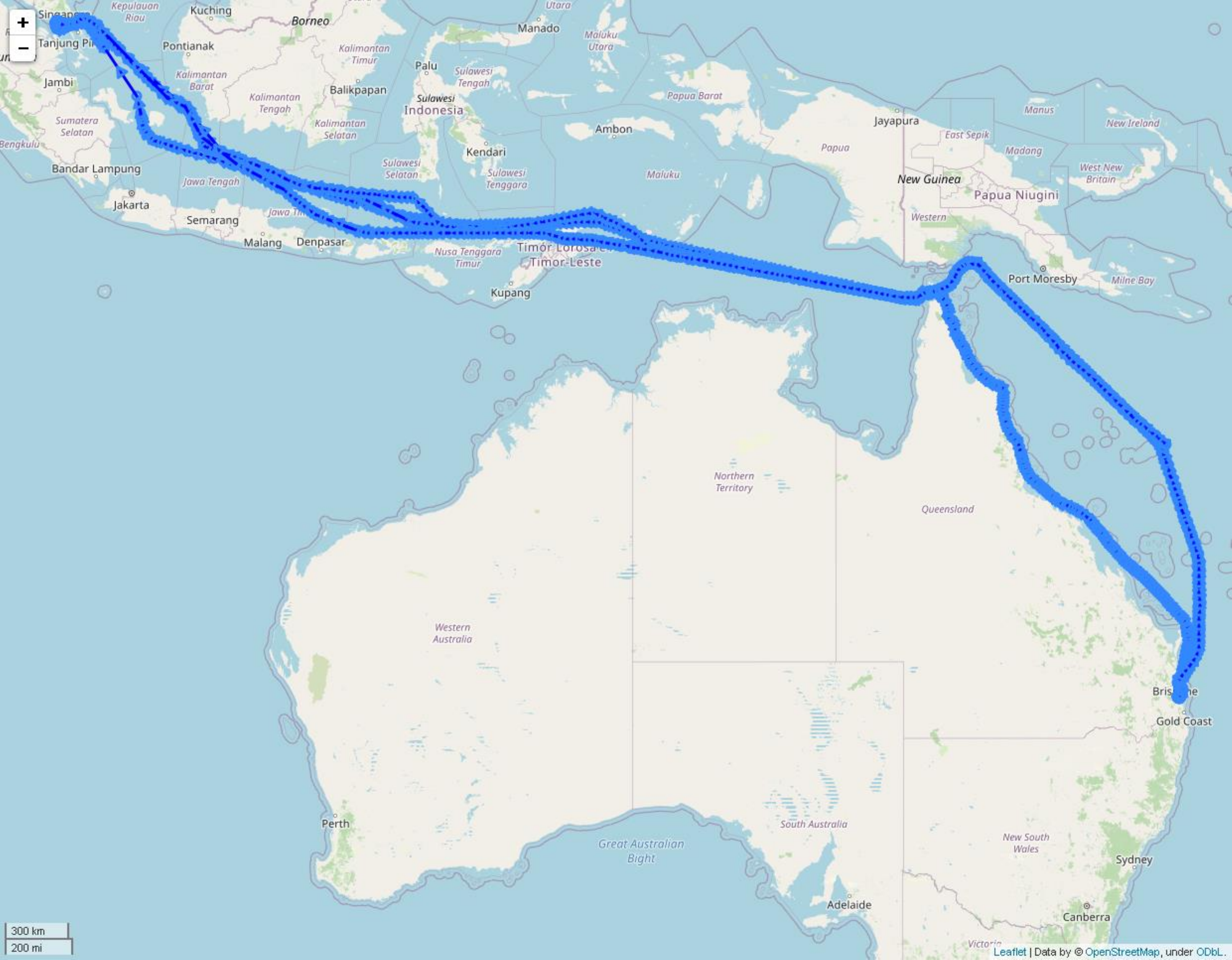}
		\caption{Actual Trajectories (Brisbane $\rightarrow$ Singapore)}
		\label{Fig: Actual Trajectories of Brisbane to Singapore}
	\end{figure}

\begin{figure}[htbp]
		\begin{subfigure}{.3\textwidth}
			\centering
			\includegraphics[width=0.98\linewidth]{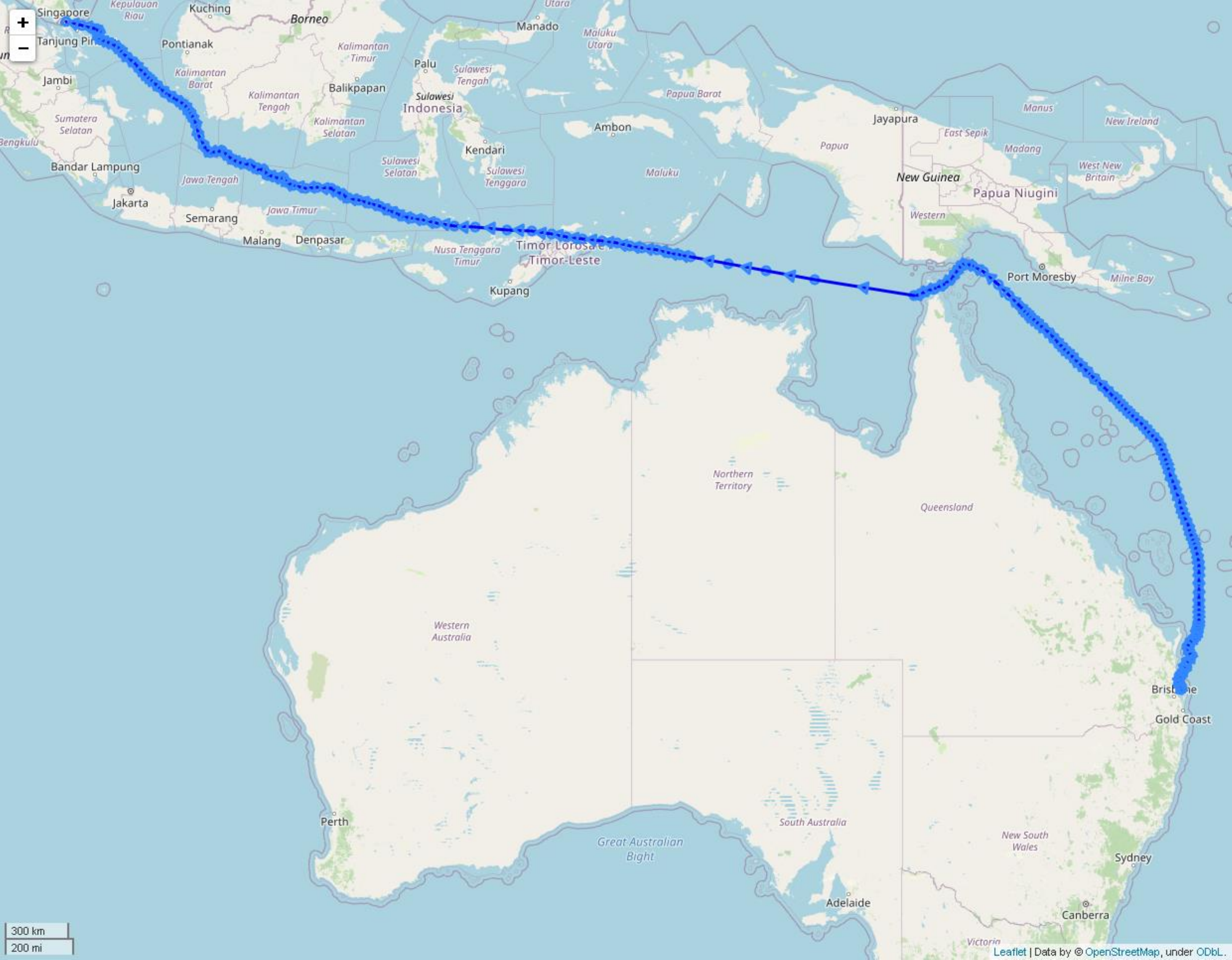}
			\caption{$\eta$=0.1, Lat-Scan}
		\end{subfigure}%
		\begin{subfigure}{.3\textwidth}
			\centering
			\includegraphics[width=0.98\linewidth]{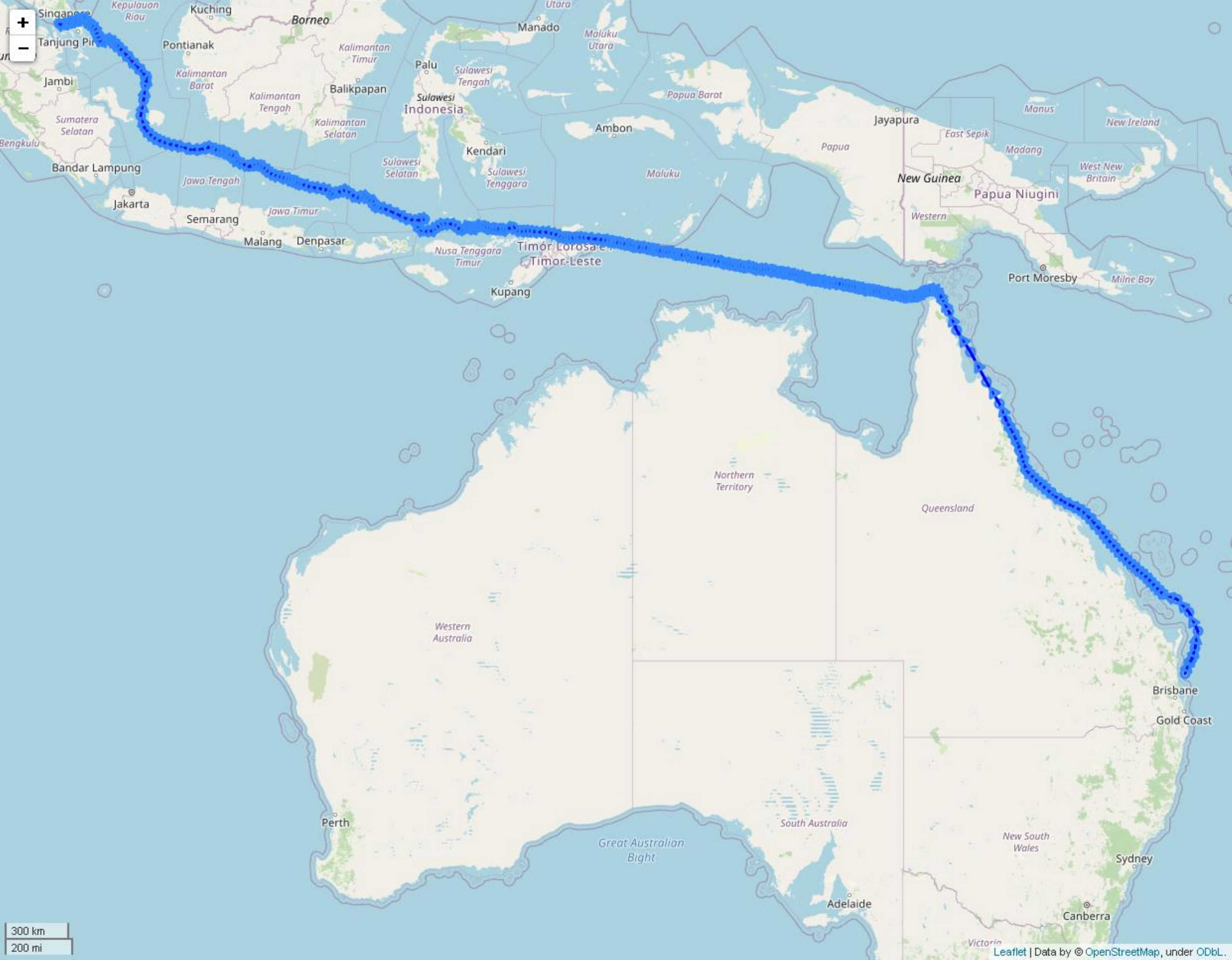}
			\caption{$\eta$=0.1, Lon-Scan}
		\end{subfigure}%
		\begin{subfigure}{.3\textwidth}
			\centering
			\includegraphics[width=0.98\linewidth]{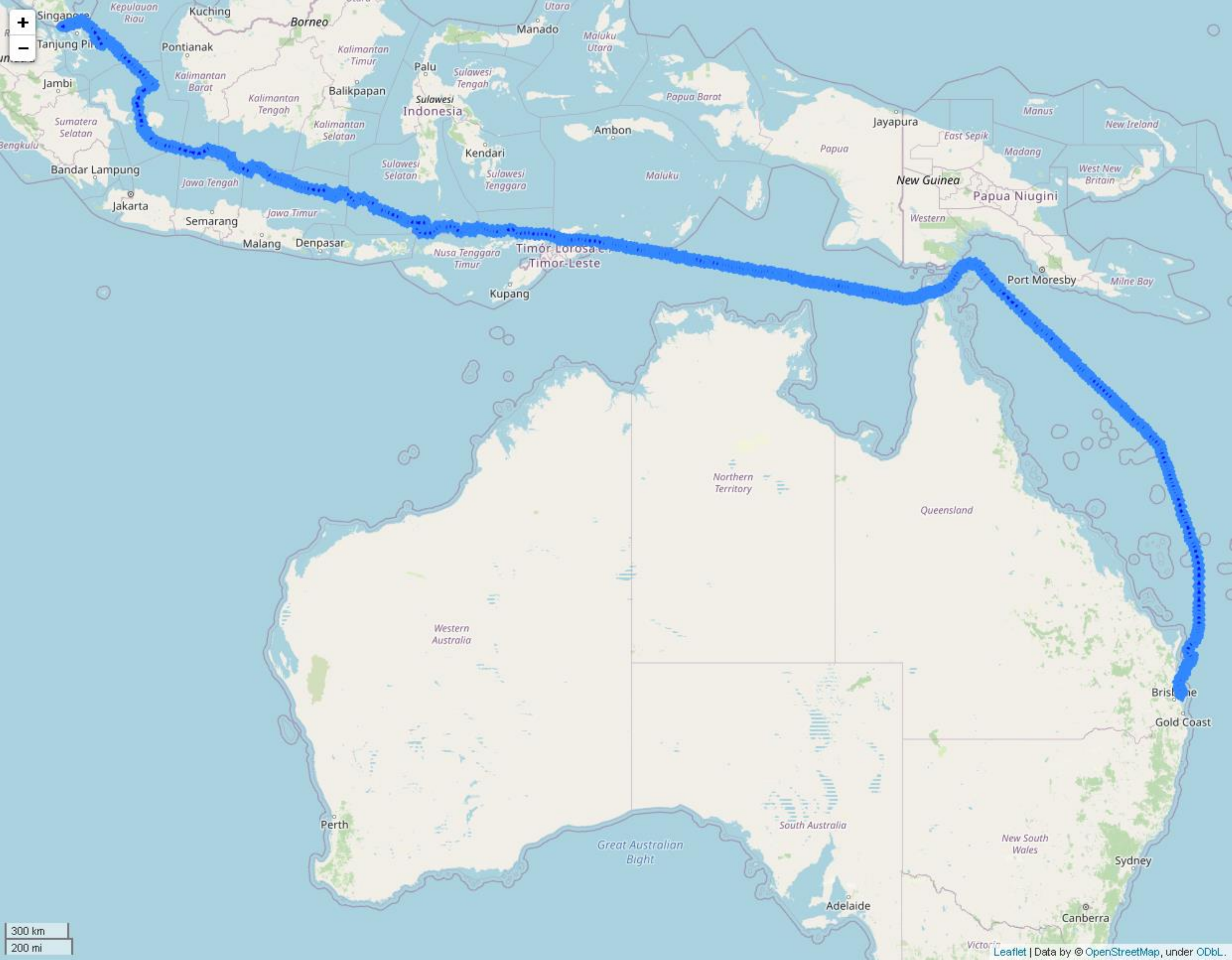}
			\caption{$\eta$=0.1, LatLon-Scan}
		\end{subfigure}

		\begin{subfigure}{.3\textwidth}
			\centering
			\includegraphics[width=0.98\linewidth]{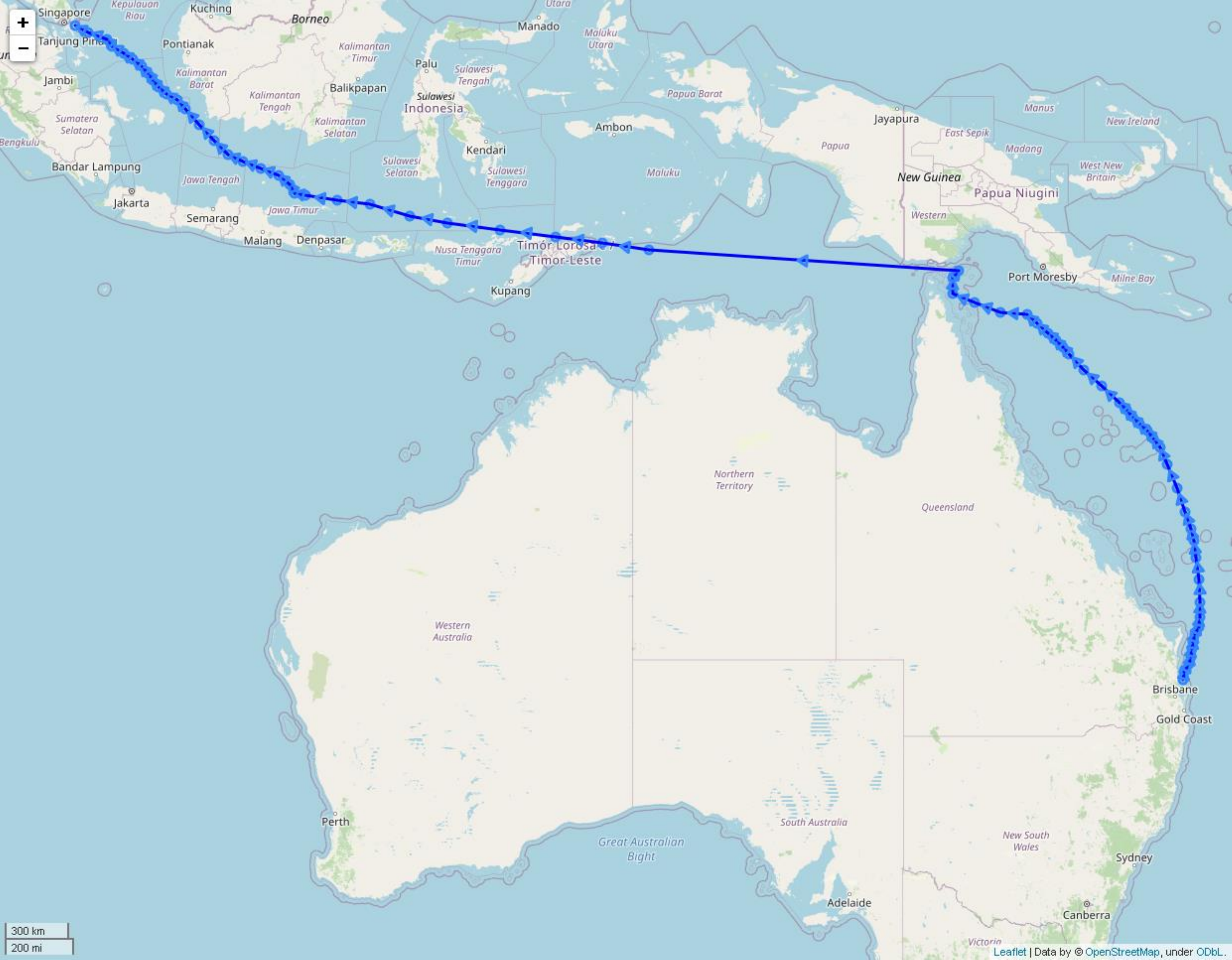}
			\caption{$\eta$=0.3, Lat-Scan}
		\end{subfigure}%
		\begin{subfigure}{.3\textwidth}
			\centering
			\includegraphics[width=0.98\linewidth]{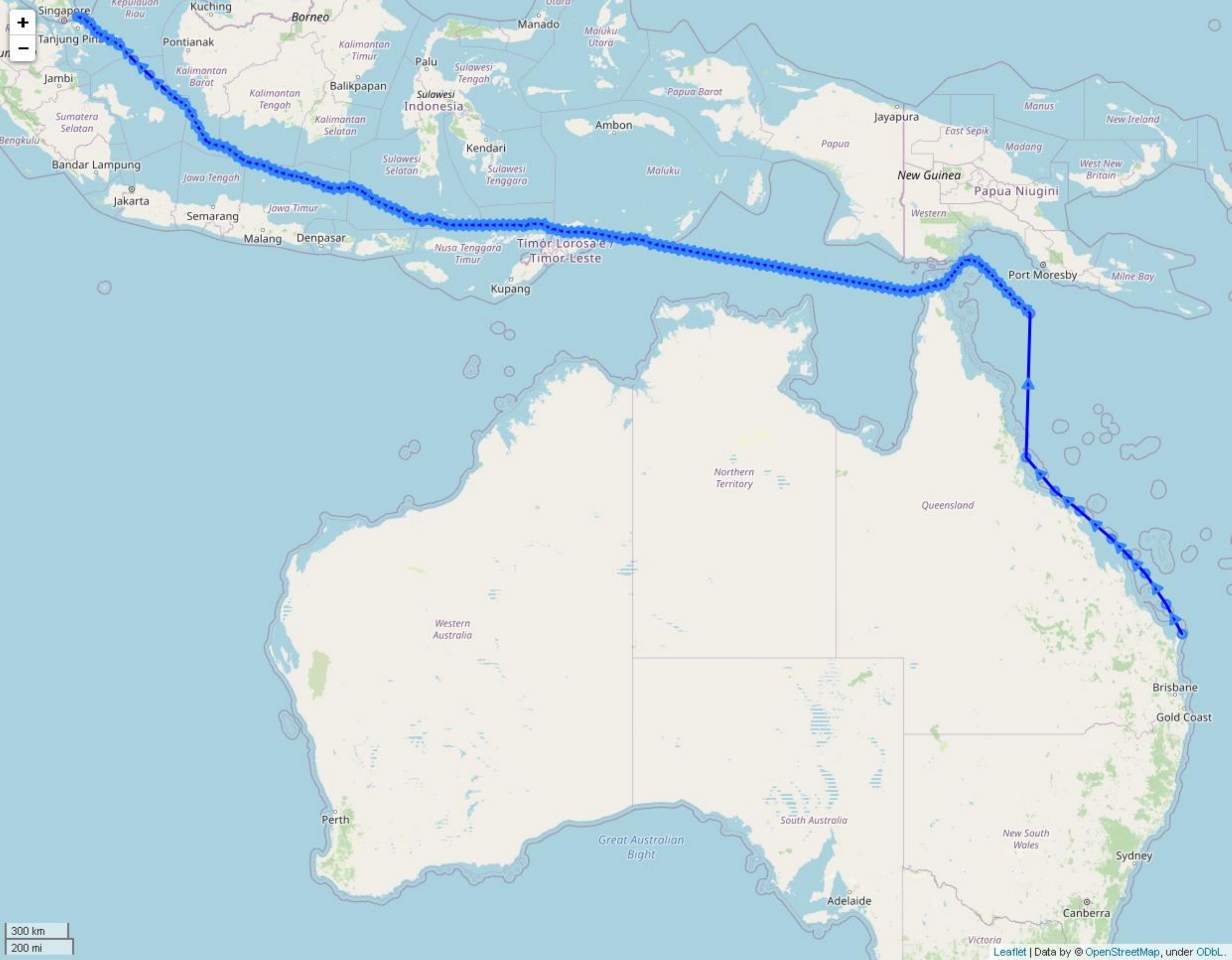}
			\caption{$\eta$=0.3, Lon-Scan}
		\end{subfigure}%
		\begin{subfigure}{.3\textwidth}
			\centering
			\includegraphics[width=0.98\linewidth]{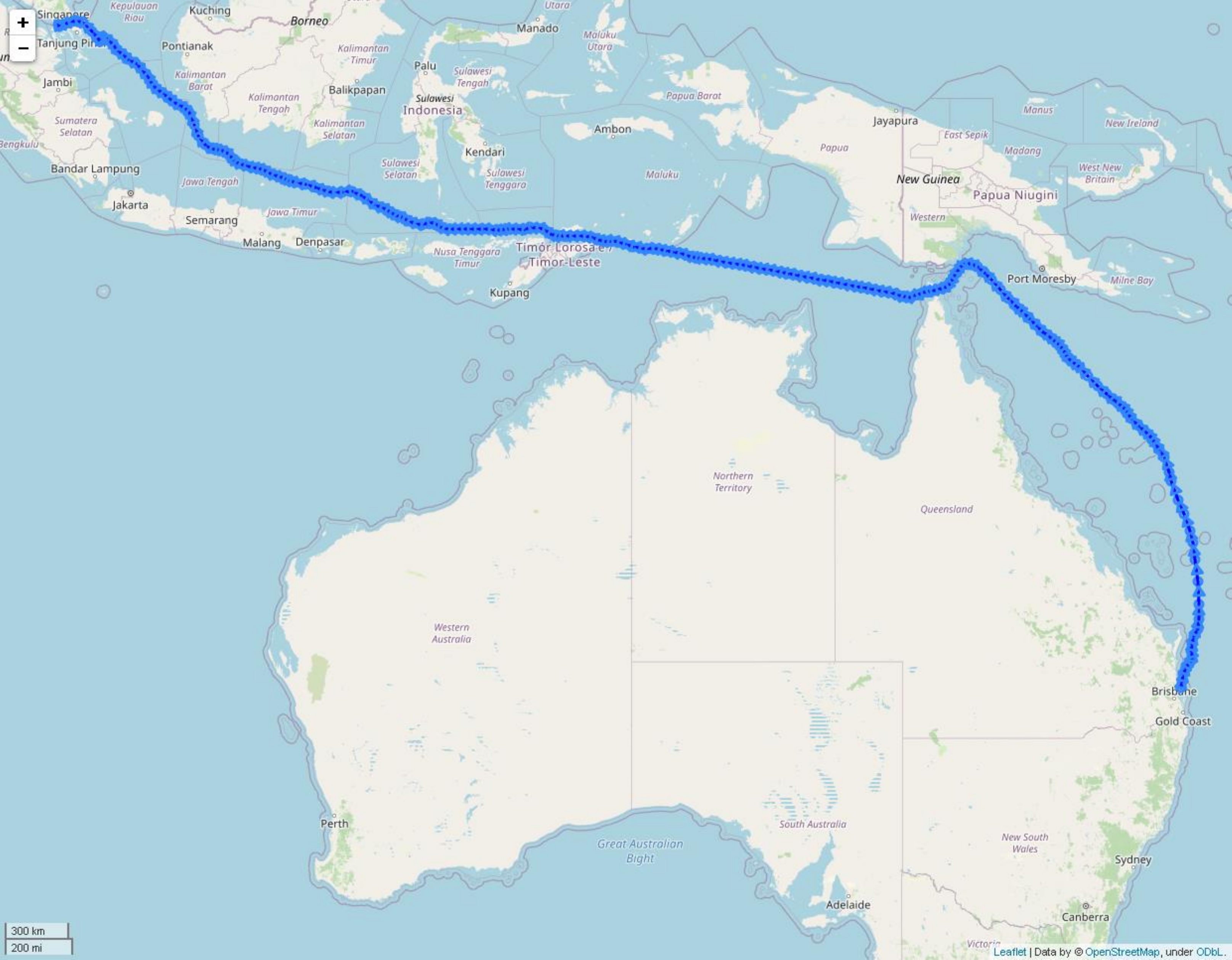}
			\caption{$\eta$=0.3, LatLon-Scan}
		\end{subfigure}

		\begin{subfigure}{.3\textwidth}
			\centering
			\includegraphics[width=0.98\linewidth]{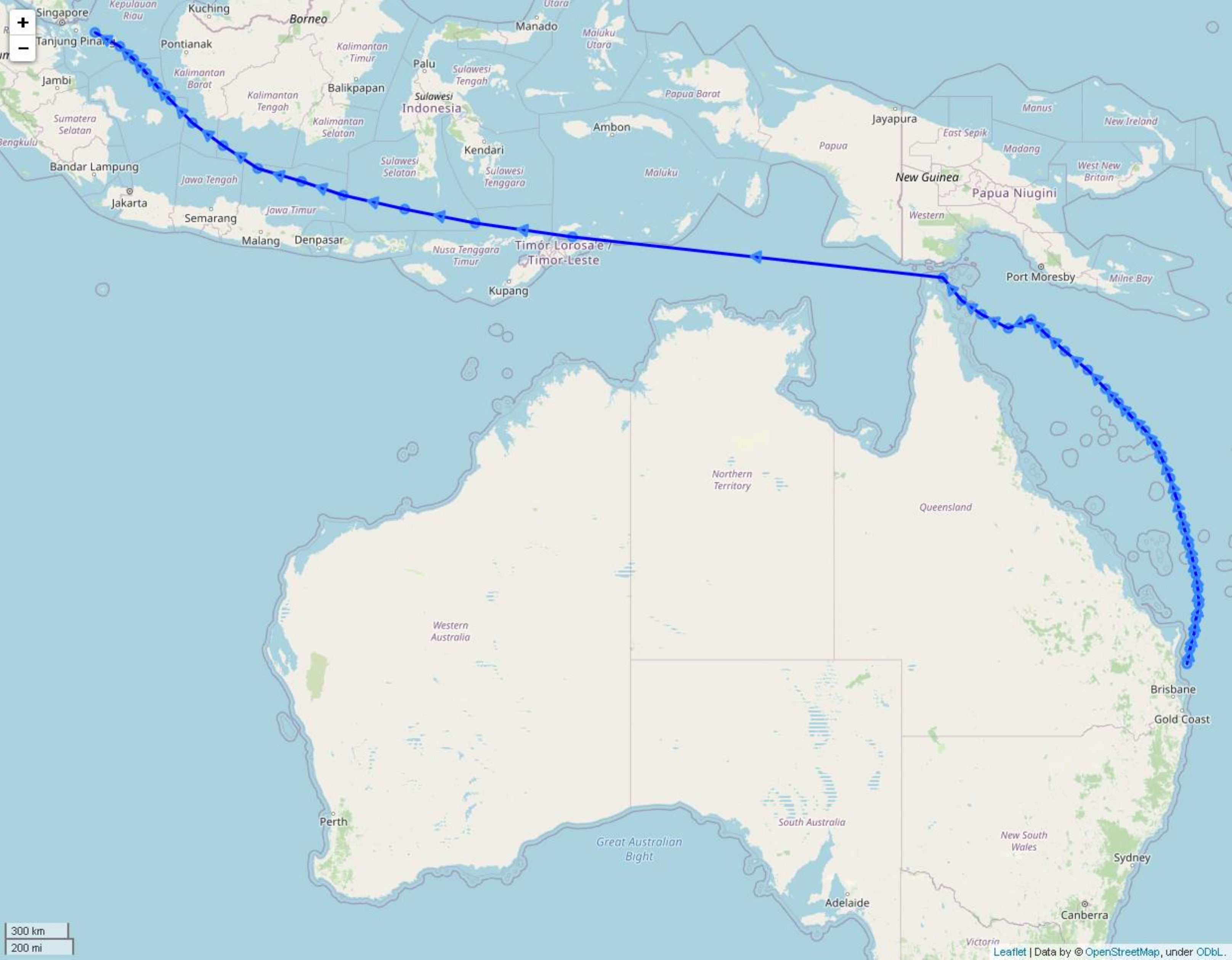}
			\caption{$\eta$=0.6, Lat-Scan}
		\end{subfigure}%
		\begin{subfigure}{.3\textwidth}
			\centering
			\includegraphics[width=0.98\linewidth]{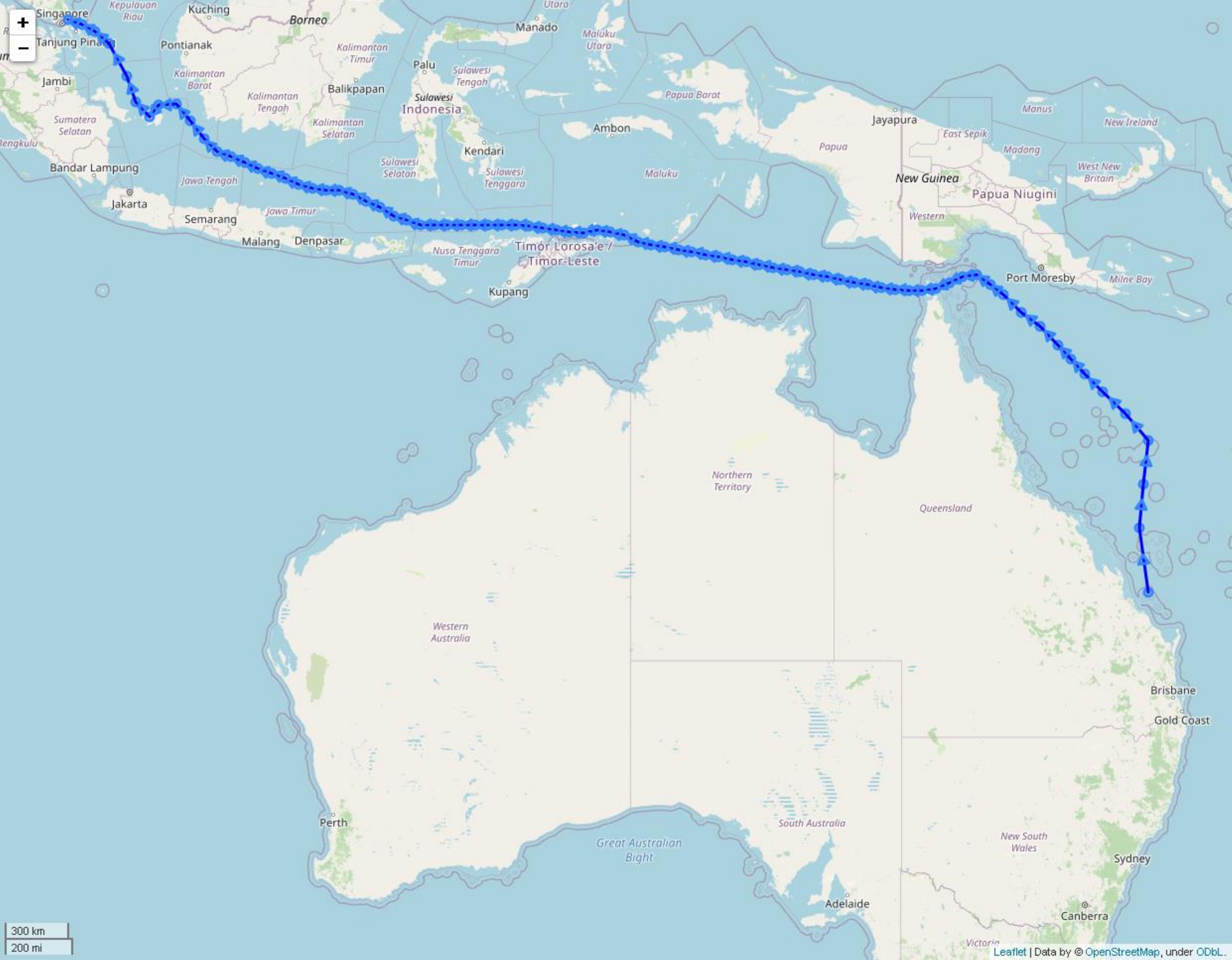}
			\caption{$\eta$=0.6, Lon-Scan}
		\end{subfigure}%
		\begin{subfigure}{.3\textwidth}
			\centering
			\includegraphics[width=0.98\linewidth]{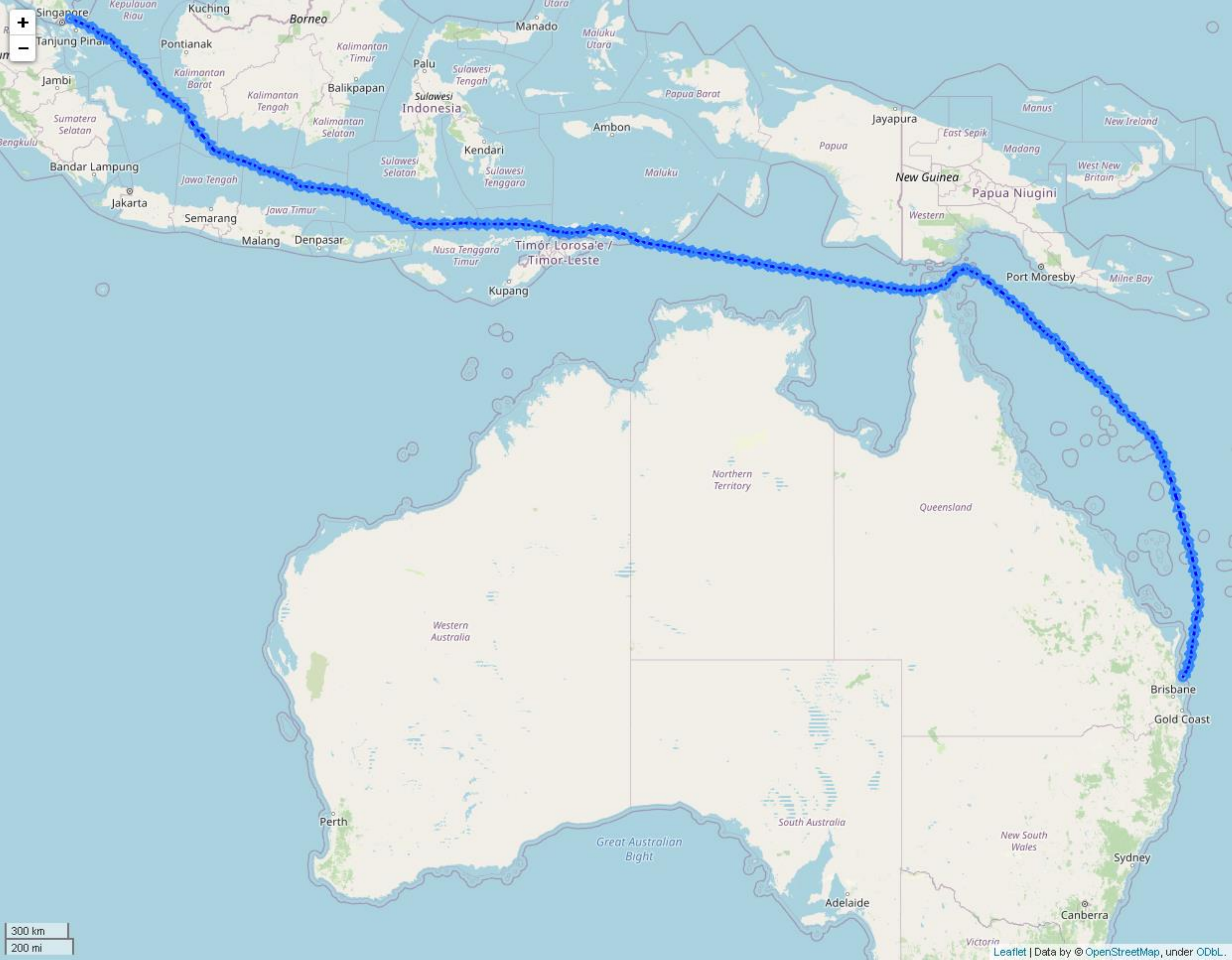}
			\caption{$\eta$=0.6, LatLon-Scan}
		\end{subfigure}

		\begin{subfigure}{.3\textwidth}
			\centering
			\includegraphics[width=0.98\linewidth]{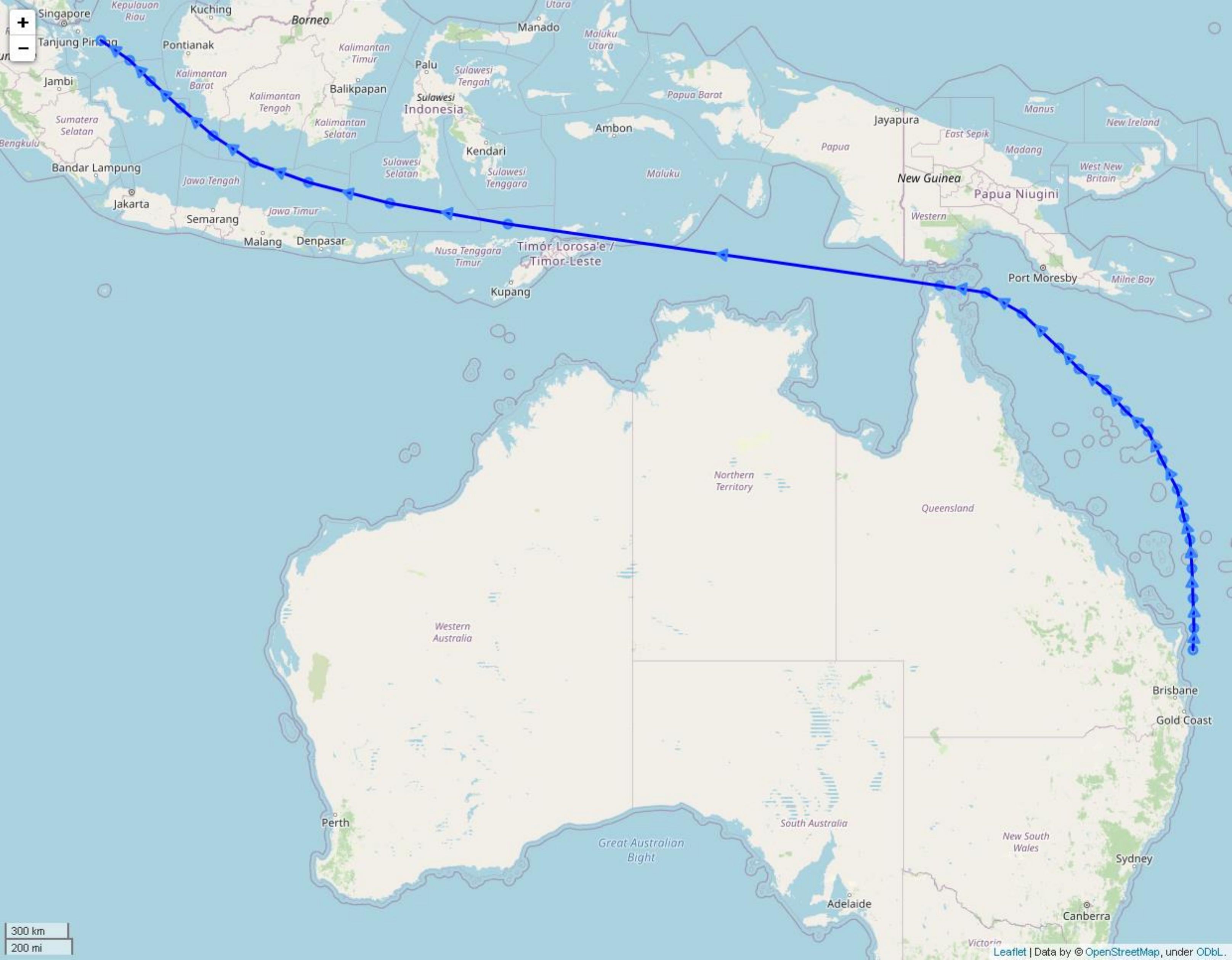}
			\caption{$\eta$=0.9, Lat-Scan}
		\end{subfigure}%
		\begin{subfigure}{.3\textwidth}
			\centering
			\includegraphics[width=0.98\linewidth]{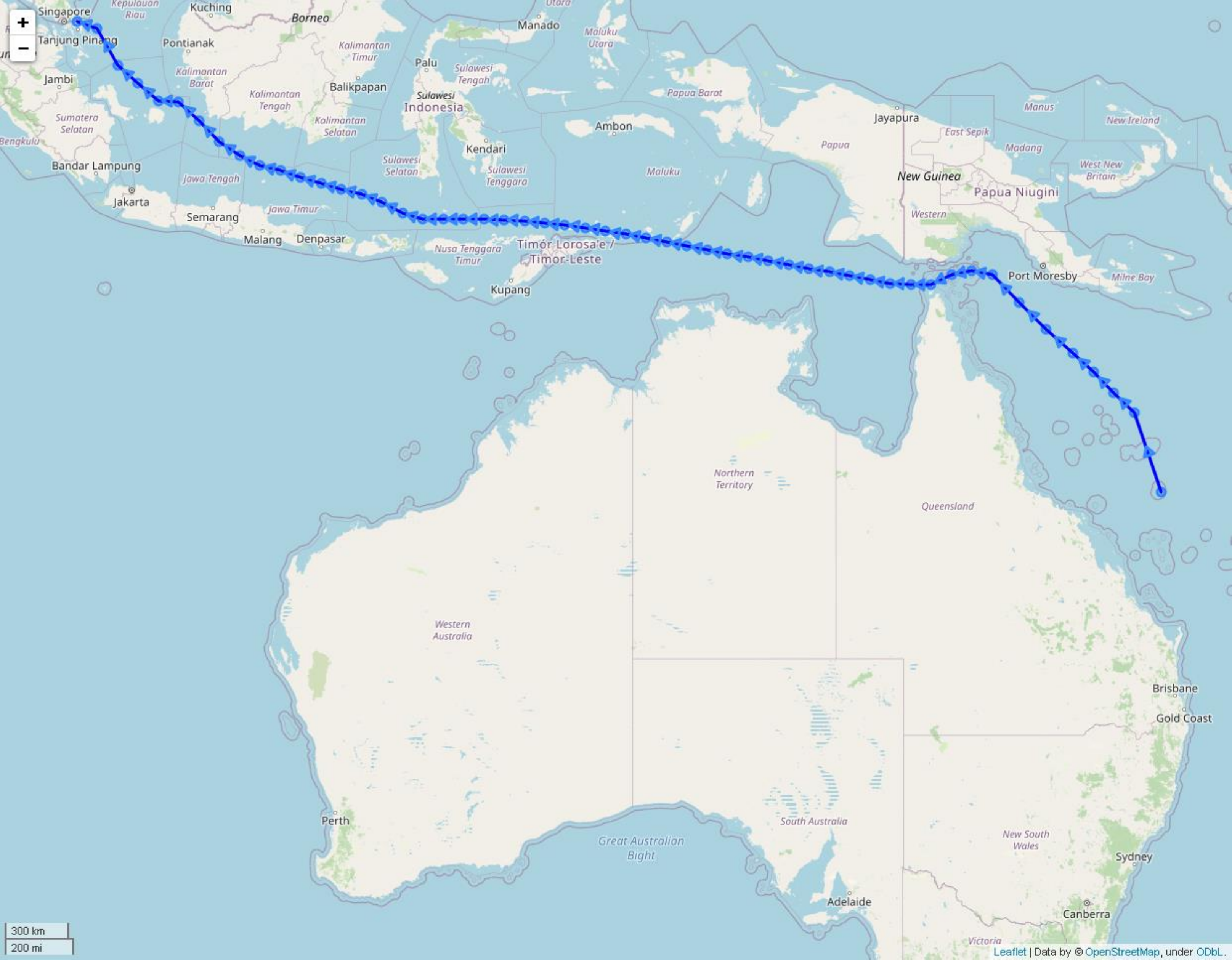}
			\caption{$\eta$=0.9, Lon-Scan}
		\end{subfigure}%
		\begin{subfigure}{.3\textwidth}
			\centering
			\includegraphics[width=0.98\linewidth]{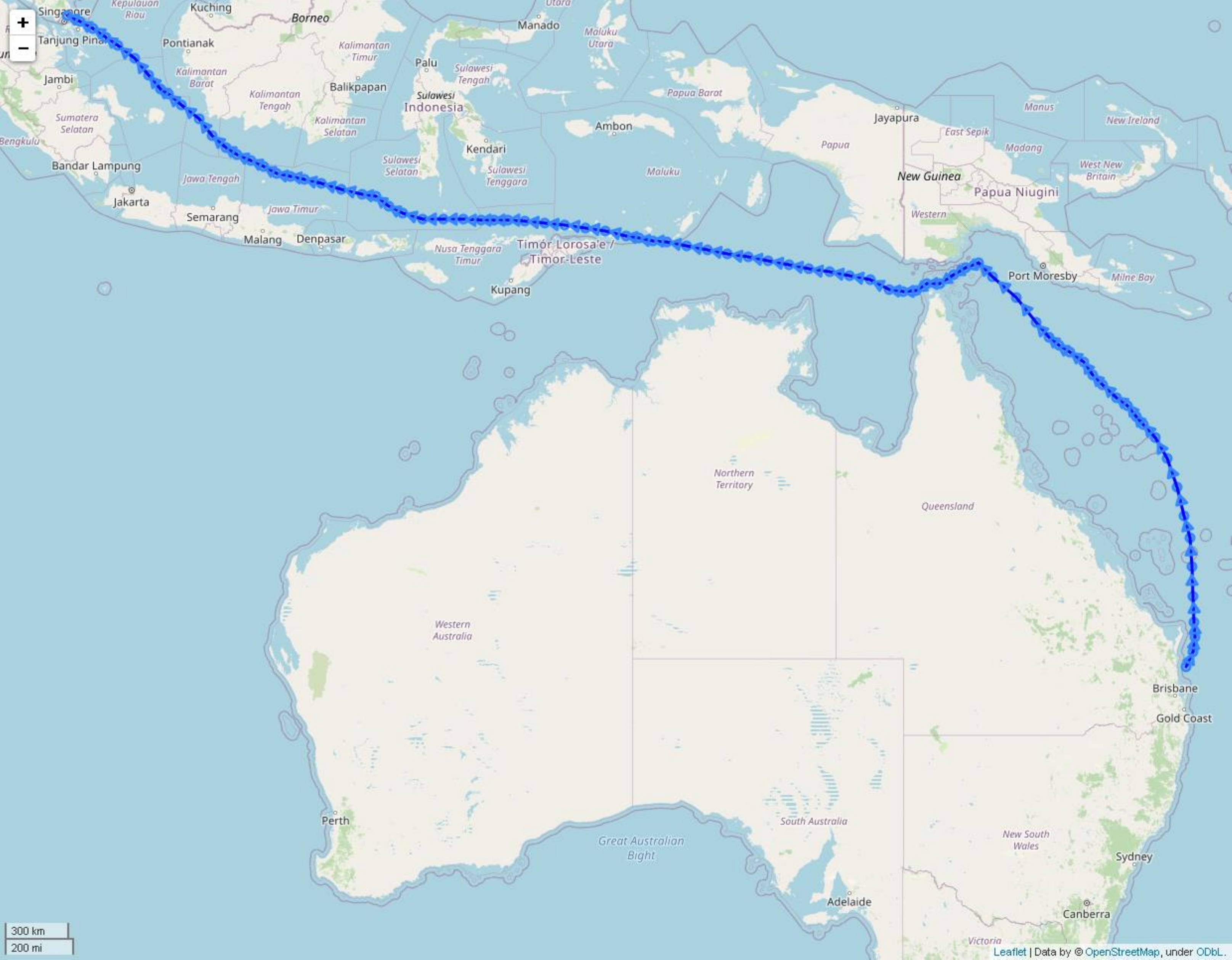}
			\caption{$\eta$=0.9, LatLon-Scan}
		\end{subfigure}

		\begin{subfigure}{.3\textwidth}
			\centering
			\includegraphics[width=0.98\linewidth]{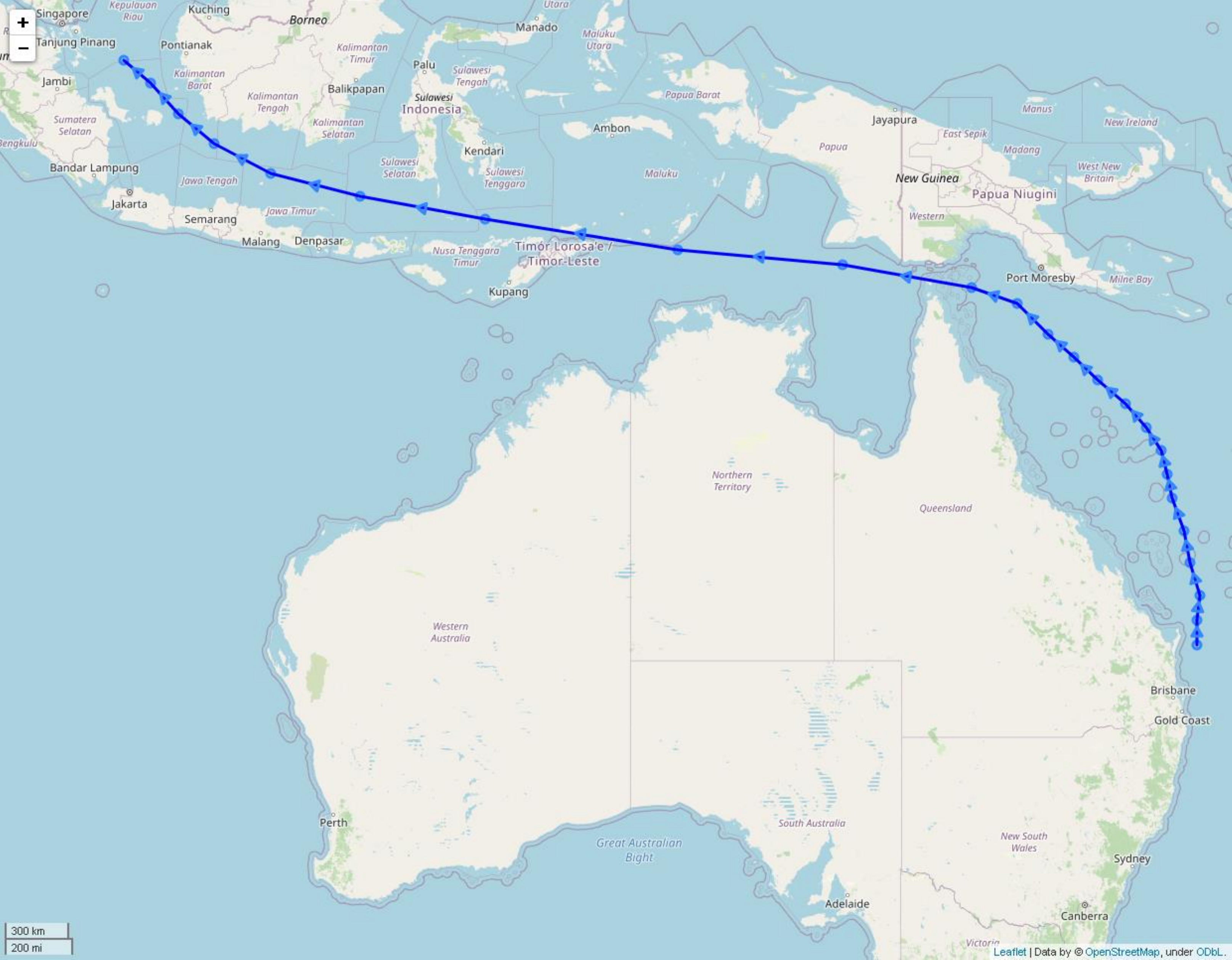}
			\caption{$\eta$=1.0, Lat-Scan}
		\end{subfigure}%
		\begin{subfigure}{.3\textwidth}
			\centering
			\includegraphics[width=0.98\linewidth]{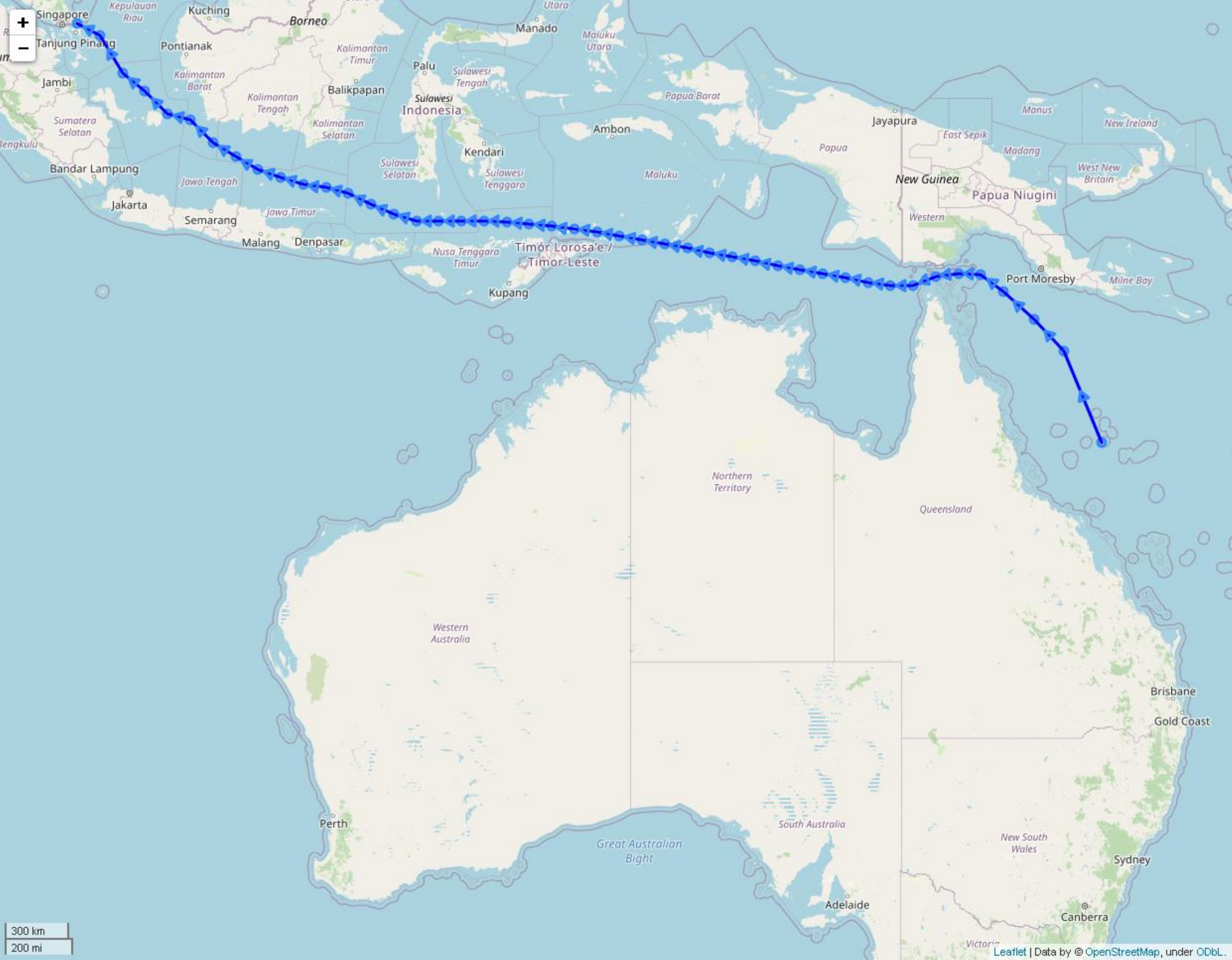}
			\caption{$\eta$=1.0, Lon-Scan}
		\end{subfigure}%
		\begin{subfigure}{.3\textwidth}
			\centering
			\includegraphics[width=0.98\linewidth]{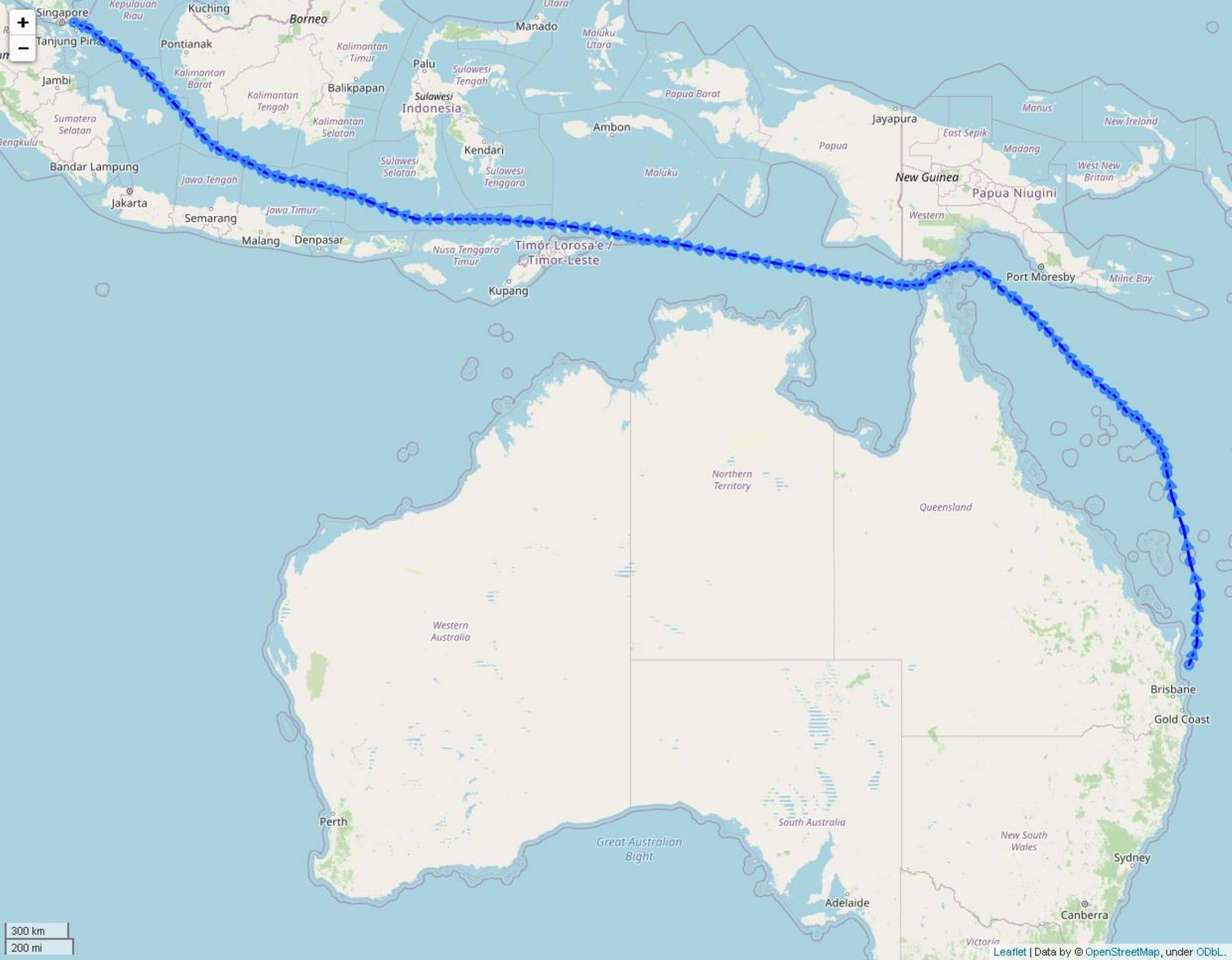}
			\caption{$\eta$=1.0, LatLon-Scan}
		\end{subfigure}

		\caption{Constructed Trajectories (Brisbane $\rightarrow$ Singapore)}
		\label{Fig: Constructed Trajectories of Brisbane to Singapore}
	\end{figure}

\newpage
Here are the historical (Figure \ref{Fig: Actual Trajectories of Brisbane to Singapore}) constructed (Figure \ref{Fig: Constructed Trajectories of Brisbane to Singapore}) trajectories by Lat-scan, Lon-Scan and LatLon-Scan for Brisbane-Singapore journey with different scanning internals ($\eta$) in degree.

According to the historical trajectories from Brisbane to Singapore in Figure \ref{Fig: Actual Trajectories of Brisbane to Singapore}, it is clear to illustrate that there is a direct route. In this case, it could be more straight forward and accurate to predict ETA of vessels.

According to the constructed trajectories from Brisbane to Singapore in Figure \ref{Fig: Constructed Trajectories of Brisbane to Singapore} across different scanning internals ($\eta$) and scanning methods, it is clear to note that LatLon-scanning can perform better in latitude and longitude directions, compared to merely Lat-scanning or Lon-scanning. Similarly to the trajectories from Singapore to Brisbane, there are also missing movements near Singapore and Brisbane for both Lat-scanning and Lon-scanning.
	
\begin{figure}[htbp]
		\centering
		\includegraphics[width=\linewidth]{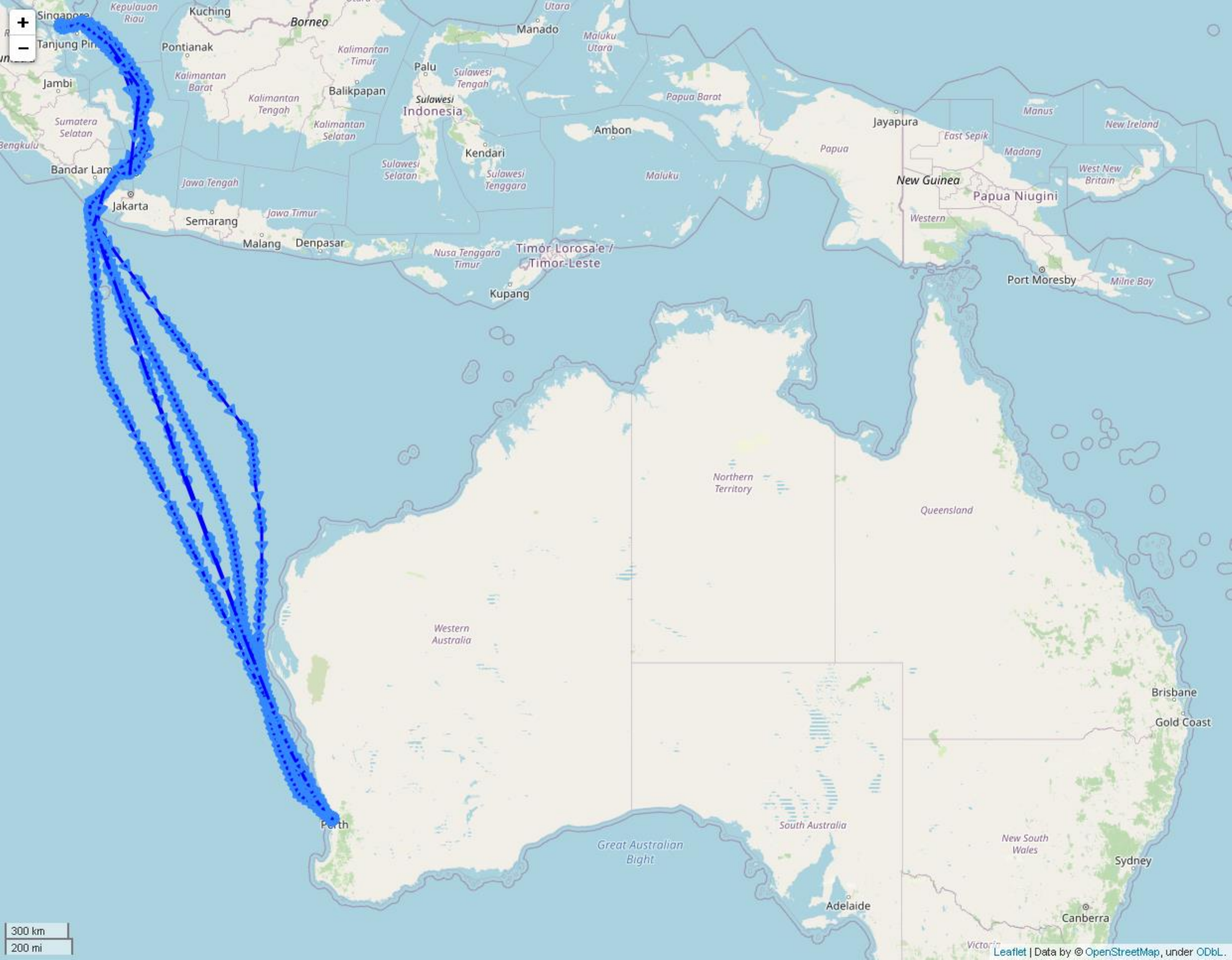}
		\caption{Actual Trajectories (Singapore $\rightarrow$ Perth)}
		\label{Fig: Actual Trajectories of Singapore to Perth}
	\end{figure}

\begin{figure}[htbp]
		\begin{subfigure}{.3\textwidth}
			\centering
			\includegraphics[width=0.98\linewidth]{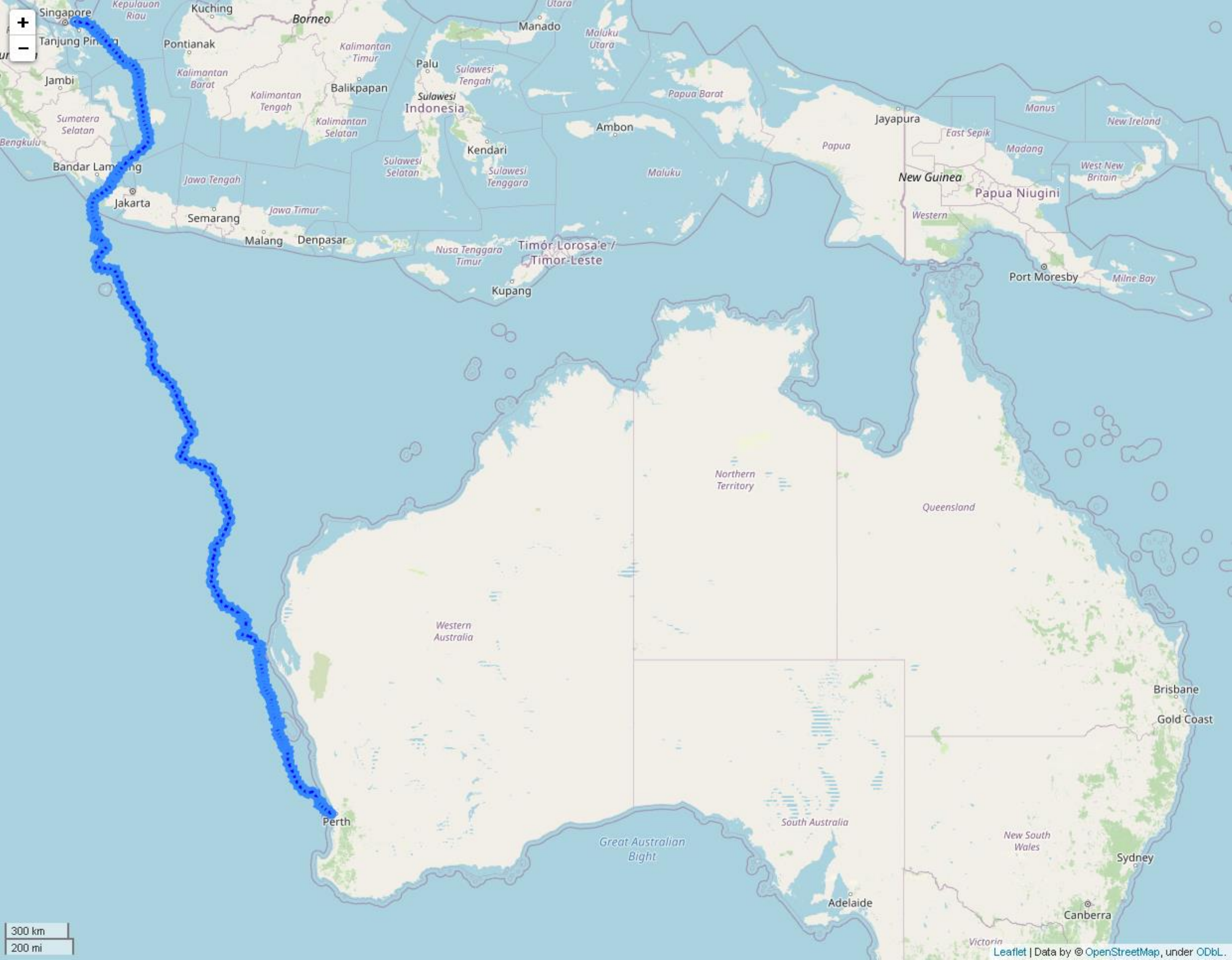}
			\caption{$\eta$=0.1, Lat-Scan}
		\end{subfigure}%
		\begin{subfigure}{.3\textwidth}
			\centering
			\includegraphics[width=0.98\linewidth]{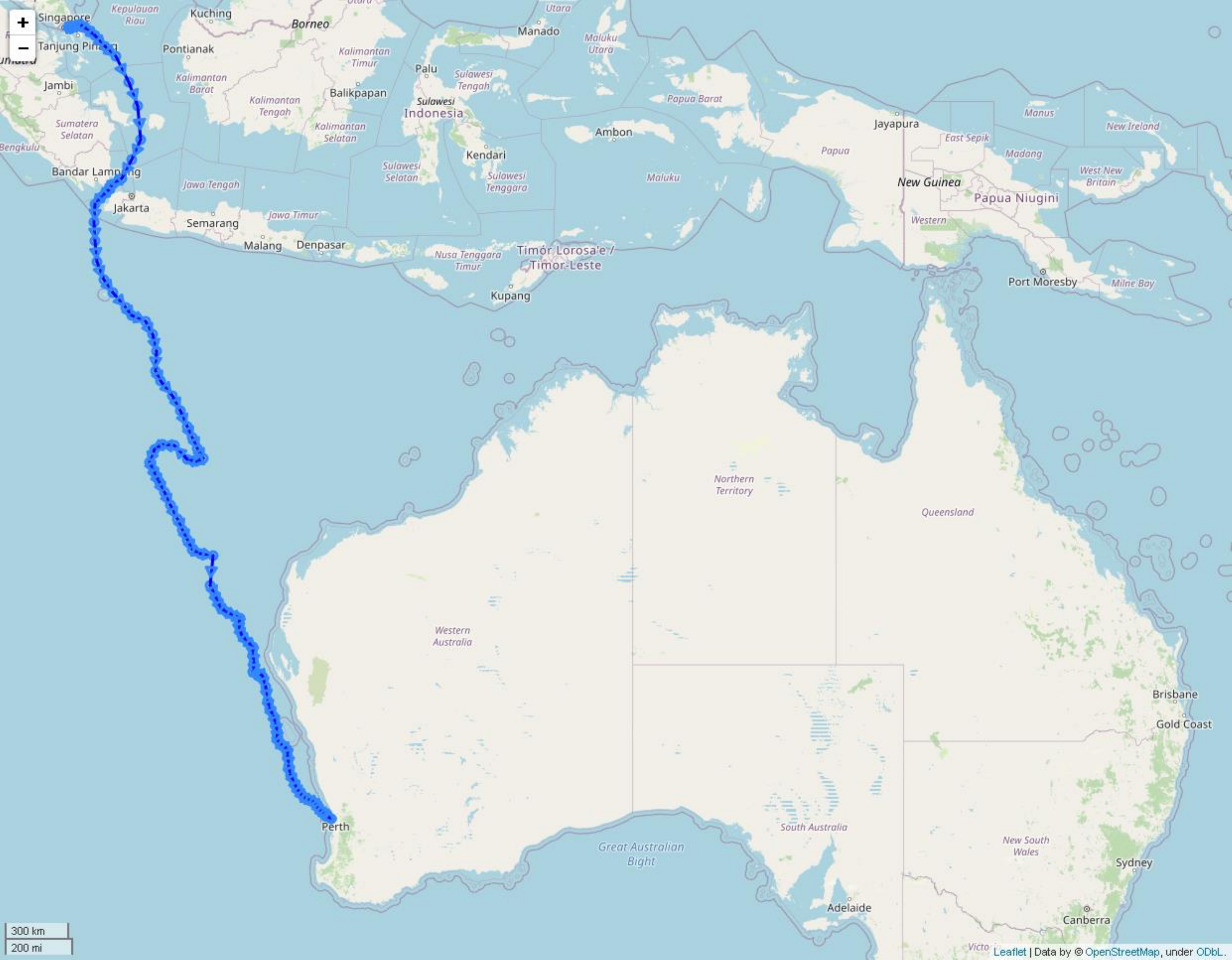}
			\caption{$\eta$=0.1, Lon-Scan}
		\end{subfigure}%
		\begin{subfigure}{.3\textwidth}
			\centering
			\includegraphics[width=0.98\linewidth]{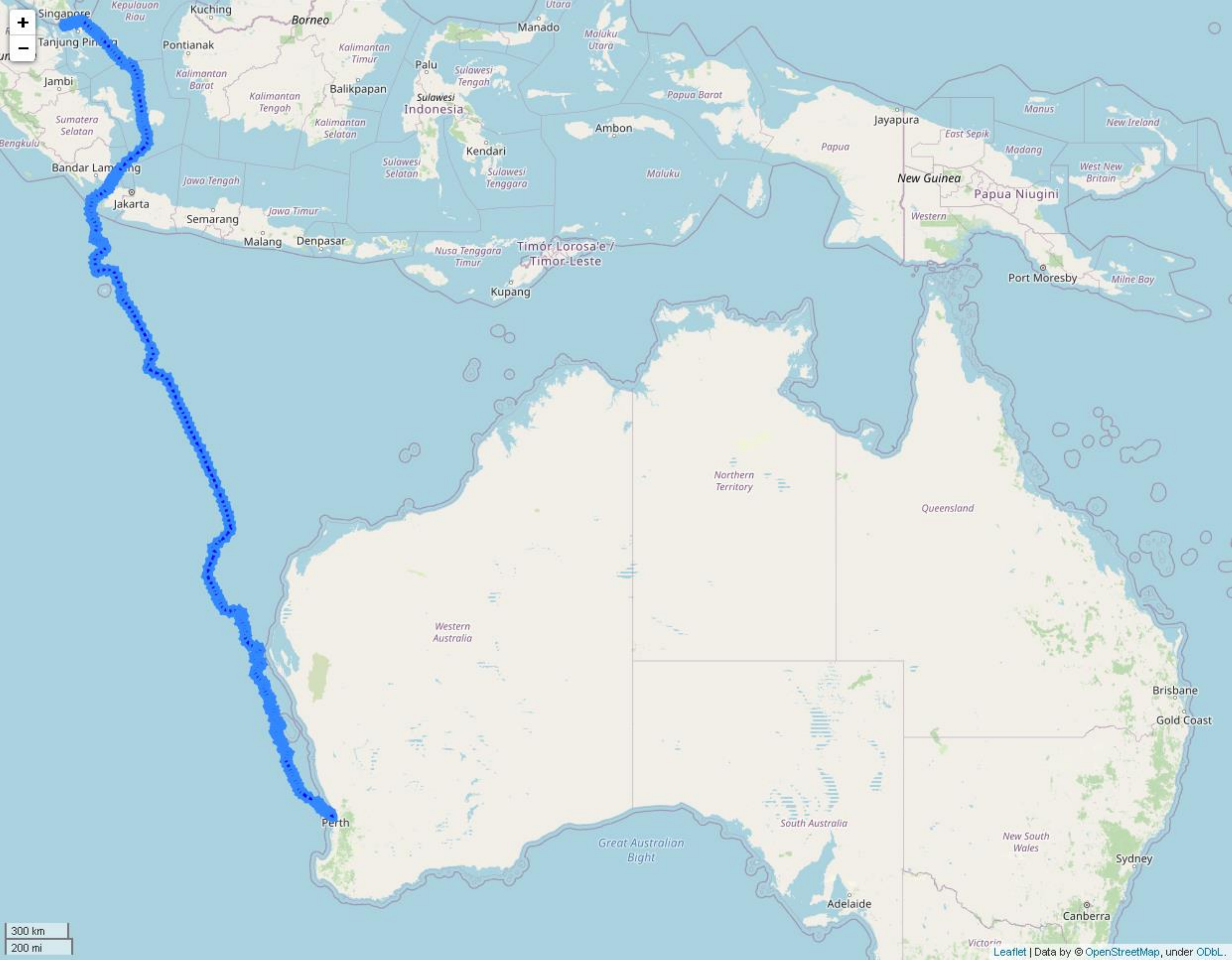}
			\caption{$\eta$=0.1, LatLon-Scan}
		\end{subfigure}

		\begin{subfigure}{.3\textwidth}
			\centering
			\includegraphics[width=0.98\linewidth]{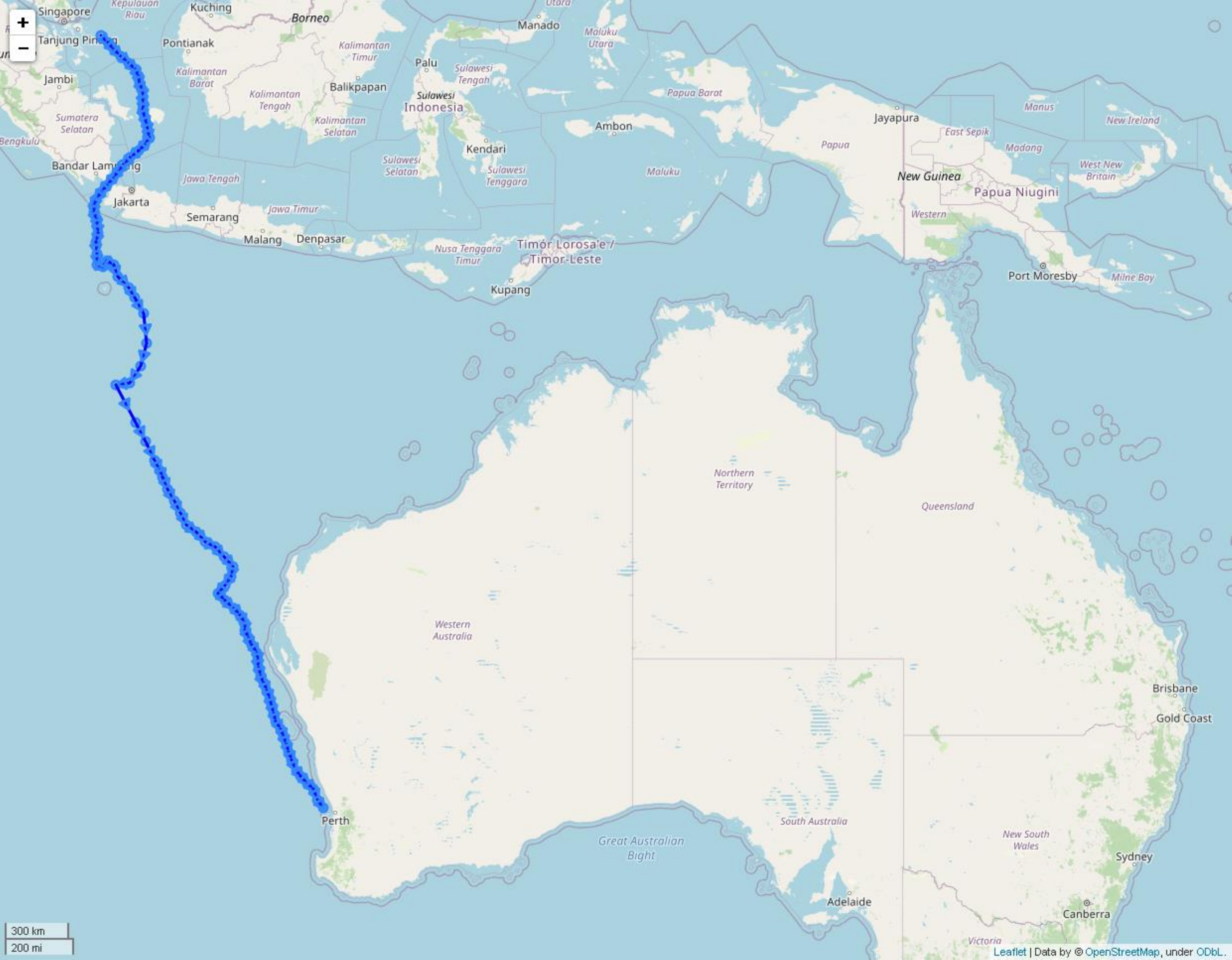}
			\caption{$\eta$=0.3, Lat-Scan}
		\end{subfigure}%
		\begin{subfigure}{.3\textwidth}
			\centering
			\includegraphics[width=0.98\linewidth]{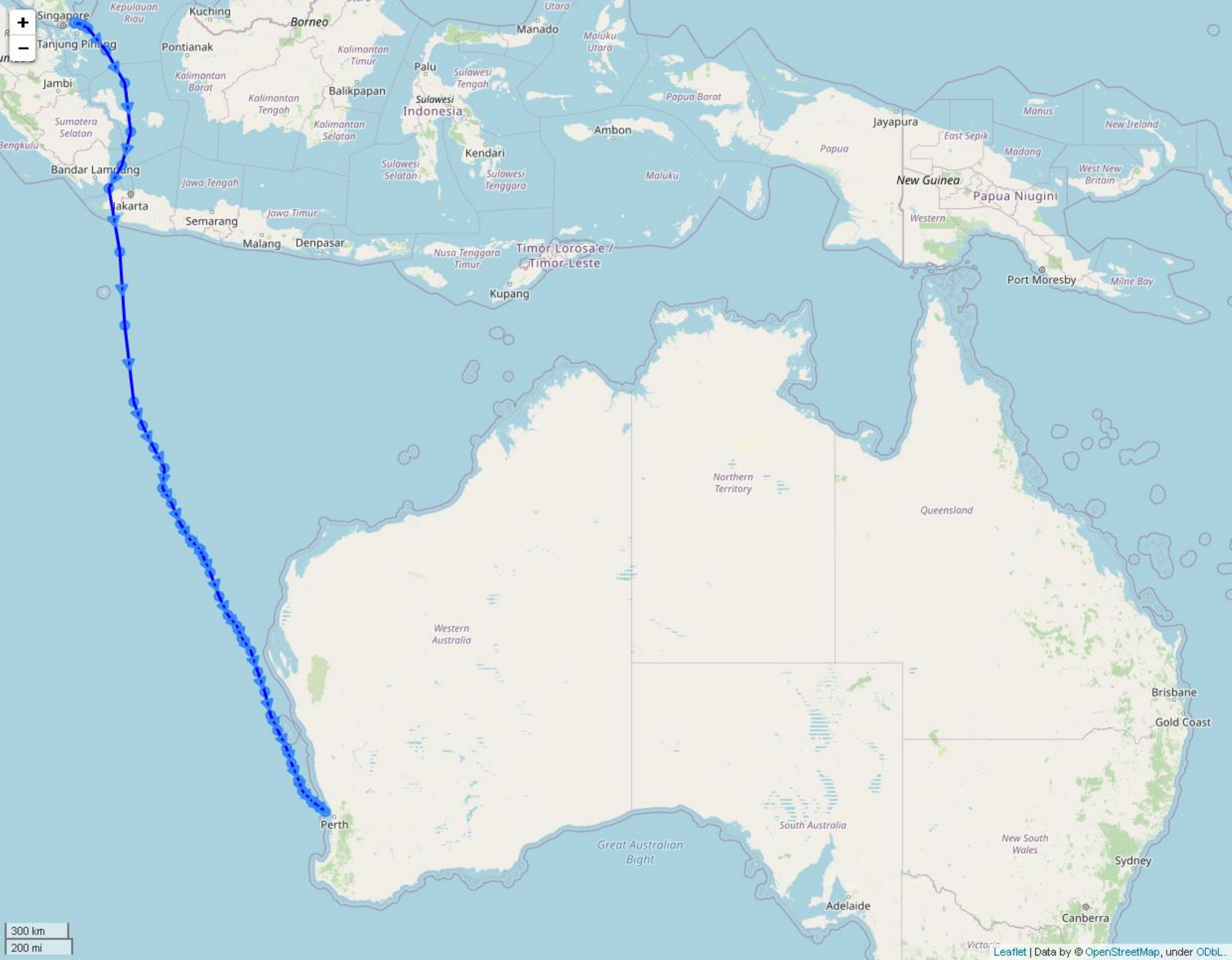}
			\caption{$\eta$=0.3, Lon-Scan}
		\end{subfigure}%
		\begin{subfigure}{.3\textwidth}
			\centering
			\includegraphics[width=0.98\linewidth]{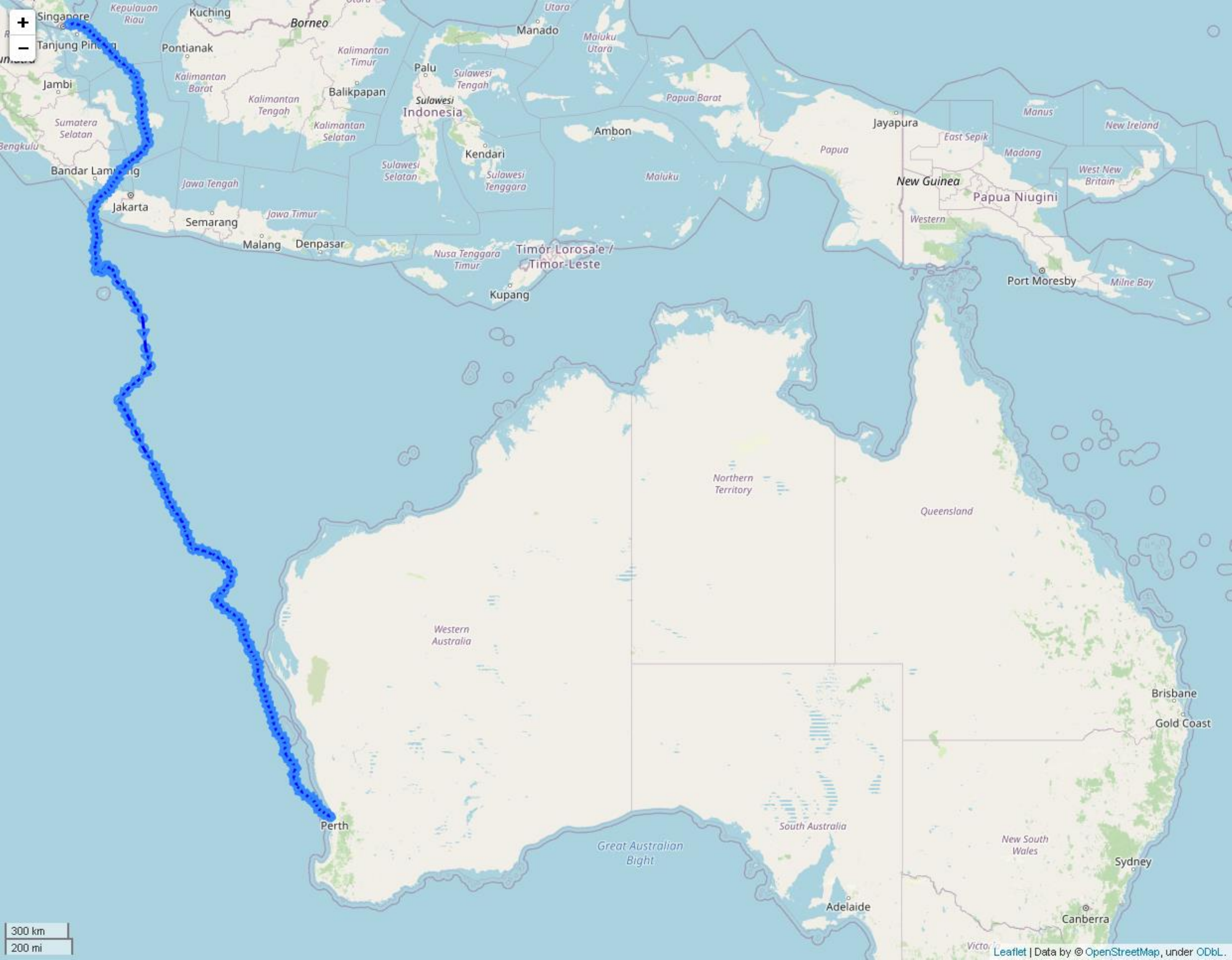}
			\caption{$\eta$=0.3, LatLon-Scan}
		\end{subfigure}

		\begin{subfigure}{.3\textwidth}
			\centering
			\includegraphics[width=0.98\linewidth]{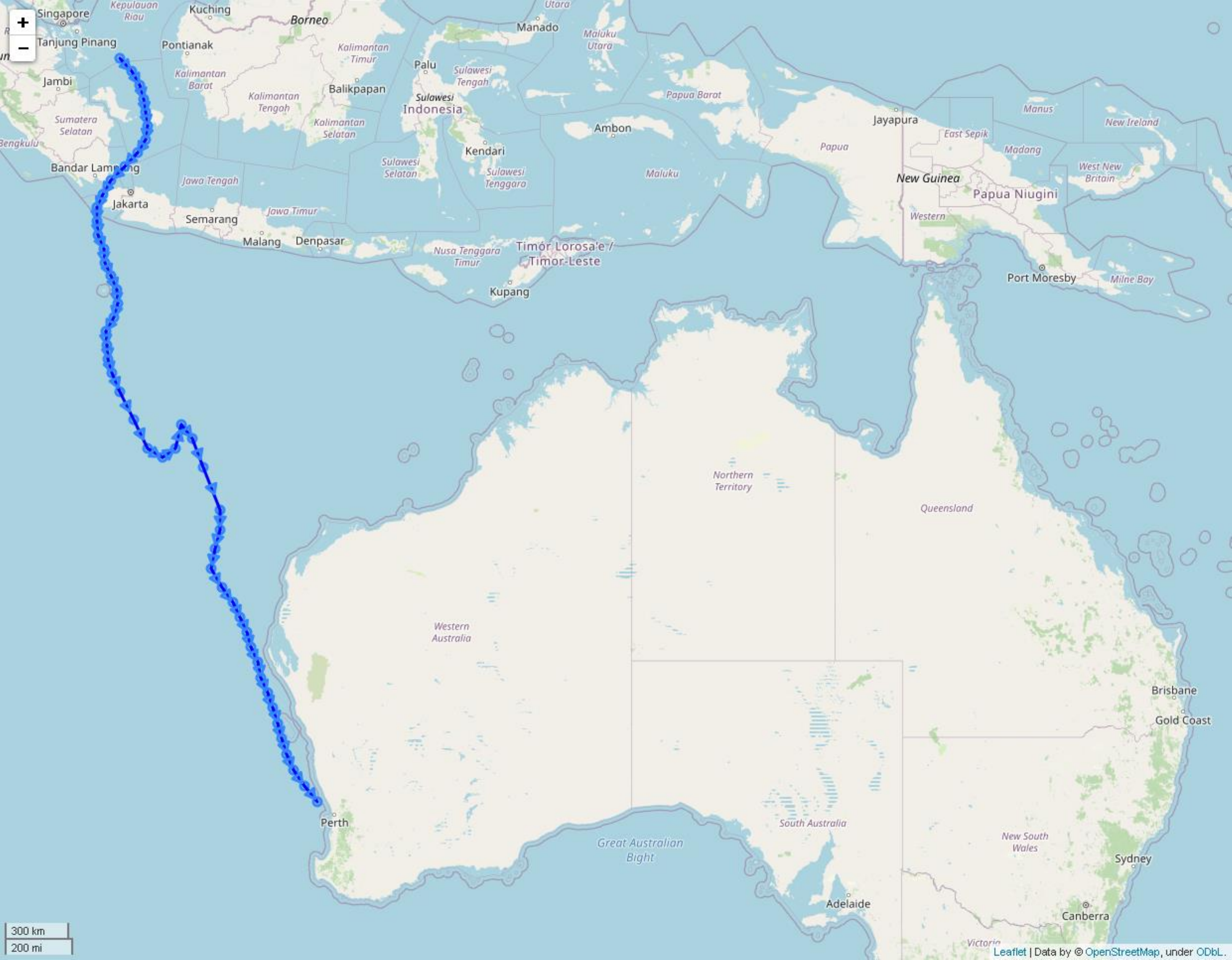}
			\caption{$\eta$=0.6, Lat-Scan}
		\end{subfigure}%
		\begin{subfigure}{.3\textwidth}
			\centering
			\includegraphics[width=0.98\linewidth]{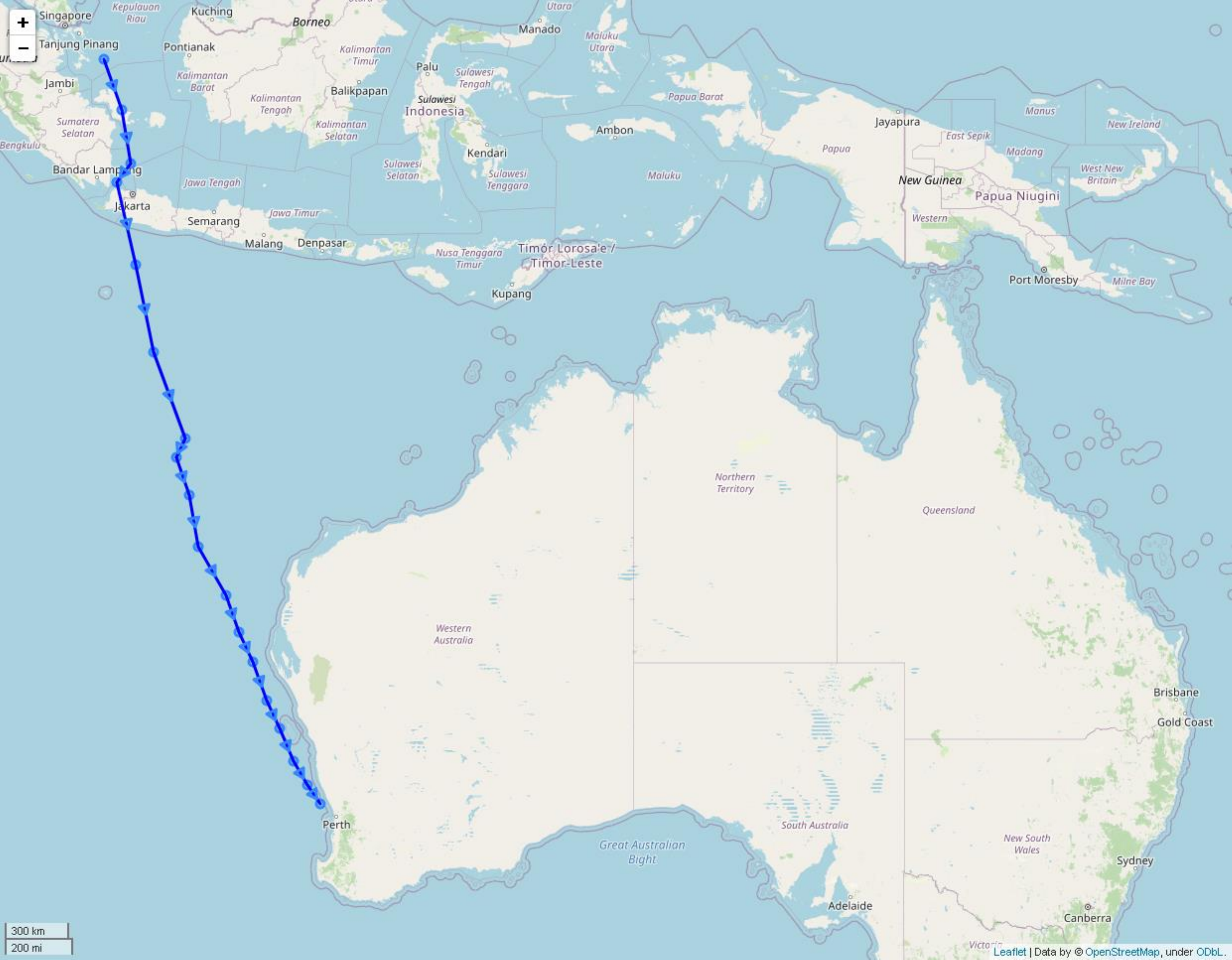}
			\caption{$\eta$=0.6, Lon-Scan}
		\end{subfigure}%
		\begin{subfigure}{.3\textwidth}
			\centering
			\includegraphics[width=0.98\linewidth]{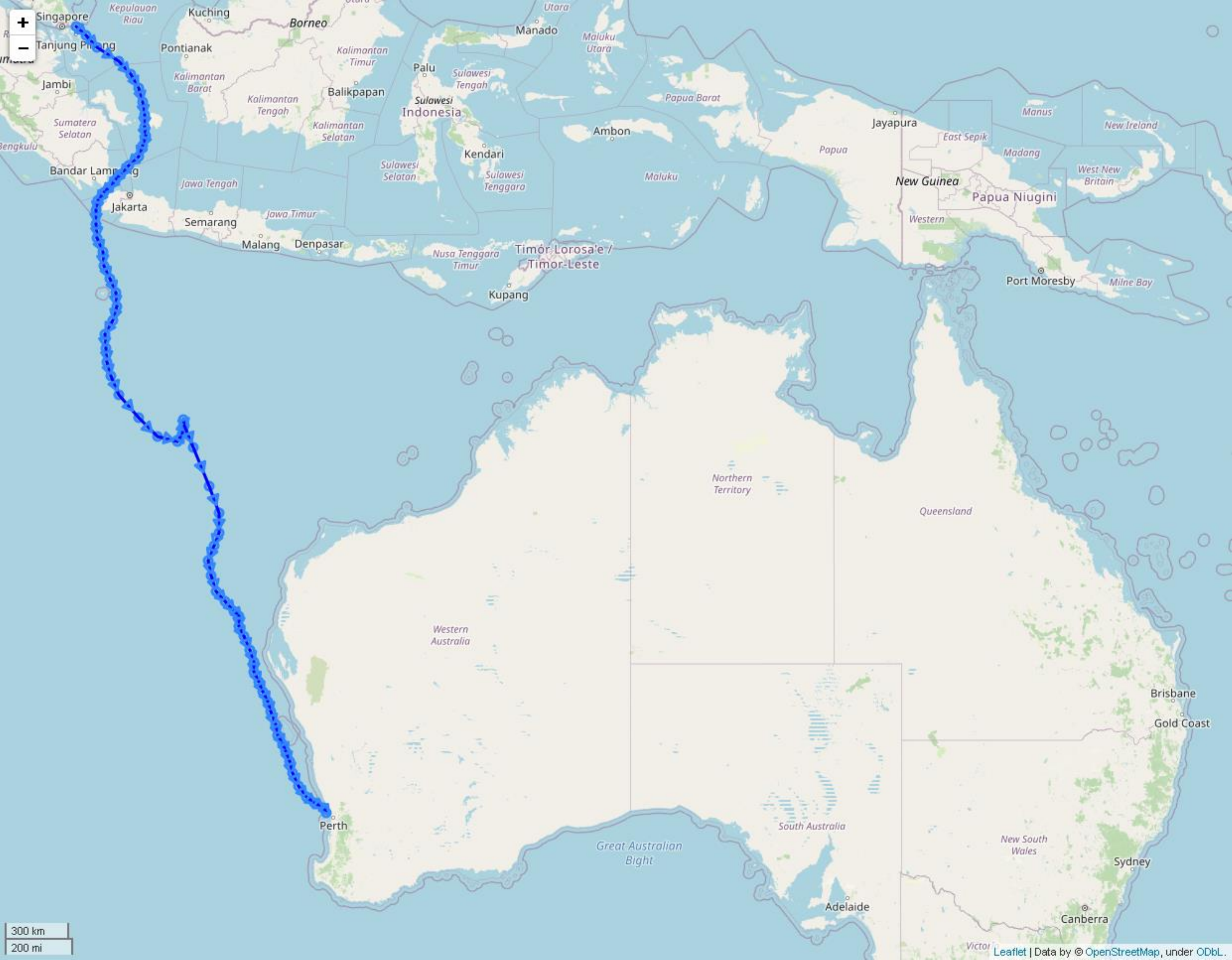}
			\caption{$\eta$=0.6, LatLon-Scan}
		\end{subfigure}

		\begin{subfigure}{.3\textwidth}
			\centering
			\includegraphics[width=0.98\linewidth]{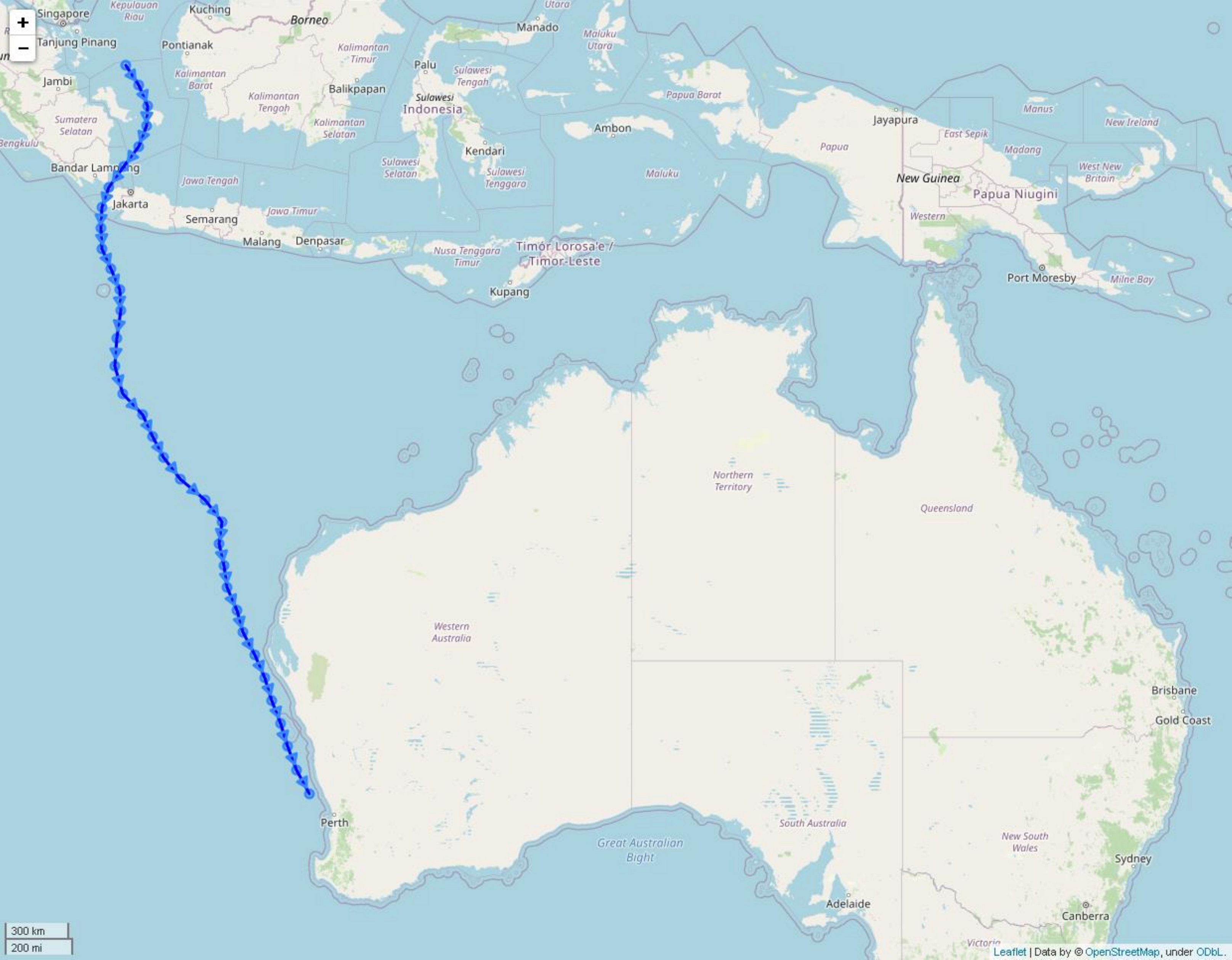}
			\caption{$\eta$=0.9, Lat-Scan}
		\end{subfigure}%
		\begin{subfigure}{.3\textwidth}
			\centering
			\includegraphics[width=0.98\linewidth]{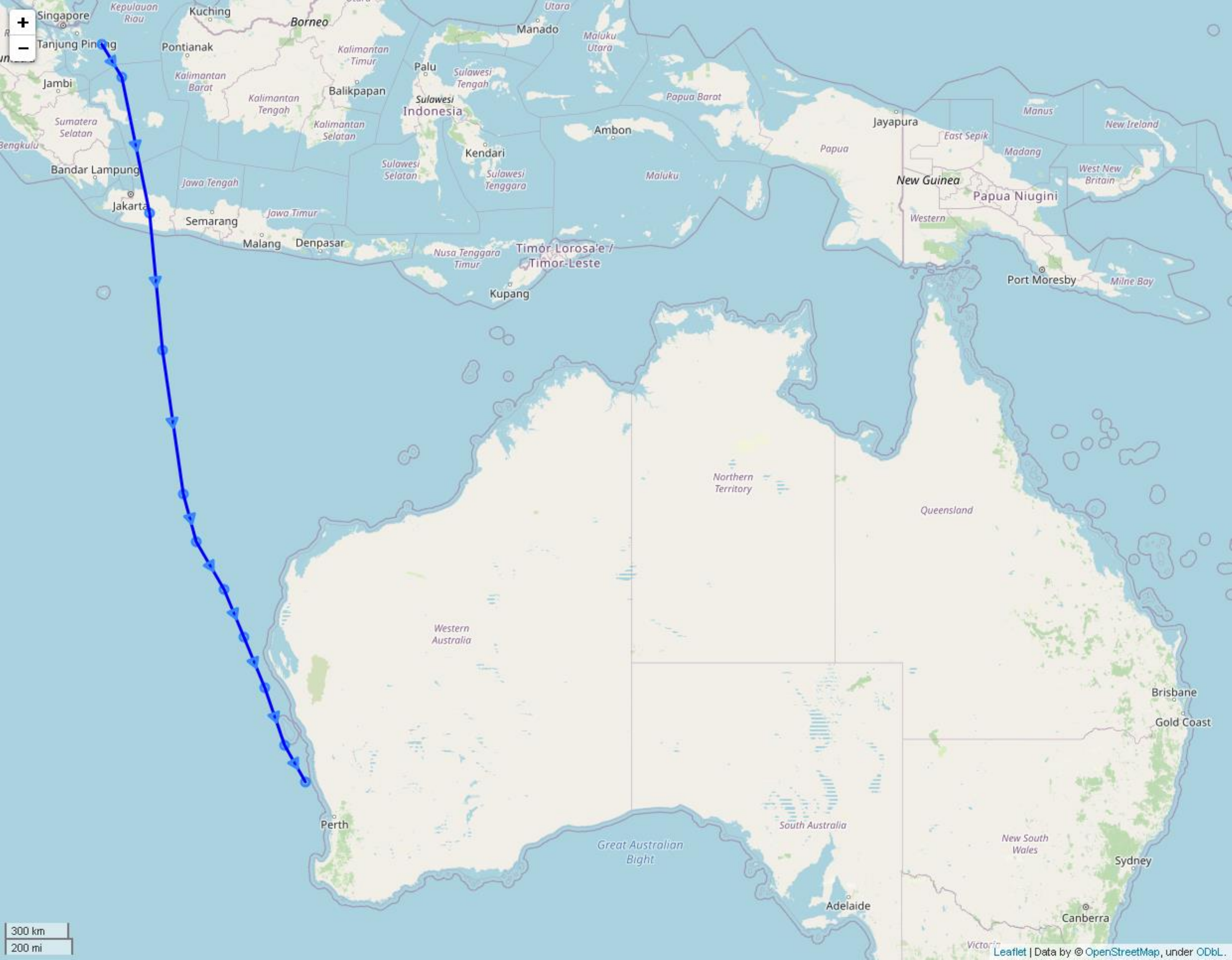}
			\caption{$\eta$=0.9, Lon-Scan}
		\end{subfigure}%
		\begin{subfigure}{.3\textwidth}
			\centering
			\includegraphics[width=0.98\linewidth]{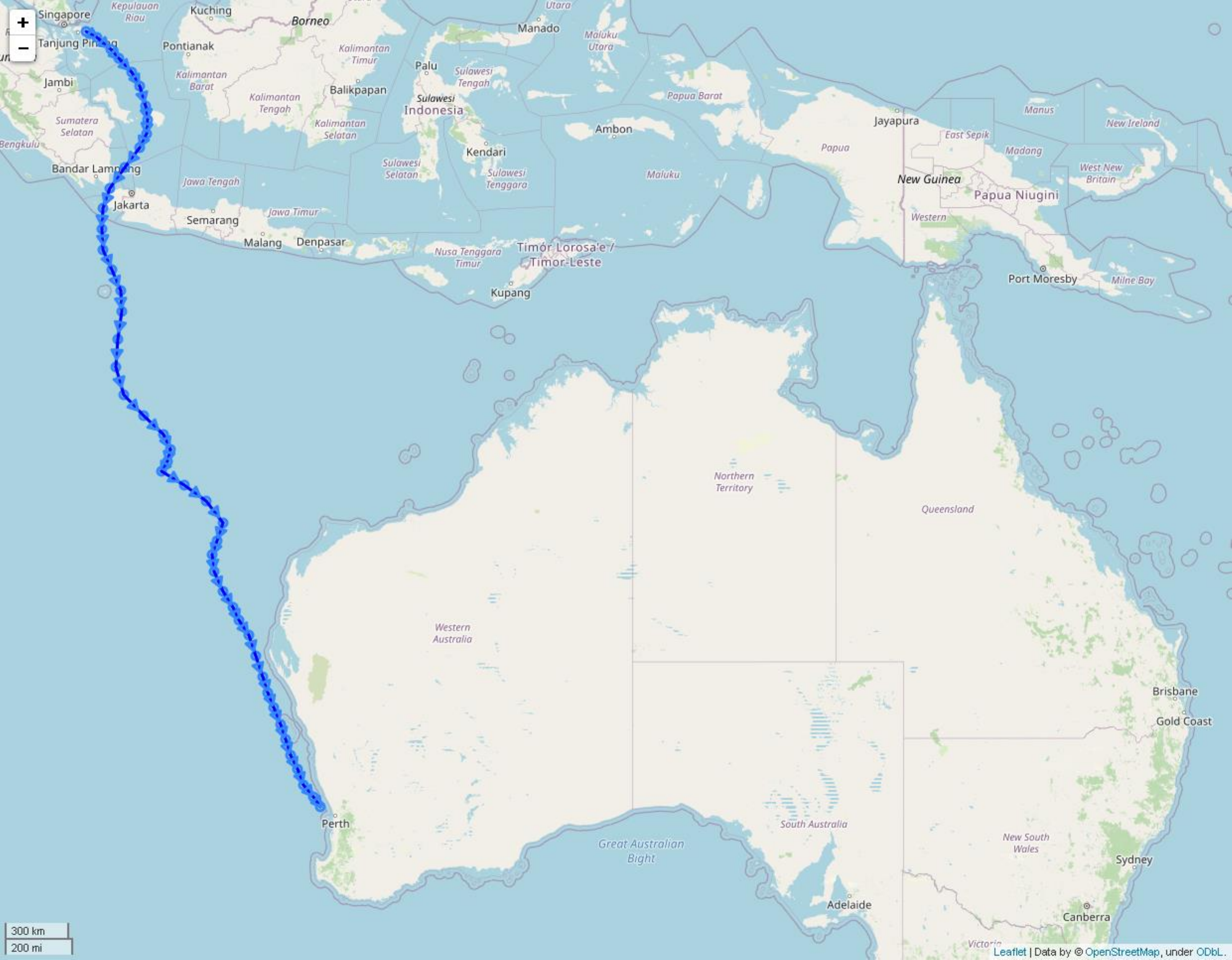}
			\caption{$\eta$=0.9, LatLon-Scan}
		\end{subfigure}

		\begin{subfigure}{.3\textwidth}
			\centering
			\includegraphics[width=0.98\linewidth]{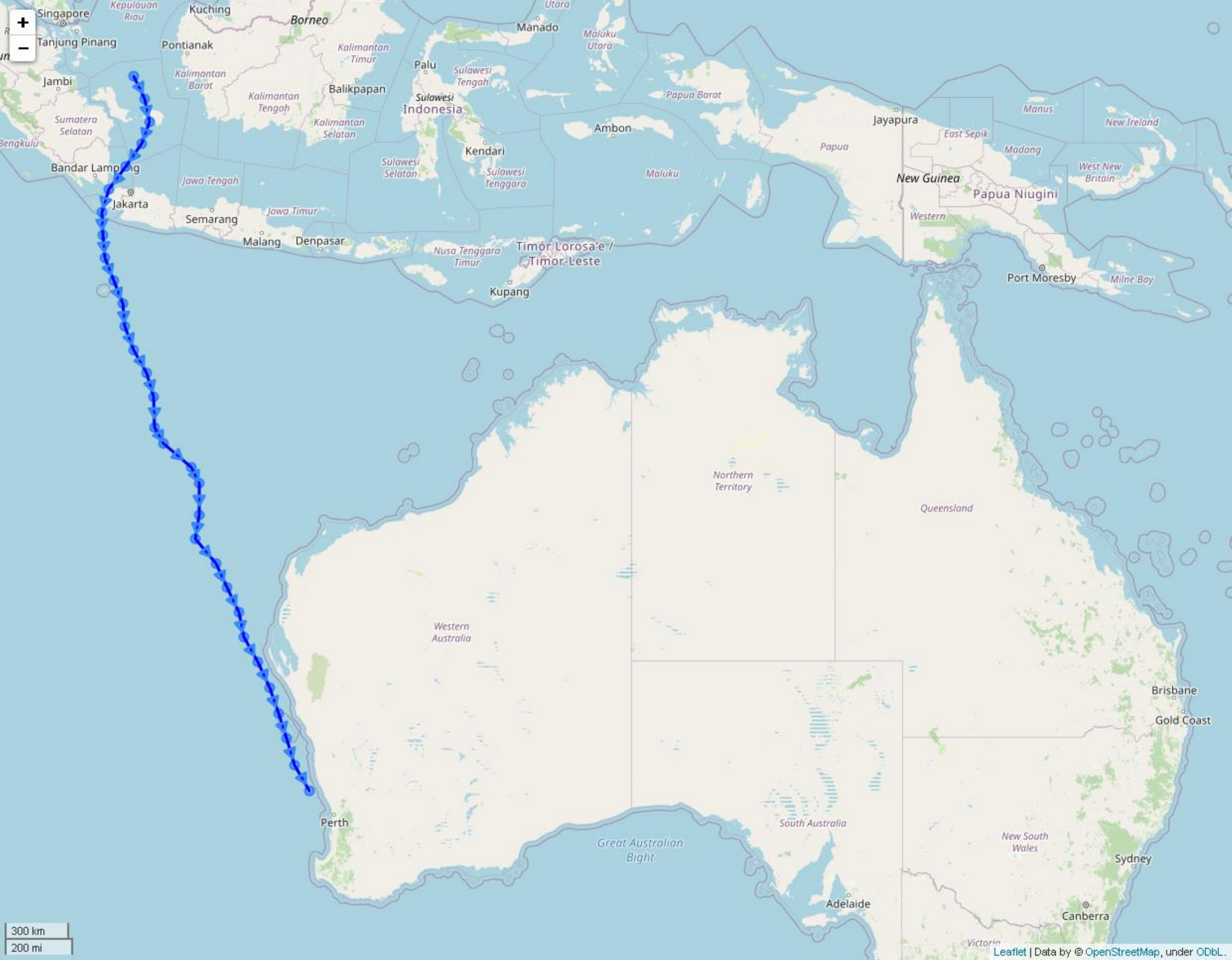}
			\caption{$\eta$=1.0, Lat-Scan}
		\end{subfigure}%
		\begin{subfigure}{.3\textwidth}
			\centering
			\includegraphics[width=0.98\linewidth]{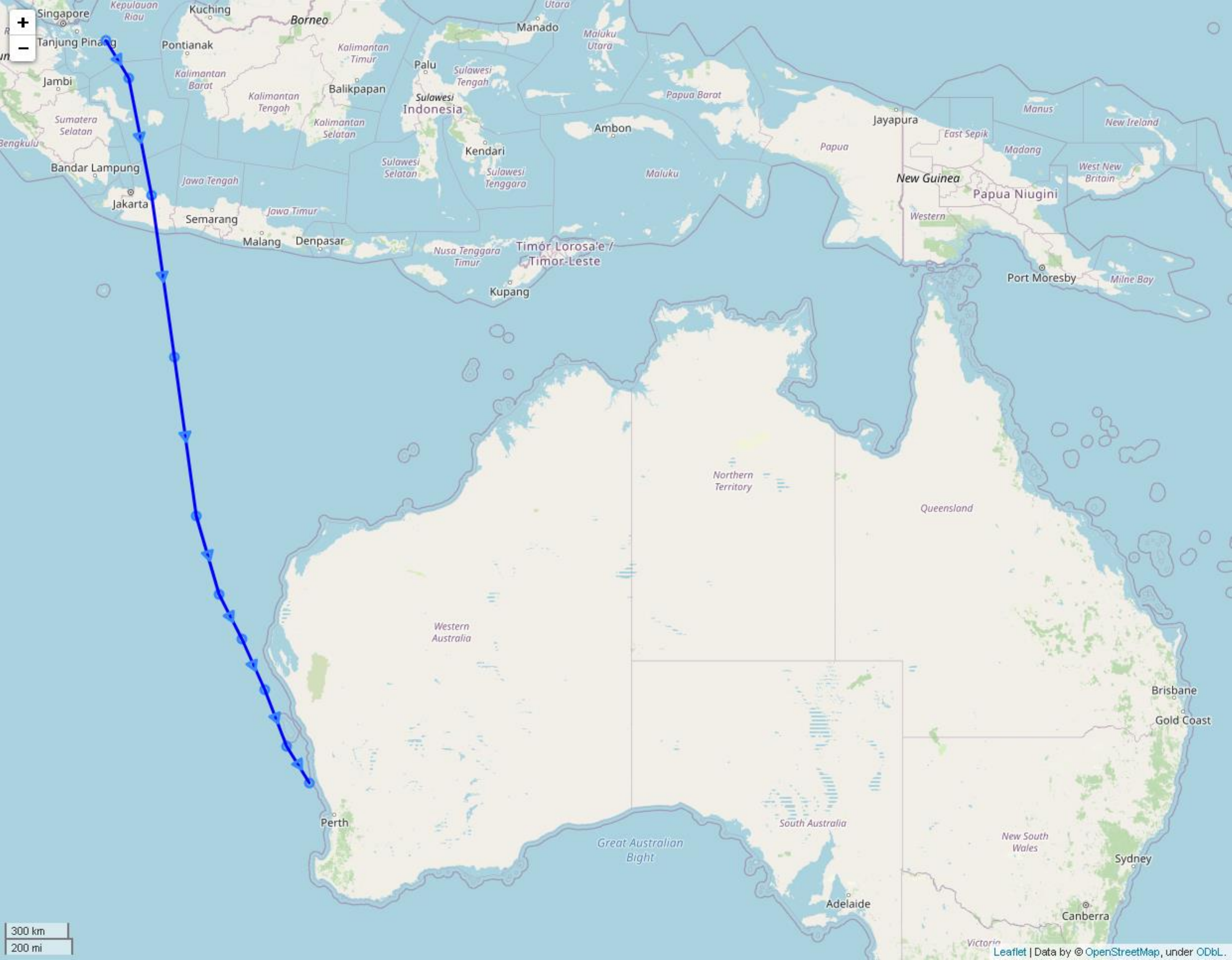}
			\caption{$\eta$=1.0, Lon-Scan}
		\end{subfigure}%
		\begin{subfigure}{.3\textwidth}
			\centering
			\includegraphics[width=0.98\linewidth]{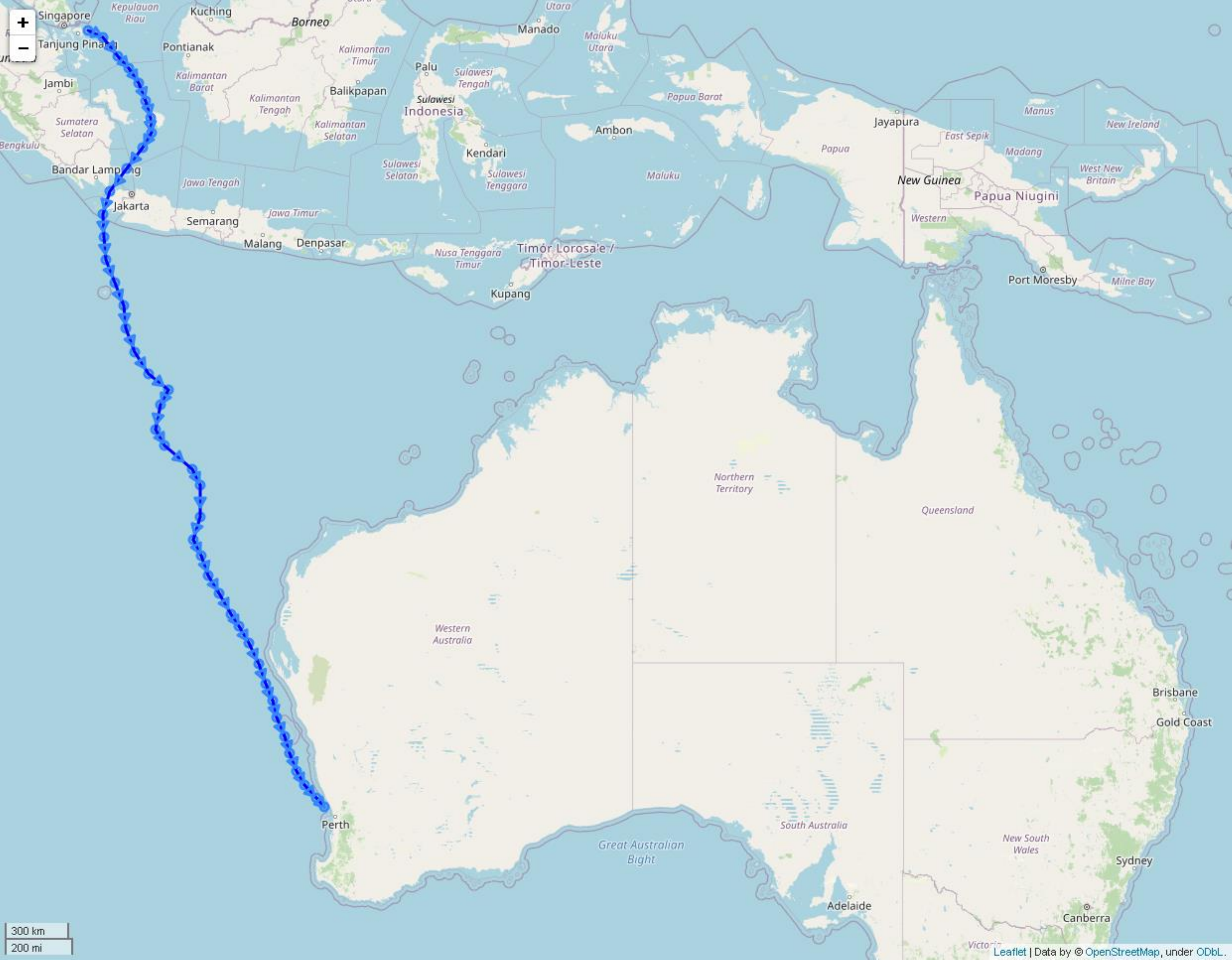}
			\caption{$\eta$=1.0, LatLon-Scan}
		\end{subfigure}

		\caption{Constructed Trajectories (Singapore $\rightarrow$ Perth)}
		\label{Fig: Constructed Trajectories of Singapore to Perth}
	\end{figure}

\newpage
Here are the historical (Figure \ref{Fig: Actual Trajectories of Singapore to Perth}) and constructed (Figure \ref{Fig: Constructed Trajectories of Singapore to Perth}) trajectories by Lat-scan, Lon-Scan and LatLon-Scan for Singapore-Perth journey with different scanning internals ($\eta$) in degree.

According to the historical trajectories from Singapore to Perth in Figure \ref{Fig: Actual Trajectories of Singapore to Perth}, it is clear to show that there is a direct route. In this case, it could be more straight forward and accurate to predict ETA of vessels.

According to the constructed trajectories from Singapore to Perth in Figure \ref{Fig: Constructed Trajectories of Singapore to Perth} across different scanning internals ($\eta$) and scanning methods, it shows that Lat-scanning loses movements near Singapore, and Lon-scanning eliminates the movement details on passing Indonesia and near Perth.
	
\begin{figure}[htbp]
		\centering
		\includegraphics[width=\linewidth]{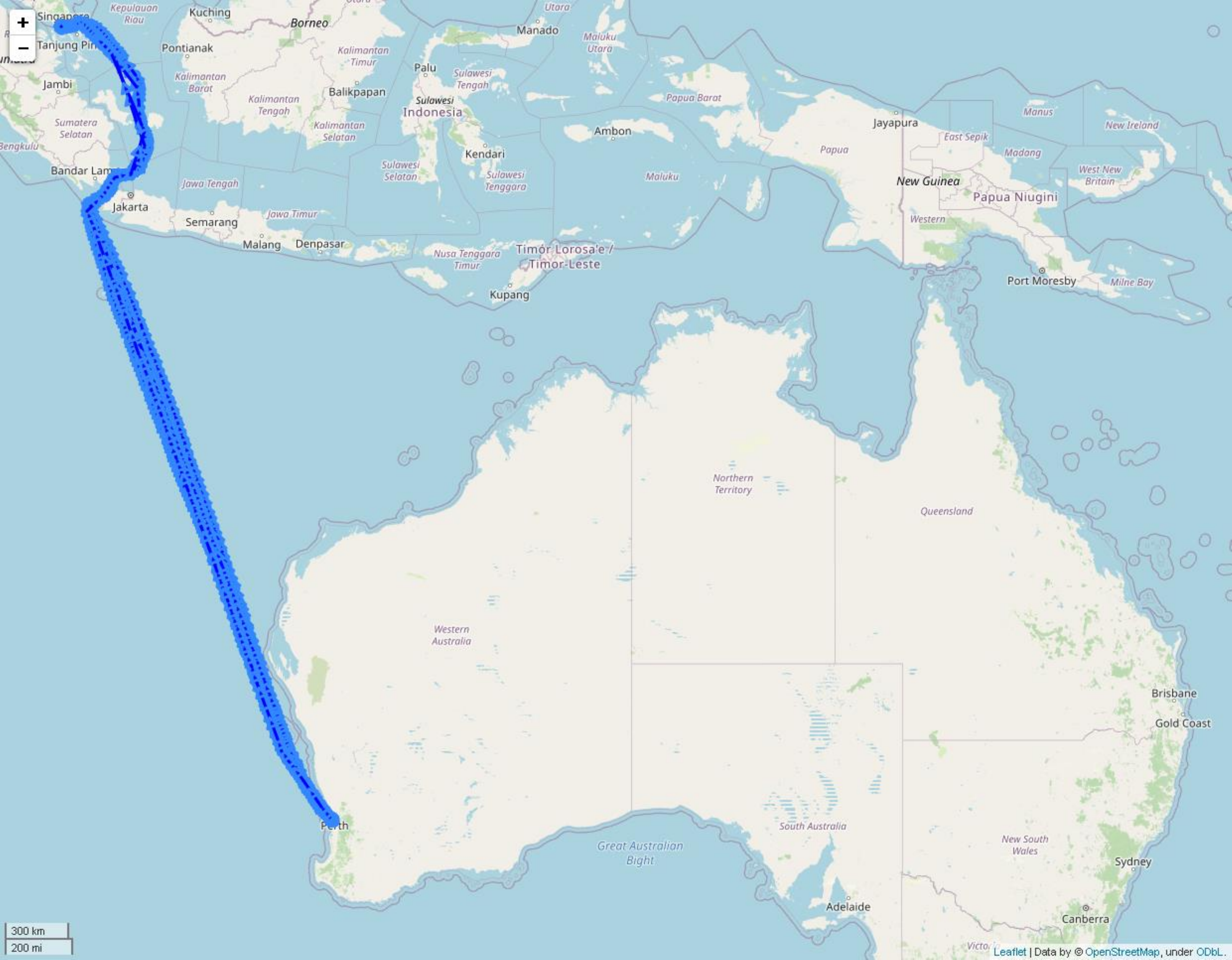}
		\caption{Actual Trajectories (Perth $\rightarrow$ Singapore)}
		\label{Fig: Actual Trajectories of Perth to Singapore}
	\end{figure}

\begin{figure}[htbp]
		\begin{subfigure}{.3\textwidth}
			\centering
			\includegraphics[width=0.98\linewidth]{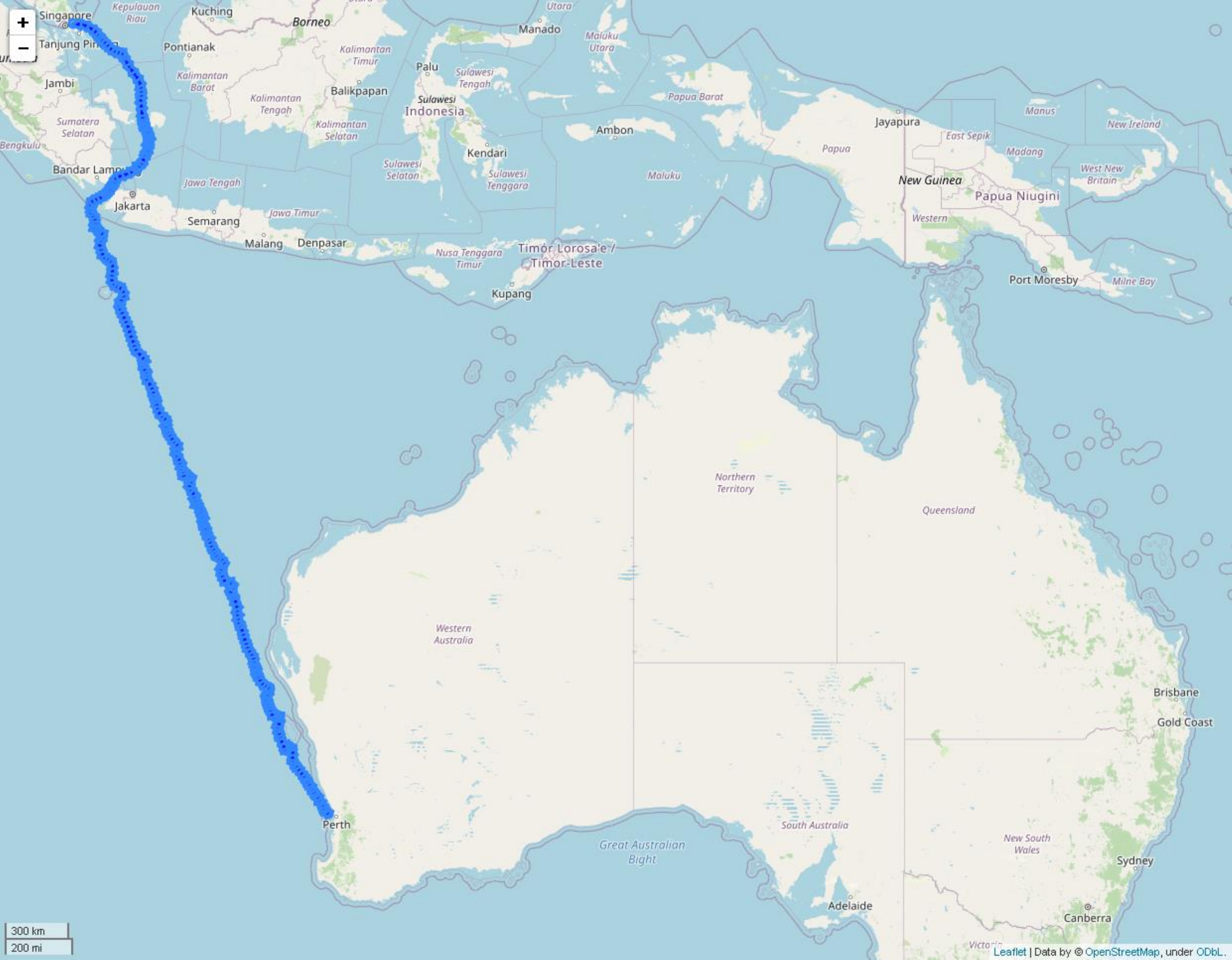}
			\caption{$\eta$=0.1, Lat-Scan}
		\end{subfigure}%
		\begin{subfigure}{.3\textwidth}
			\centering
			\includegraphics[width=0.98\linewidth]{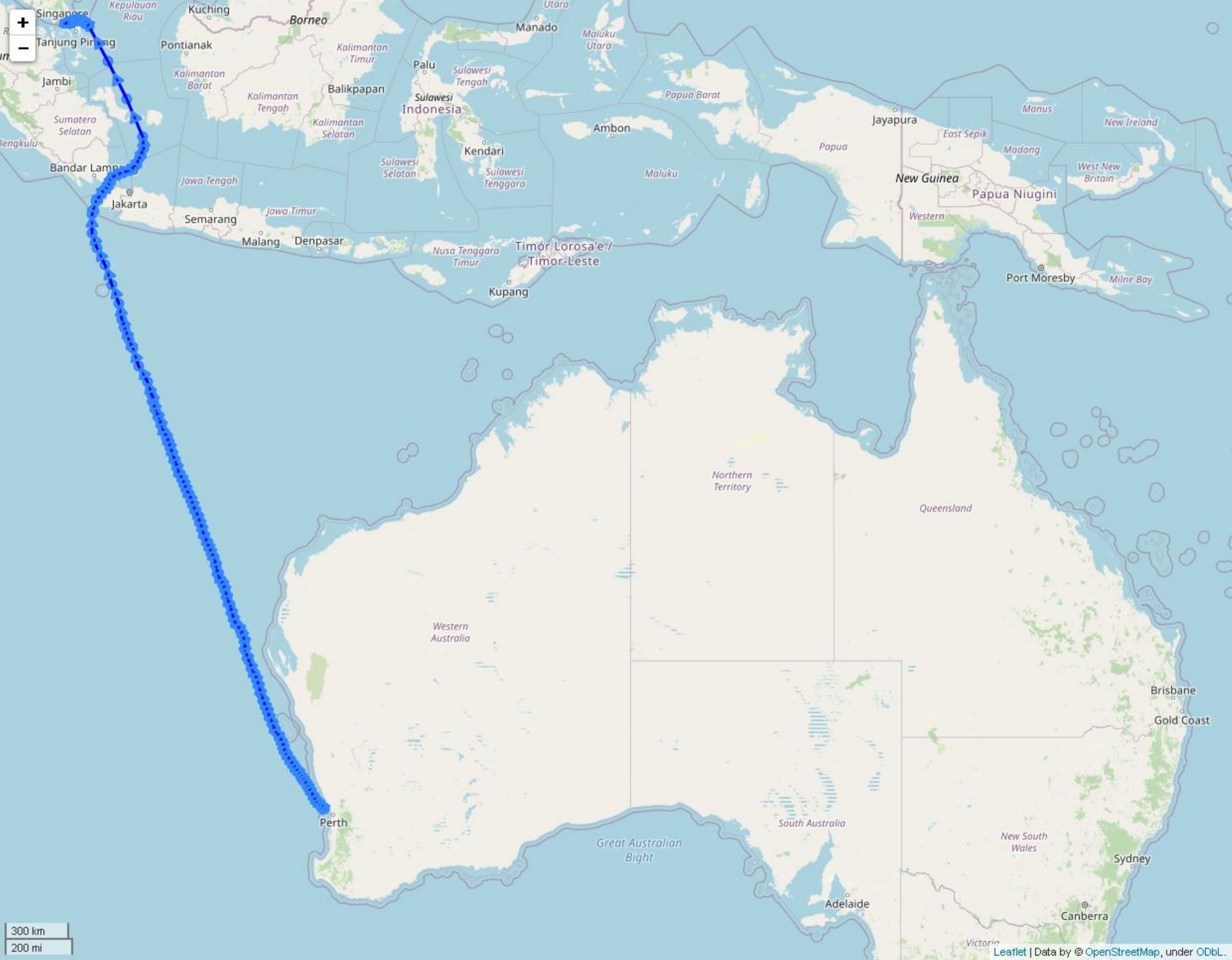}
			\caption{$\eta$=0.1, Lon-Scan}
		\end{subfigure}%
		\begin{subfigure}{.3\textwidth}
			\centering
			\includegraphics[width=0.98\linewidth]{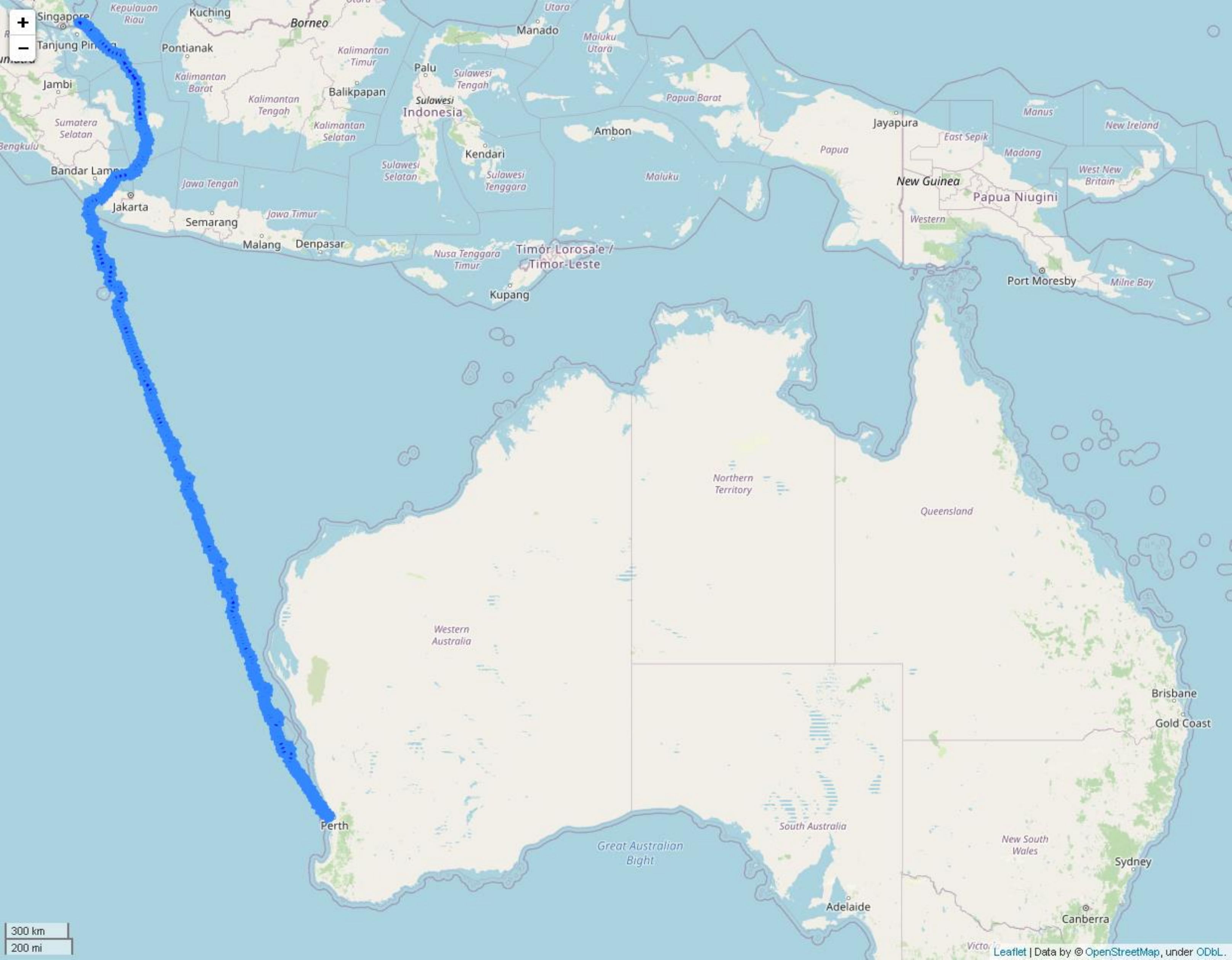}
			\caption{$\eta$=0.1, LatLon-Scan}
		\end{subfigure}

		\begin{subfigure}{.3\textwidth}
			\centering
			\includegraphics[width=0.98\linewidth]{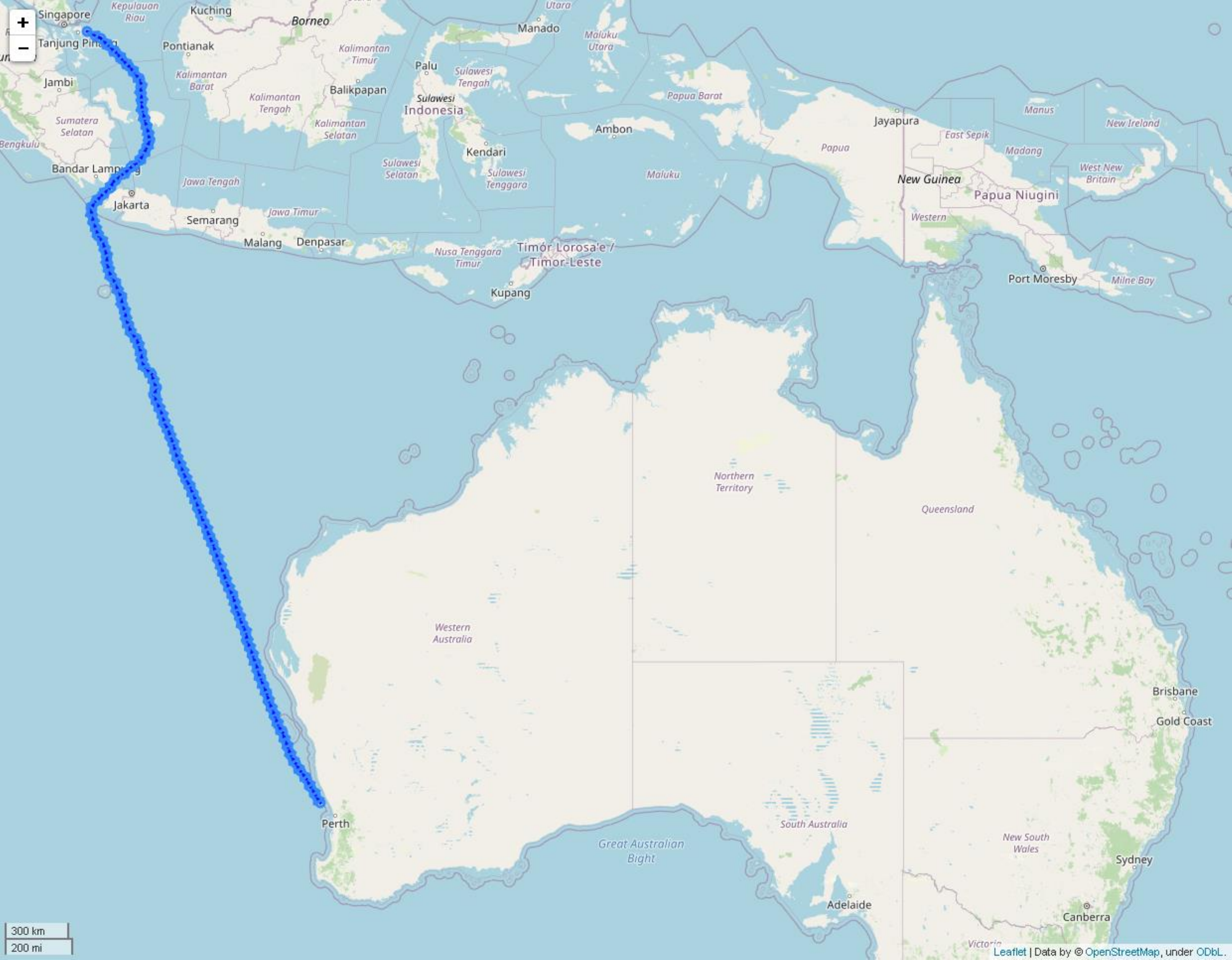}
			\caption{$\eta$=0.3, Lat-Scan}
		\end{subfigure}%
		\begin{subfigure}{.3\textwidth}
			\centering
			\includegraphics[width=0.98\linewidth]{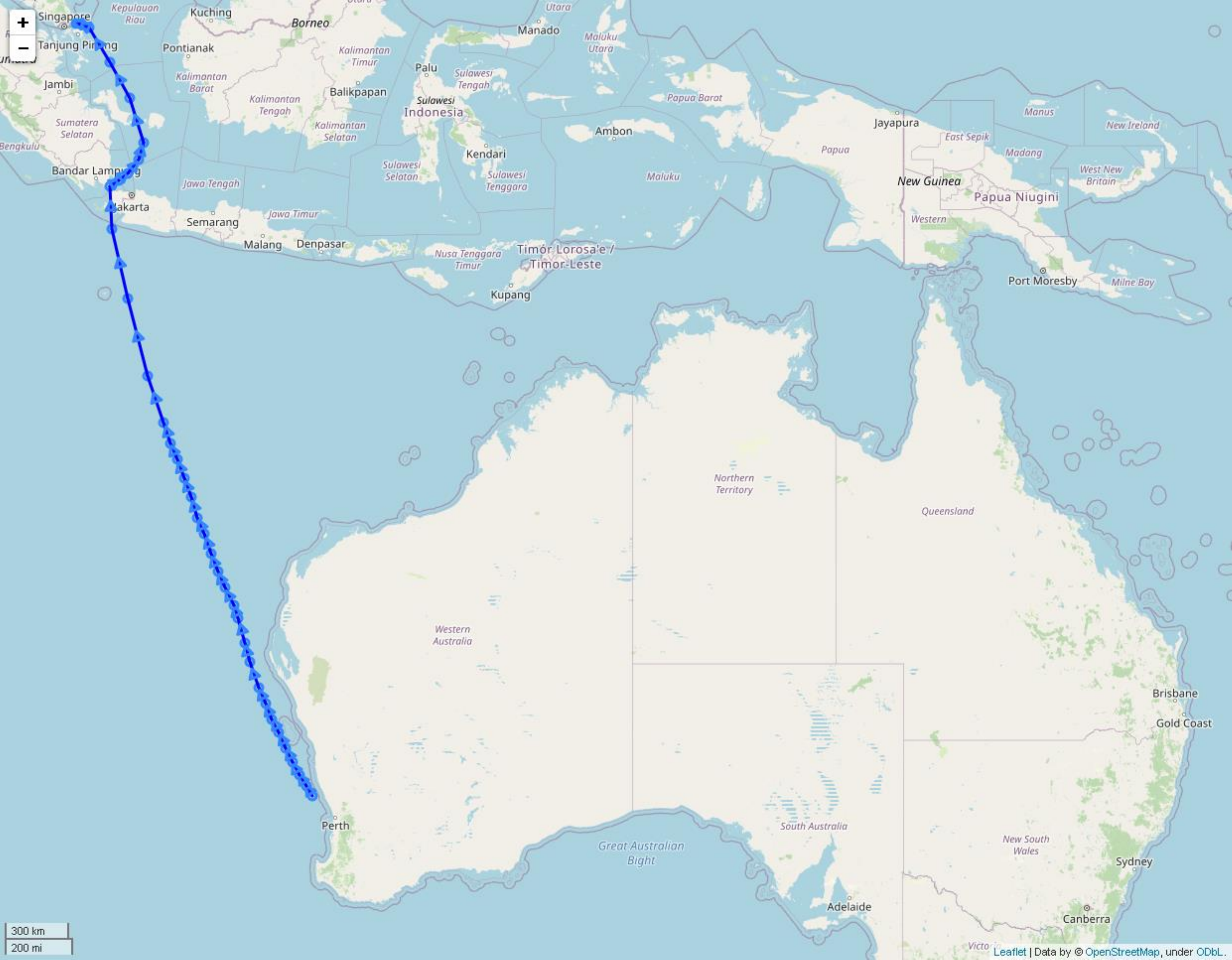}
			\caption{$\eta$=0.3, Lon-Scan}
		\end{subfigure}%
		\begin{subfigure}{.3\textwidth}
			\centering
			\includegraphics[width=0.98\linewidth]{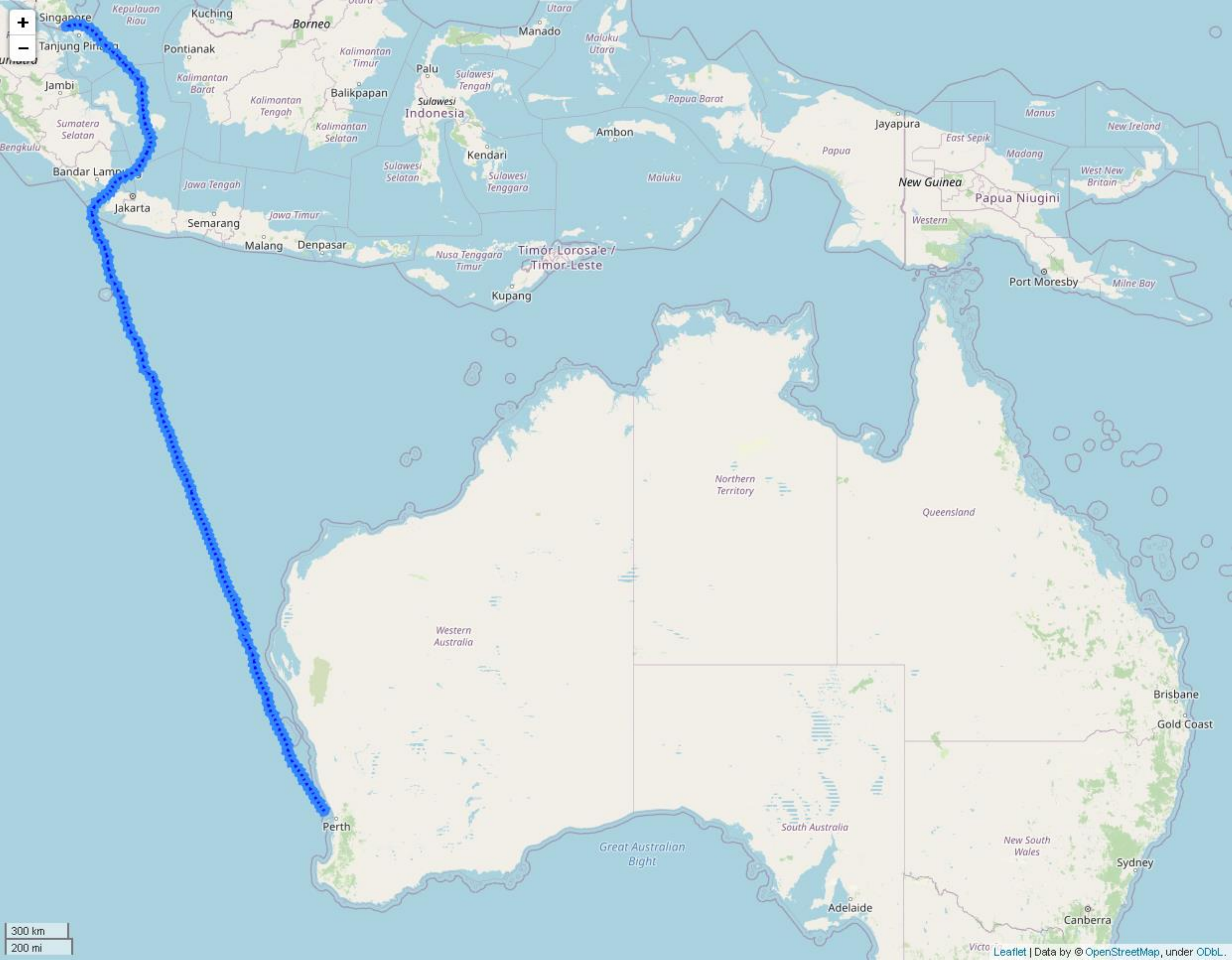}
			\caption{$\eta$=0.3, LatLon-Scan}
		\end{subfigure}

		\begin{subfigure}{.3\textwidth}
			\centering
			\includegraphics[width=0.98\linewidth]{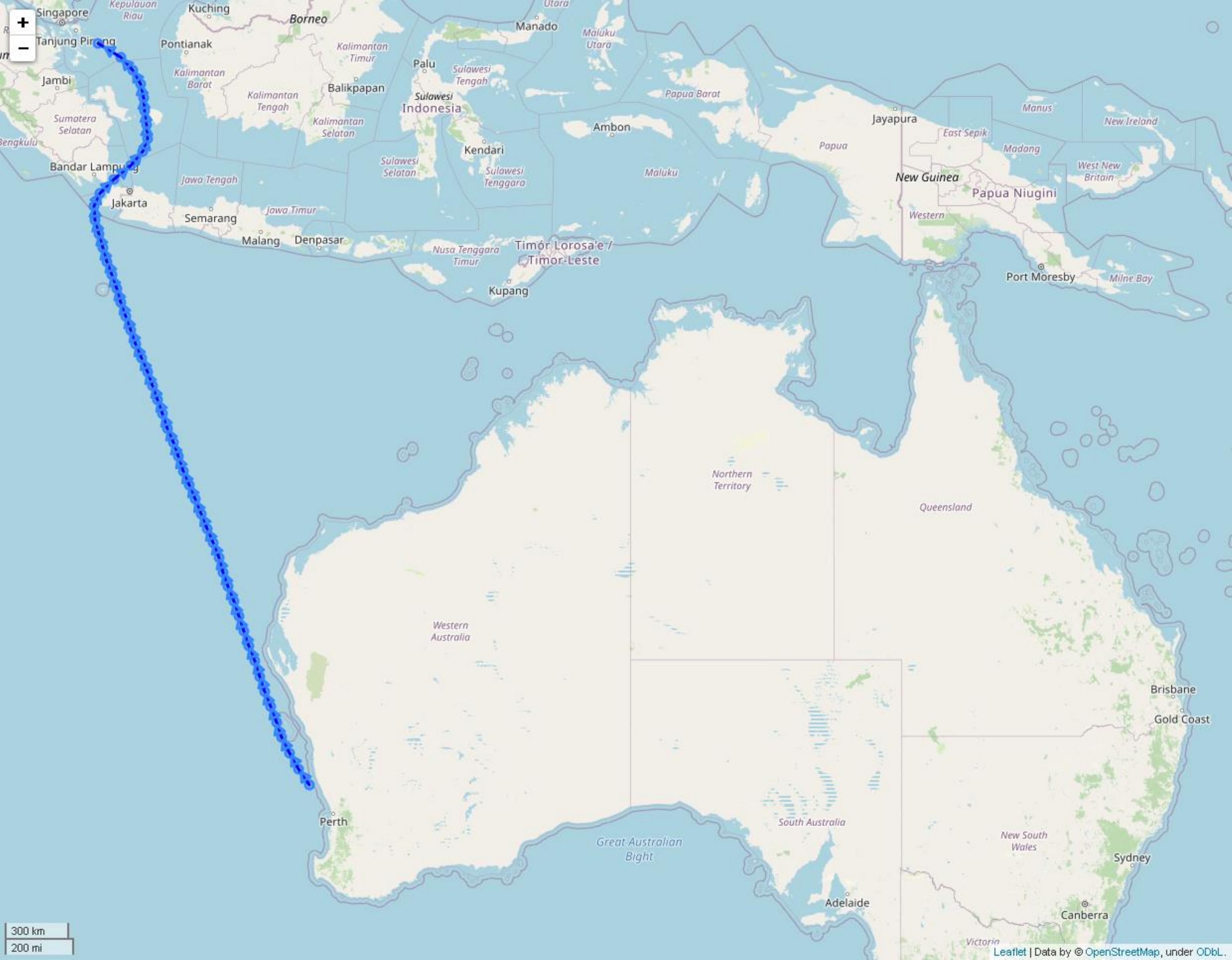}
			\caption{$\eta$=0.6, Lat-Scan}
		\end{subfigure}%
		\begin{subfigure}{.3\textwidth}
			\centering
			\includegraphics[width=0.98\linewidth]{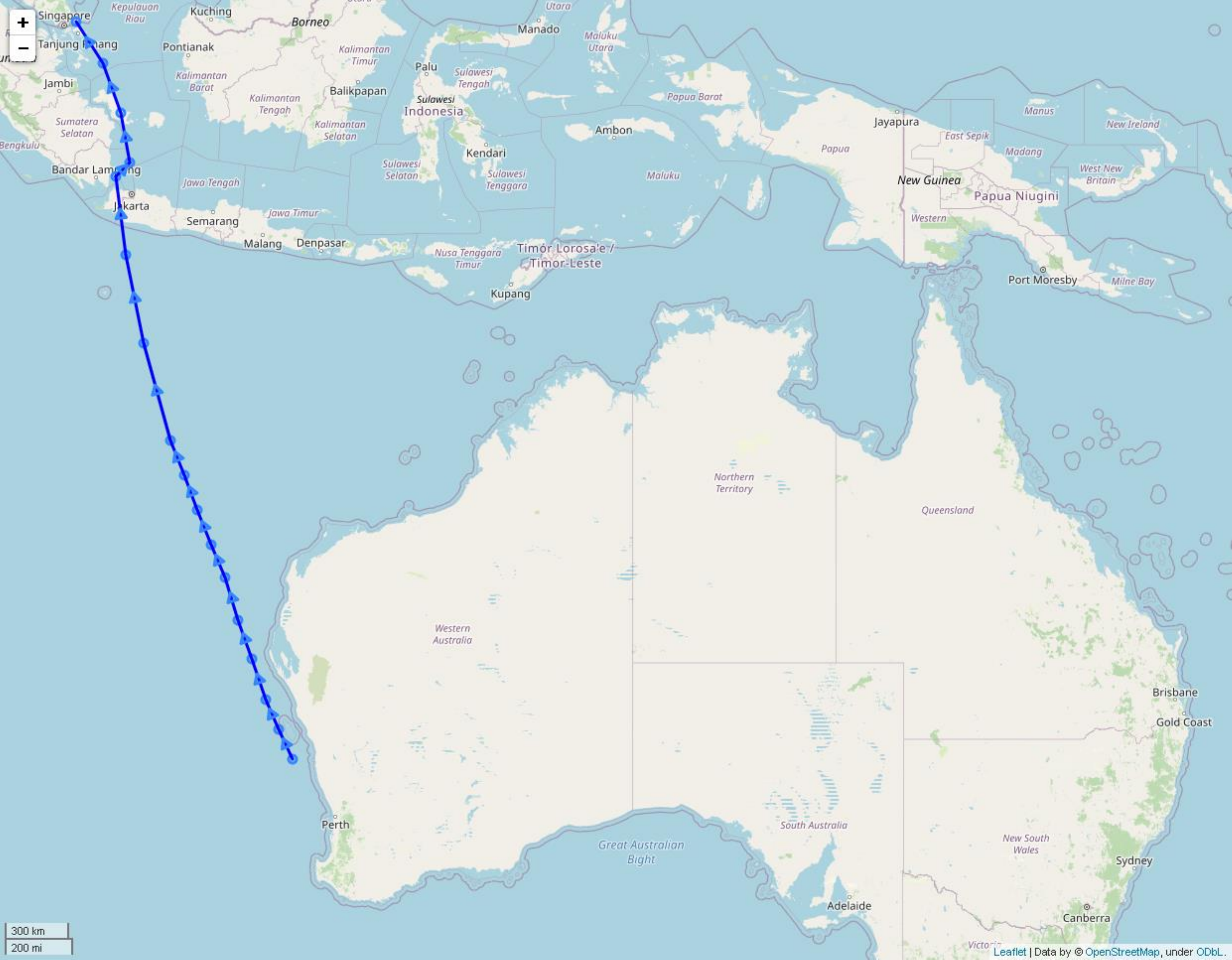}
			\caption{$\eta$=0.6, Lon-Scan}
		\end{subfigure}%
		\begin{subfigure}{.3\textwidth}
			\centering
			\includegraphics[width=0.98\linewidth]{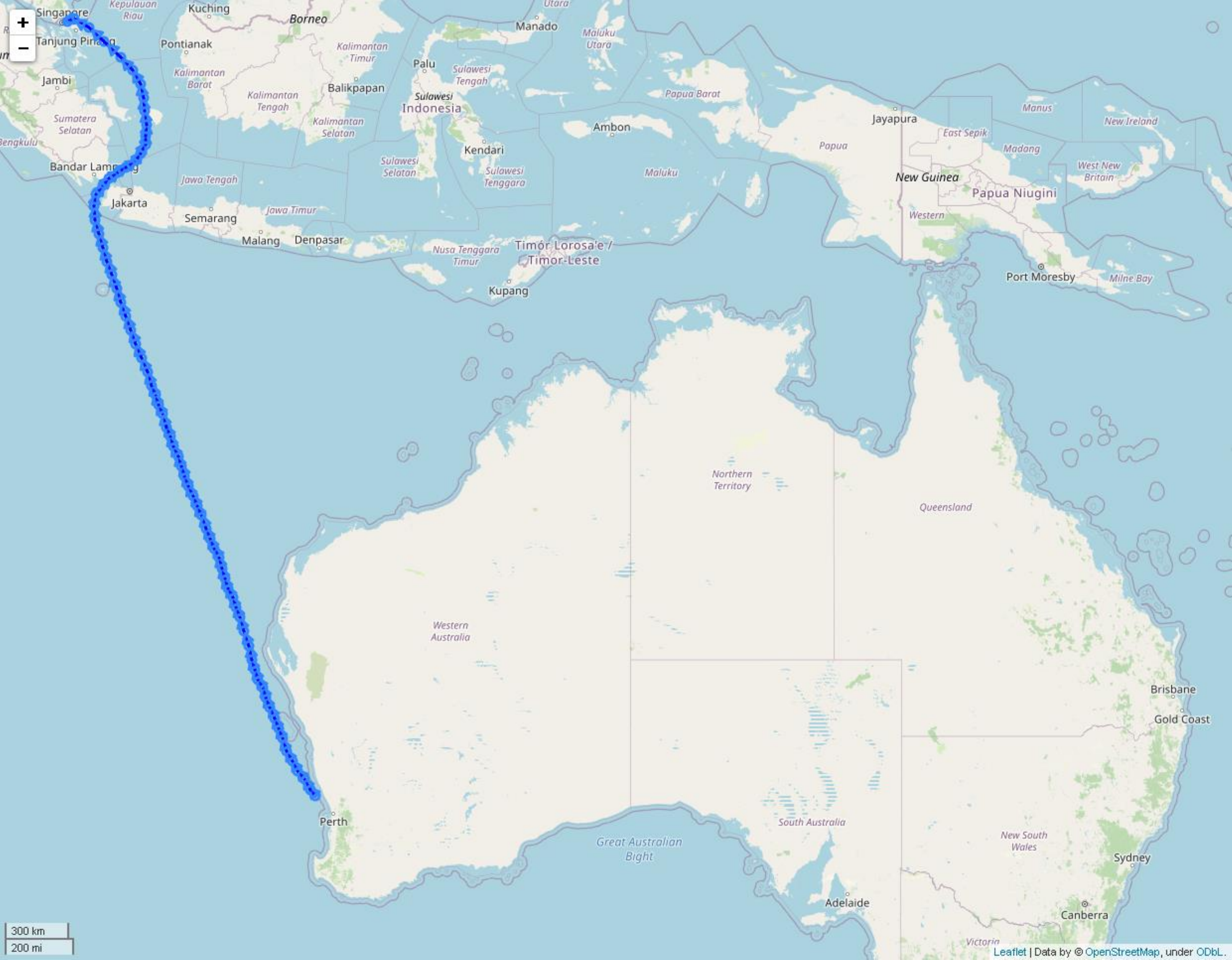}
			\caption{$\eta$=0.6, LatLon-Scan}
		\end{subfigure}

		\begin{subfigure}{.3\textwidth}
			\centering
			\includegraphics[width=0.98\linewidth]{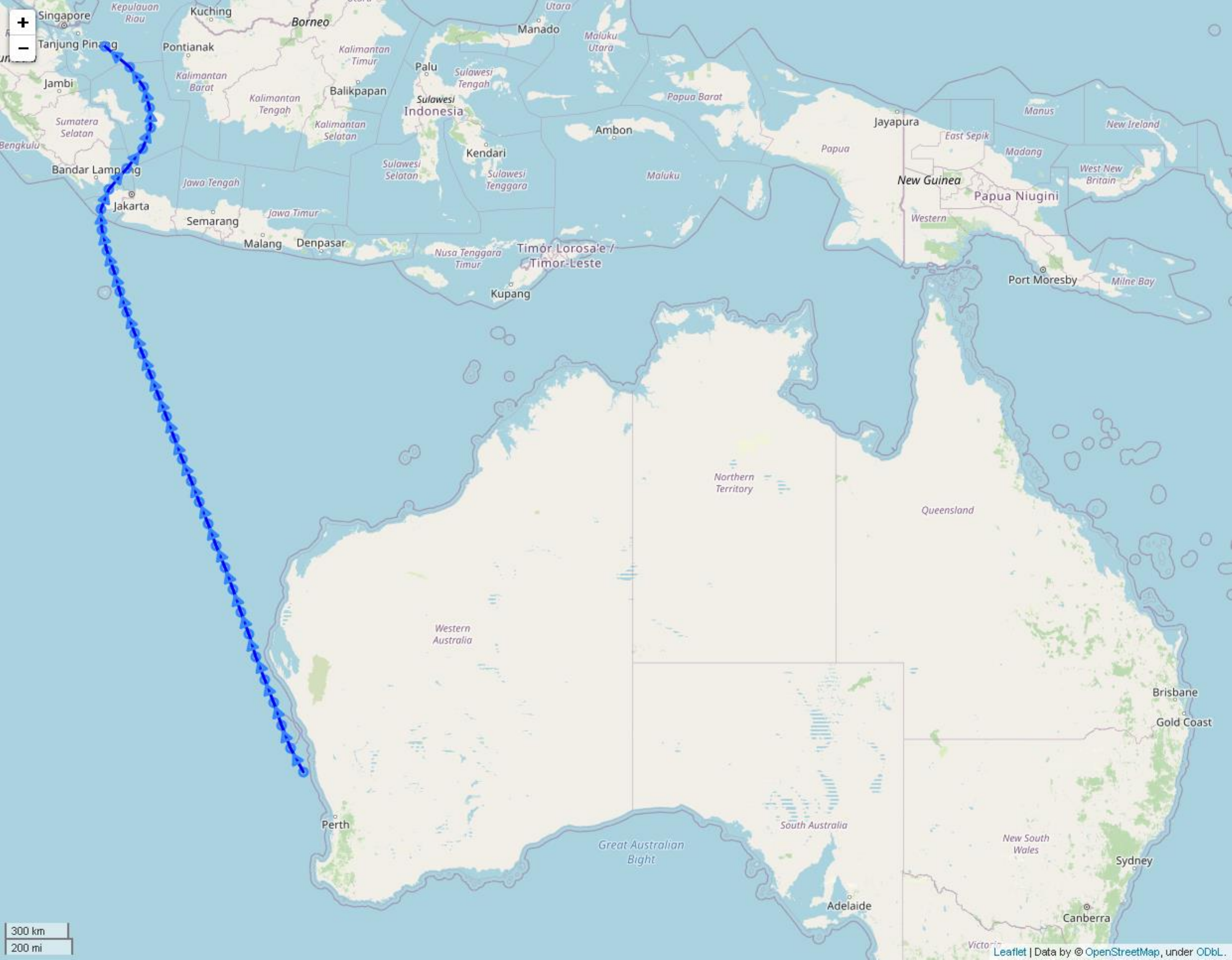}
			\caption{$\eta$=0.9, Lat-Scan}
		\end{subfigure}%
		\begin{subfigure}{.3\textwidth}
			\centering
			\includegraphics[width=0.98\linewidth]{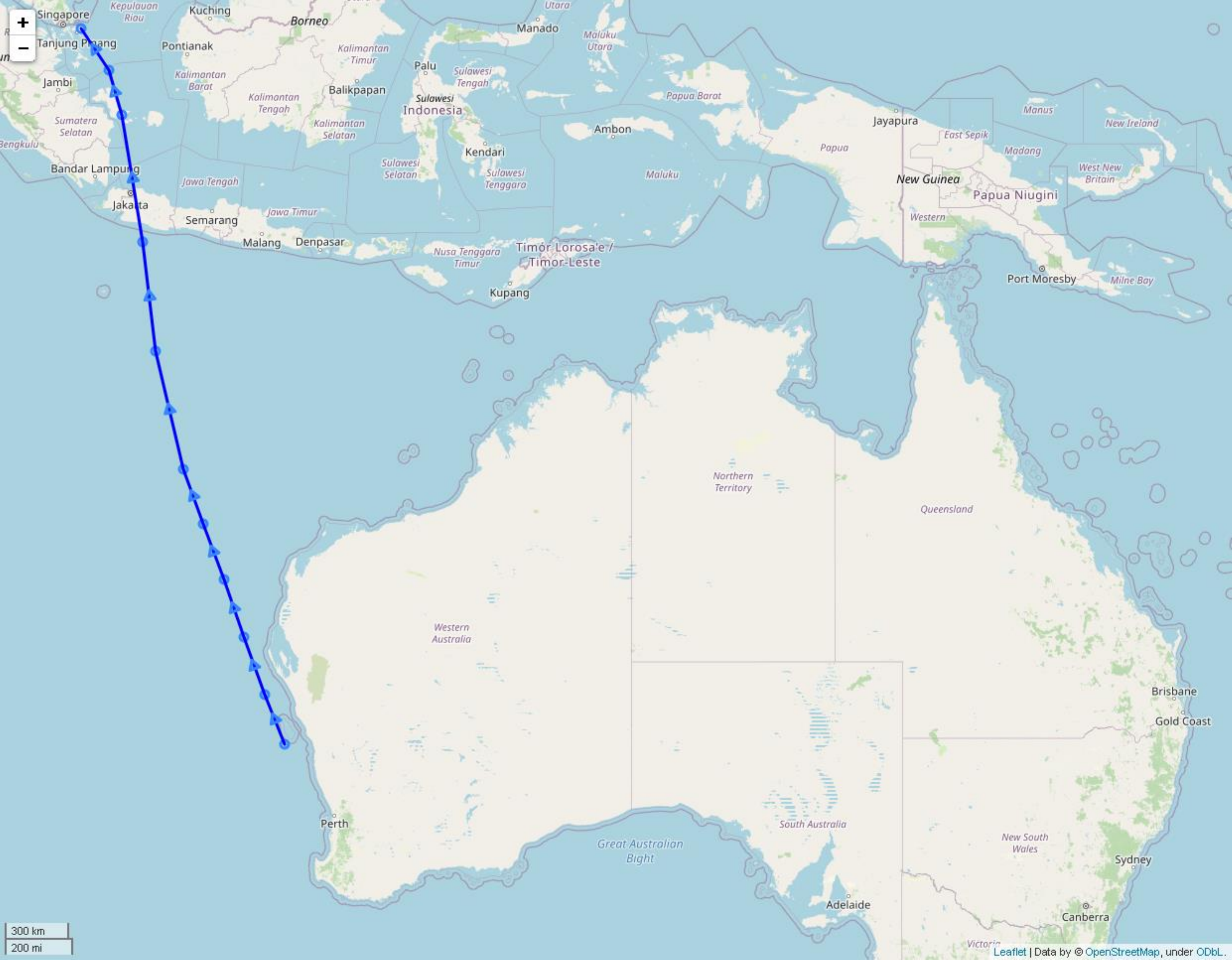}
			\caption{$\eta$=0.9, Lon-Scan}
		\end{subfigure}%
		\begin{subfigure}{.3\textwidth}
			\centering
			\includegraphics[width=0.98\linewidth]{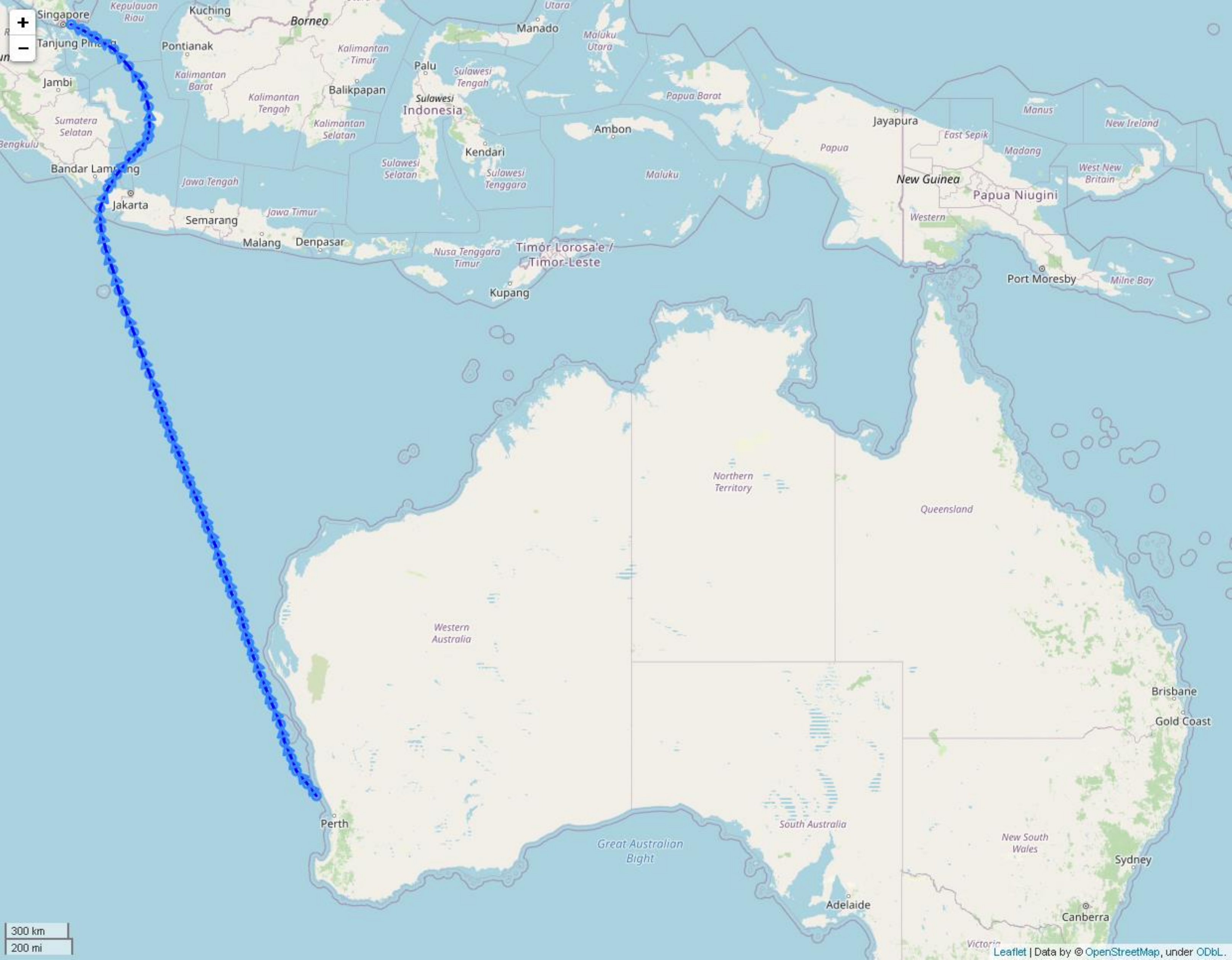}
			\caption{$\eta$=0.9, LatLon-Scan}
		\end{subfigure}

		\begin{subfigure}{.3\textwidth}
			\centering
			\includegraphics[width=0.98\linewidth]{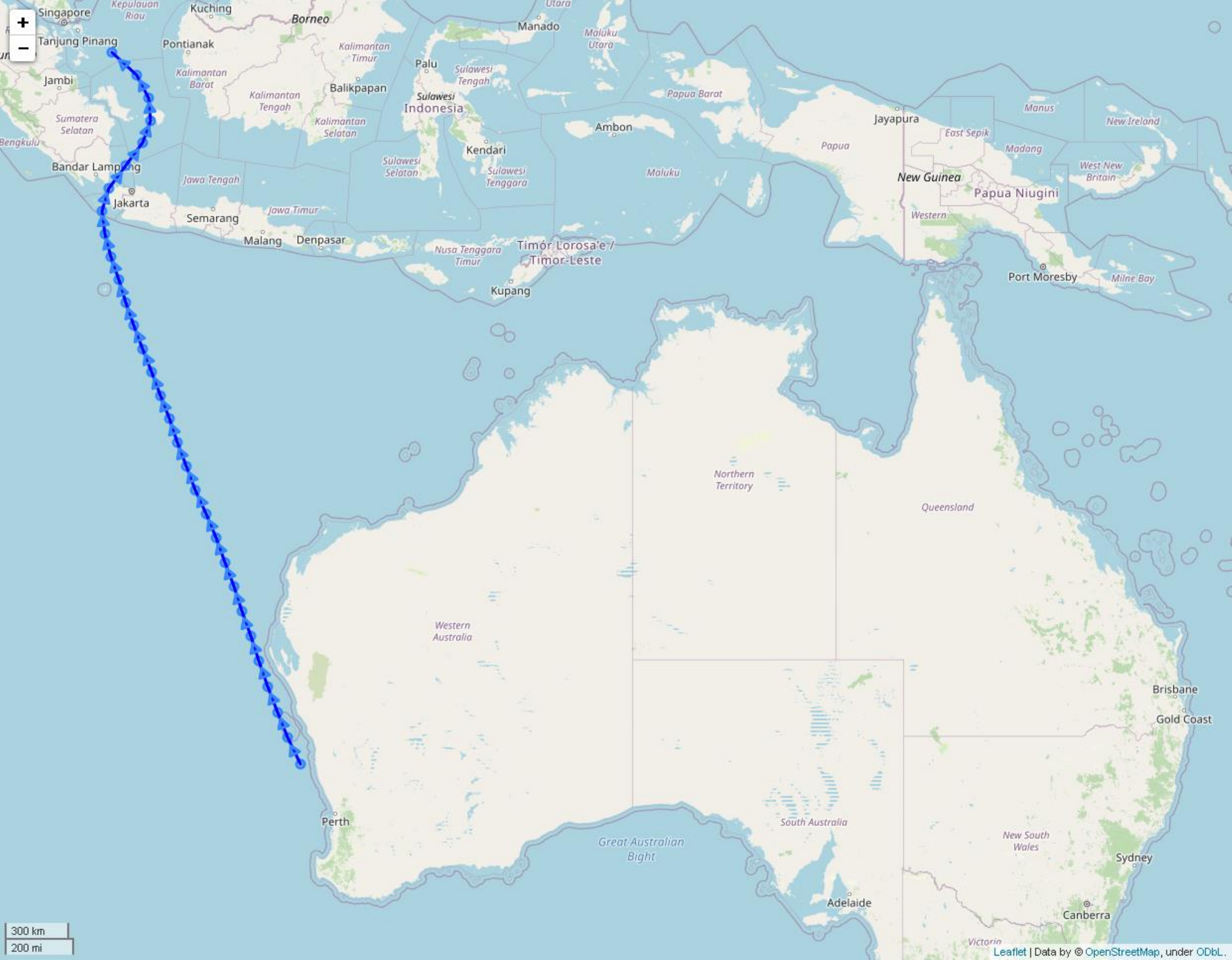}
			\caption{$\eta$=1.0, Lat-Scan}
		\end{subfigure}%
		\begin{subfigure}{.3\textwidth}
			\centering
			\includegraphics[width=0.98\linewidth]{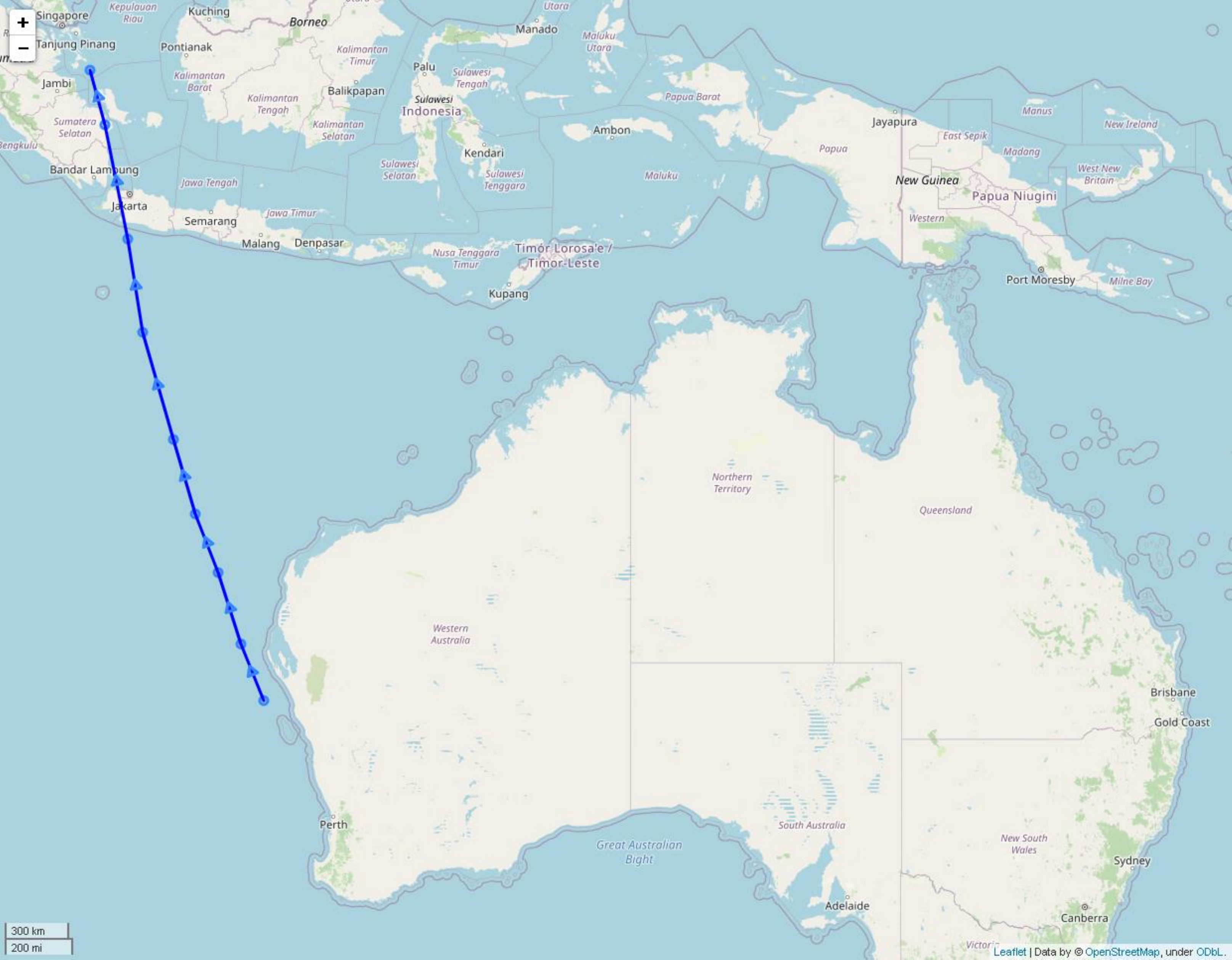}
			\caption{$\eta$=1.0, Lon-Scan}
		\end{subfigure}%
		\begin{subfigure}{.3\textwidth}
			\centering
			\includegraphics[width=0.98\linewidth]{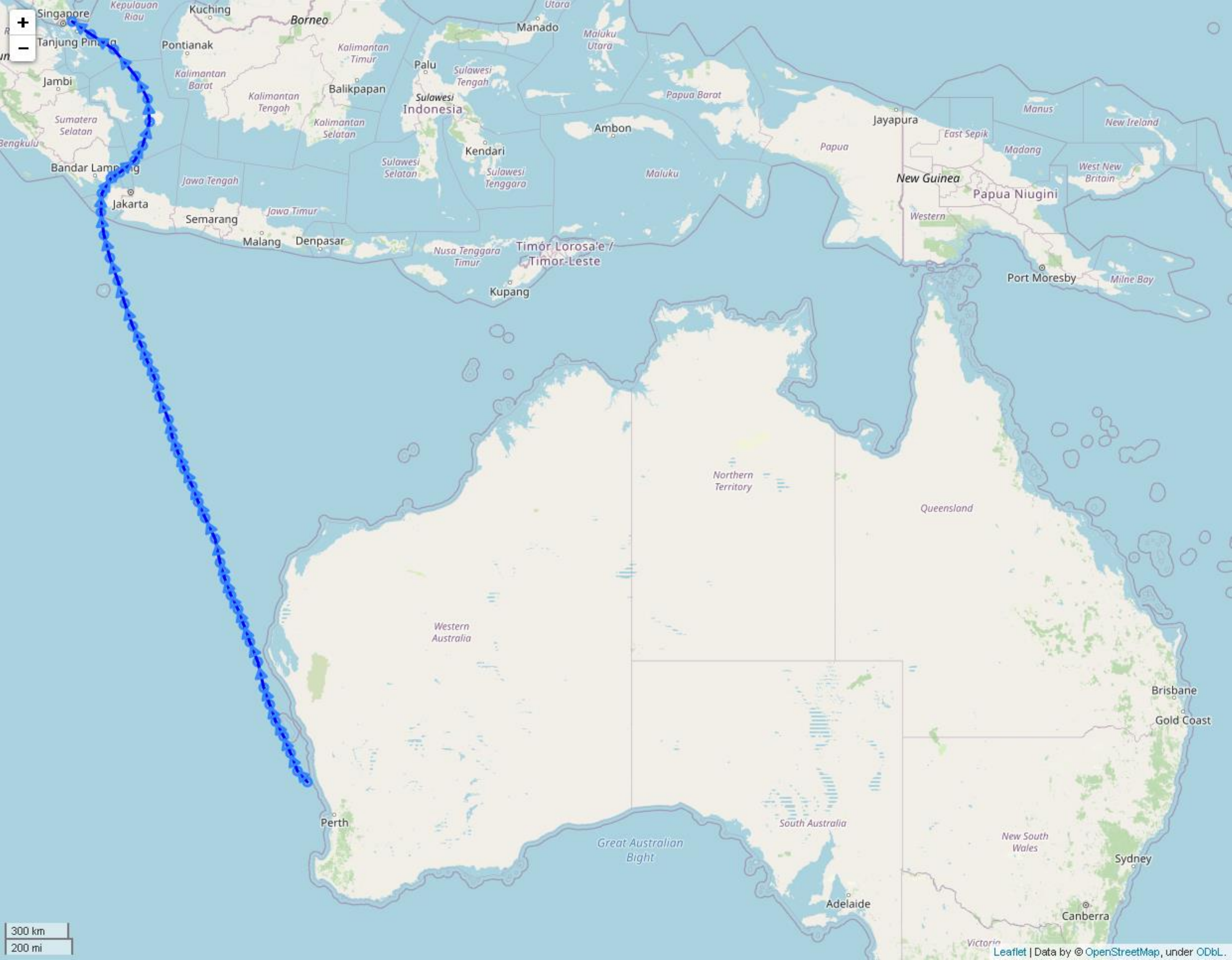}
			\caption{$\eta$=1.0, LatLon-Scan}
		\end{subfigure}

		\caption{Constructed Trajectories (Perth $\rightarrow$ Singapore)}
		\label{Fig: Constructed Trajectories of Perth to Singapore}
	\end{figure}

\newpage
Here are the historical (Figure \ref{Fig: Actual Trajectories of Perth to Singapore}) and constructed (Figure \ref{Fig: Constructed Trajectories of Perth to Singapore}) trajectories by Lat-scan, Lon-Scan and LatLon-Scan for Perth-Singapore journey with different scanning internals ($\eta$) in degree.

According to the historical trajectories from Perth to Singapore in Figure \ref{Fig: Actual Trajectories of Perth to Singapore}, it is clear to show that there is a direct route. In this case, it could be more straight forward and accurate to predict ETA of vessels.

According to the constructed trajectories from Perth to Singapore in Figure \ref{Fig: Constructed Trajectories of Perth to Singapore} across different scanning internals ($\eta$) and scanning methods, similarly, it presents that there are missing details on passing Indonesia, near Singapore and Perth for Lat-scanning and Lon-scanning. latLon-scanning provides most holistic and detailed movements.
	
\begin{figure}[htbp]
		\begin{subfigure}{0.5\textwidth}
			\centering
			\includegraphics[width=0.98\linewidth]{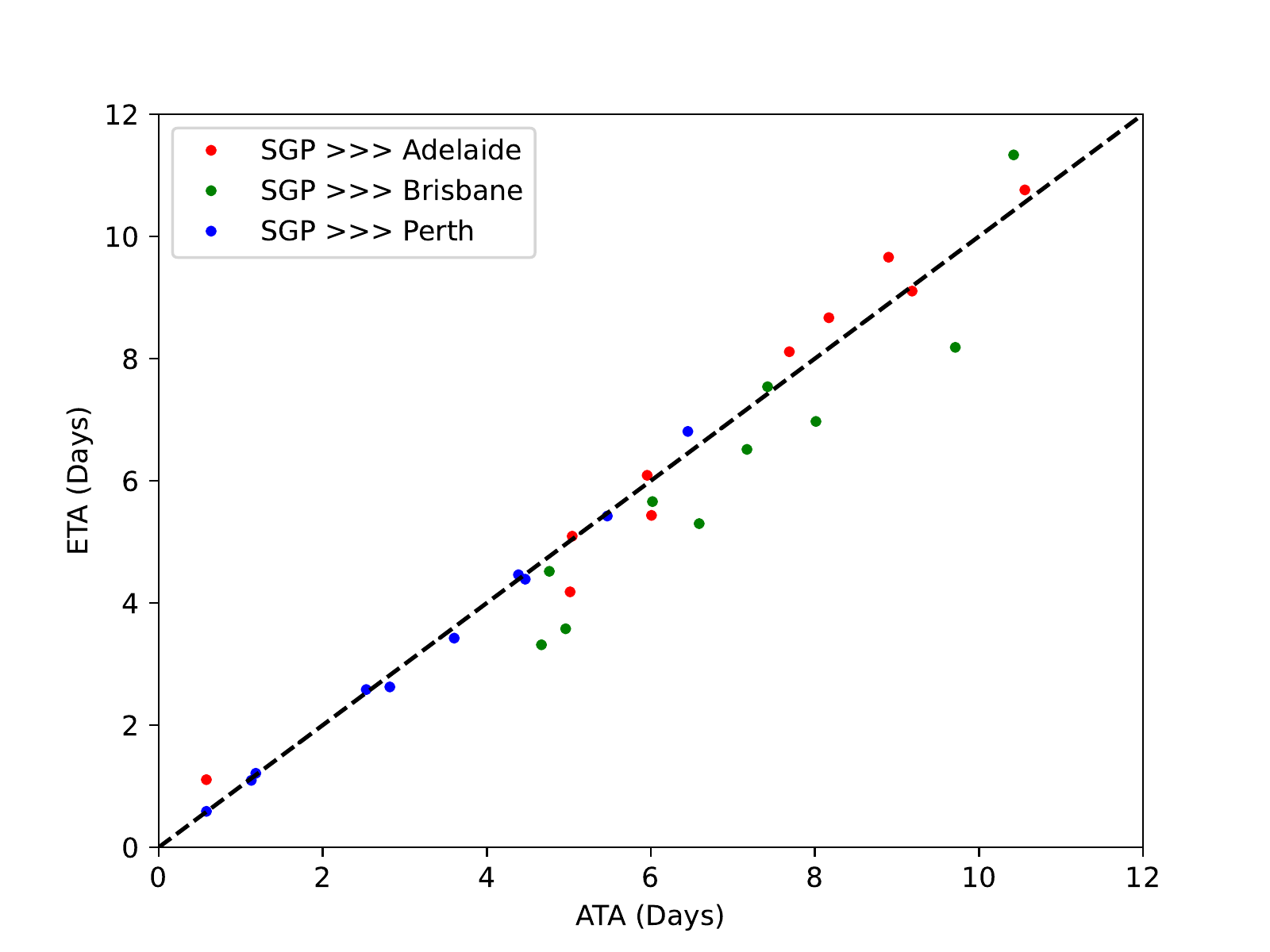}
			\caption{Trajectories (From Singapore Port)}
		\end{subfigure}%
		\begin{subfigure}{0.5\textwidth}
			\centering
			\includegraphics[width=0.98\linewidth]{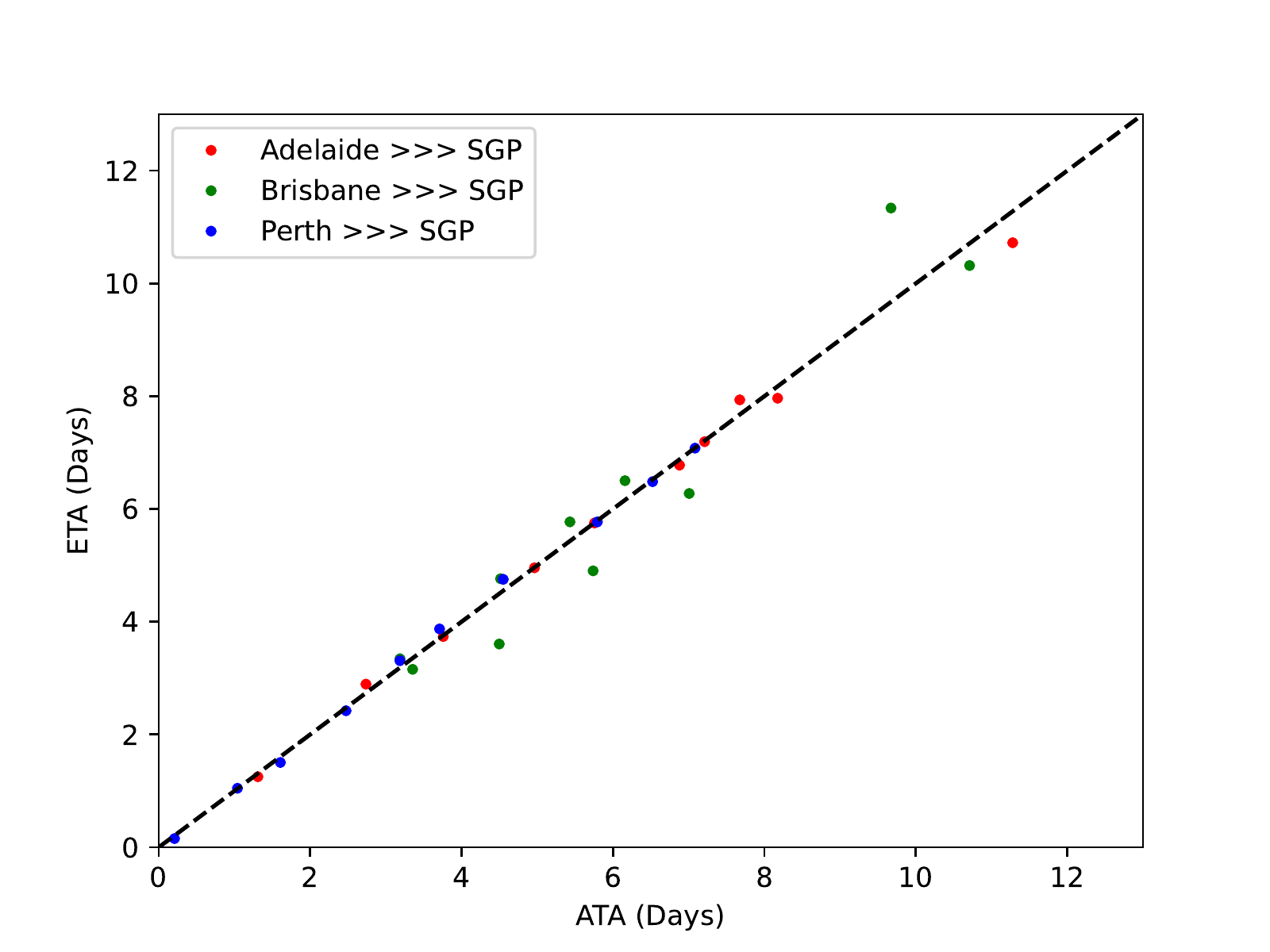}
			\caption{Trajectories (To Singapore Port)}
		\end{subfigure}
		\caption{Visualization of ATA and ETA Predictions}
		\label{Figure: Visualization of ATA and ETA Predictions}
	\end{figure}

\newpage
The evaluation metrics are Mean Absolute Error (MAE), Mean Squared Error (MSE), Root Mean Squared Error (RMSE), Mean Absolute Percentage Error (MAPE), Accuracy (ACC), R-Squared, Mean of Error ($\boldsymbol{\mu_{e}}$) and Standard Deviation of Error ($\boldsymbol{\sigma_{e}}$) according to actual time of arrival (ATA) and ETA predicted. The evaluations of ETA prediction experimental results are tabulated as follows in Table \ref{Table: Evaluations of ETA Prediction Experimental Results}. The evaluations of ATA and ETA are also visualized in Figure \ref{Figure: Visualization of ATA and ETA Predictions}. 

	\begin{table}[htbp]
		\resizebox{\linewidth}{!}{%
		\begin{tabular}{l|cccrcccc}
			\hline
			\textbf{Trajectory} & \textbf{MAE} & \textbf{MSE} & \textbf{RMSE} & \textbf{MAPE} & \textbf{ACC} & \textbf{R-Squared} & $\boldsymbol{\mu_{e}}$ & $\boldsymbol{\sigma_{e}}$ \\
			\hline
			SGP-Adelaide        & 0.409   & 0.238  & 0.488   & 14.31\%       & 85.69\%      & 0.967        & 0.111    & 0.475       \\
			SGP-Brisbane        & 0.889   & 1.029  & 1.015   & 13.58\%       & 86.42\%      & 0.714        & 0.683    & 0.750       \\
			SGP-Perth           & 0.105   & 0.021  & 0.147   & 3.02\%        & 96.98\%      & 0.994        & 0.001    & 0.147       \\
			\textbf{From SGP}   & 0.467   & 0.430  & 0.655   & 10.30\%       & 89.70\%      & 0.944        & 0.191    & 0.627       \\
			\hline
			Adelaide-SGP        & 0.139   & 0.046  & 0.215   & 2.36\%        & 97.64\%      & 0.994        & 0.056    & 0.208       \\
			Brisbane-SGP        & 0.580   & 0.531  & 0.729   & 9.39\%        & 90.61\%      & 0.906        & 0.030    & 0.728       \\
			Perth-SGP           & 0.077   & 0.010  & 0.100   & 4.88\%        & 95.12\%      & 0.998        & 0.022    & 0.098       \\
			\textbf{To SGP}     & 0.266   & 0.196  & 0.443   & 5.54\%        & 94.46\%      & 0.973        & 0.021    & 0.442       \\
			\hline
			\textbf{OVERALL}    & 0.366   & 0.313  & 0.559   & 7.92\%        & 92.08\%      & 0.959        & 0.106    & 0.549       \\
			\hline
		\end{tabular}}
		\newline\newline\footnotesize\text{$\boldsymbol{\mu_{e}}$: Mean of ETA prediction errors in days.} \\
		\footnotesize\text{$\boldsymbol{\sigma_{e}}$: Standard deviation of ETA prediction errors in days.}
		\caption{Evaluations of ETA Prediction Experimental Results}
		\label{Table: Evaluations of ETA Prediction Experimental Results}
	\end{table}

From Table \ref{Table: Evaluations of ETA Prediction Experimental Results} and Figure \ref{Figure: Visualization of ATA and ETA Predictions}, the ETA predictions of trajectories from Singapore port to Australia ports are slightly less accurate than those of trajectories from Australia ports to Singapore port. A direct finding could be one of reasons to this, which is that the trajectories are more divergent from Singapore port to Australia ports according to illustration in Figures (\ref{Fig: Actual Trajectories of Singapore to Adelaide}, \ref{Fig: Actual Trajectories of Singapore to Brisbane} and \ref{Fig: Actual Trajectories of Singapore to Perth}) and Figures (\ref{Fig: Actual Trajectories of Adelaide to Singapore}, \ref{Fig: Actual Trajectories of Brisbane to Singapore} and \ref{Fig: Actual Trajectories of Perth to Singapore}). This inconsistency of trajectories results in the constructed trajectories being different from each particular historical trajectory. 

Numerically, the ETA predictions have 89.70\% and 94.46\% in accuracy with 0.944 and 0.973 R-Squared values for the trajectories ``from Singapore'' and ``to Singapore'', respectively. While the overall ETA prediction errors are about 0.106 days (i.e. 2.544 hours) on average with 0.549 days (i.e. 13.176 hours) standard deviation, and the proposed approach has an accuracy of 92.08\% with 0.959 R-Squared value for overall trajectories without differentiating origin and destination. Comparison among different trajectories ``from Singapore'' shows that Perth (96.98\%) as destination has the highest accuracy of ETA prediction, followed by Brisbane (86.42\%) and Adelaide (85.69\%). Similarly, the ETA prediction on trajectories ``to Singapore'' demonstrates that Adelaide (97.64\%) as origin has the highest accuracy, followed by Perth (95.12\%) and Brisbane (90.61\%).

\section{Conclusion}
In this study, a probability density-based approach for constructing trajectories is proposed and validated through an typical use-case application: Estimated Time of Arrival (ETA) prediction given origin-destination pairs. The ETA prediction is based on physics and mathematical laws given by the extracted information of probability density-based trajectories constructed. The overall ETA prediction errors are about 0.106 days (i.e. 2.544 hours) on average with 0.549 days (i.e. 13.176 hours) standard deviation, and the proposed approach has an accuracy of 92.08\% with 0.959 R-Squared value for overall trajectories selected between Singapore and Australia ports. Due to the nature of physics and mathematical laws, the computational cost of ETA prediction is extremely low and real-time response, compared with big-data approaches using AI/ML. However, the limitations of proposed approach are also considered and listed as follows:

	\begin{itemize}
		\item[(a)] Some origin-destination pairs have no direct route. Therefore, stops in between can introduce time gaps for ETA predictions. In this study, time gaps for stops are excluded before experimentation, so that the ETA predictions are mainly equivalent to consider travelling time solely.
		\item[(b)] The ETA prediction is calculated by current position and speed with the information of constructed trajectories. However, if there are unpredictable changes in routes due to bad weather or disasters etc., then ETA prediction may not be feasible to achieve with the constructed trajectories.  
	\end{itemize}

Last but not least, future direction of follow-ups on this study could be evaluating and comparing different approaches on constructing trajectories and ETA predictions, such as physics-mathematical based and big data AI/ML based approaches, in terms of accuracy, feasibility, real-time response capability, computational cost, etc.

\section*{Acknowledgment}
The authors gratefully thank for all who provide kind reviews, constructive comments and suggestions.

\newpage
\bibliography{mybibfile}

\end{document}